\documentclass[prl,amsmath,amssymb, notitlepage, twocolumn,
nofootinbib,
superscriptaddress,
longbibliography
]{revtex4-1}

\usepackage{amsmath}
\usepackage{tabularx,graphicx}
\usepackage{epstopdf}
\usepackage{graphicx}
\usepackage{latexsym}
\usepackage{amssymb}
\usepackage{amsmath}
\usepackage{color, colortbl}
\usepackage{psfrag}
\usepackage{bbm}
\usepackage{bm}
\usepackage{titlesec}
\usepackage{dsfont}
\usepackage{feynmp}
\usepackage{slashed}
\usepackage{multirow}
\usepackage{url}

\newcommand{\mc}[1]{\mathcal{#1}}

\newcommand{\beq}{\begin{eqnarray}}
	\newcommand{\eeq}{\end{eqnarray}}

\newcommand{\bsp}{\begin{split}}
	\newcommand{\esp}{\end{split}}

\newcommand{\ie}{{i.e., }}
\newcommand{\eg}{{e.g., }}

\usepackage{color}
\definecolor{darkblue}{rgb}{0.,0.,0.4}
\definecolor{darkred}{rgb}{0.5,0.,0.}
\definecolor{BlueViolet}{RGB}{138,43,226}
\definecolor{SkyBlue}{RGB}{30,144,255}
\definecolor{DarkGreen}{RGB}{0,100,0}
\usepackage[pdftex,colorlinks=true,linkcolor=darkblue,citecolor=blue,urlcolor=darkred]{hyperref}
\usepackage[normalem]{ulem}

\renewcommand{\vec}[1]{\bm{#1}}

\usepackage{mathrsfs}
\usepackage{subcaption}
\usepackage{comment}

\begin{document}

\title{
UV/IR Mixing in Marginal Fermi Liquids  
}
	
\author{Weicheng Ye}
\affiliation{Perimeter Institute for Theoretical Physics, Waterloo, Ontario, Canada N2L 2Y5}
\affiliation{Department of Physics and Astronomy, University of Waterloo, Waterloo, Ontario, Canada N2L 3G1}

\author{Sung-Sik Lee}
\affiliation{Perimeter Institute for Theoretical Physics, Waterloo, Ontario, Canada N2L 2Y5}
\affiliation{Department of Physics and Astronomy, McMaster University, Hamilton, Ontario, Canada L8S 4M1}
	
\author{Liujun Zou}
\affiliation{Perimeter Institute for Theoretical Physics, Waterloo, Ontario, Canada N2L 2Y5}

\begin{abstract}

When Fermi surfaces (FS) are subject to long-range interactions that are marginal in the renormalization-group sense, 
Landau Fermi liquids  are destroyed,  but only barely.
With the interaction further screened by 
particle-hole excitations through one-loop  quantum corrections,
it has been believed that
these marginal Fermi liquids (MFLs) are described by weakly coupled field theories at low energies.
In this paper, we point out a possibility in which higher-loop processes qualitatively change the picture through UV/IR mixing, in which the size of FS enters as a relevant scale.
The UV/IR mixing effect enhances the coupling at low energies, such that the basin of attraction for the weakly coupled fixed point of a (2+1)-dimemsional MFL shrinks to a measure-zero set in the low-energy limit.
This UV/IR mixing is caused by gapless
virtual Cooper pairs 
that spread over the
entire FS
through marginal long-range interactions.
Our finding signals a possible breakdown of the patch description for the MFL, and questions the validity of using the MFL as the base theory  in a controlled scheme for non-Fermi liquids that arise from relevant long-range interactions.

\end{abstract}

\maketitle

{\it Introduction.} 
Non-Fermi liquids (nFLs) arise ubiquitously when Fermi surfaces (FS) are coupled to gapless collective modes that mediate long-range interactions.
The physics of nFLs is central to  the strange metallic behavior and/or unconventional superconductivity in various systems, including cuprates, heavy-fermion compounds, half-filled Landau level
and pnictides \cite{Senthil2006, Lee2017, Berg2018}. However, understanding nFLs 
has been a long-standing challenge due to
strong quantum fluctuations 
amplified by abundant gapless modes near FS \cite{
Polchinski1992,
ALTSHULER,
Kim1995,
ABANOV2,
Lee2009,
Metlitski2010}.

Under renormalization group (RG) flow, most theories for (2+1)-dimensional nFLs flow to the strong-coupling regime at low energies,
and non-perturbative methods are required to understand their universal long-distance physics \cite{Sur2013,PhysRevX.7.021010}.
However, there is a special class of nFLs,
marginal Fermi liquids (MFLs)\footnote{In this paper, the terms ``metal", ``nFL" and ``MFL" may refer to that of either electrons or some emergent fermions, depending on the context.},
where interaction effects are relatively weak,
\ie marginal in the RG sense with logarithmic (log) perturbative corrections.
If the marginal interactions are further screened, the long-distance physics should be captured by weakly coupled theories.
While the MFL was first introduced for cuprates \cite{
VARMALI},
it is relevant in rather broad contexts.
First, metallic states realized at the half-filled Landau level and exotic Mott transitions may be related to MFLs \cite{HLR,Senthil2008}. 
Second,  MFLs have been used as a foothold to gain a controlled access to strongly coupled nFLs in an expansion scheme,
where the exponent with which long-range interactions decay in space is used as a control parameter \cite{Nayak1994,Mross2010}.

\begin{figure}[h]
\centering
\includegraphics[width=0.36\textwidth]{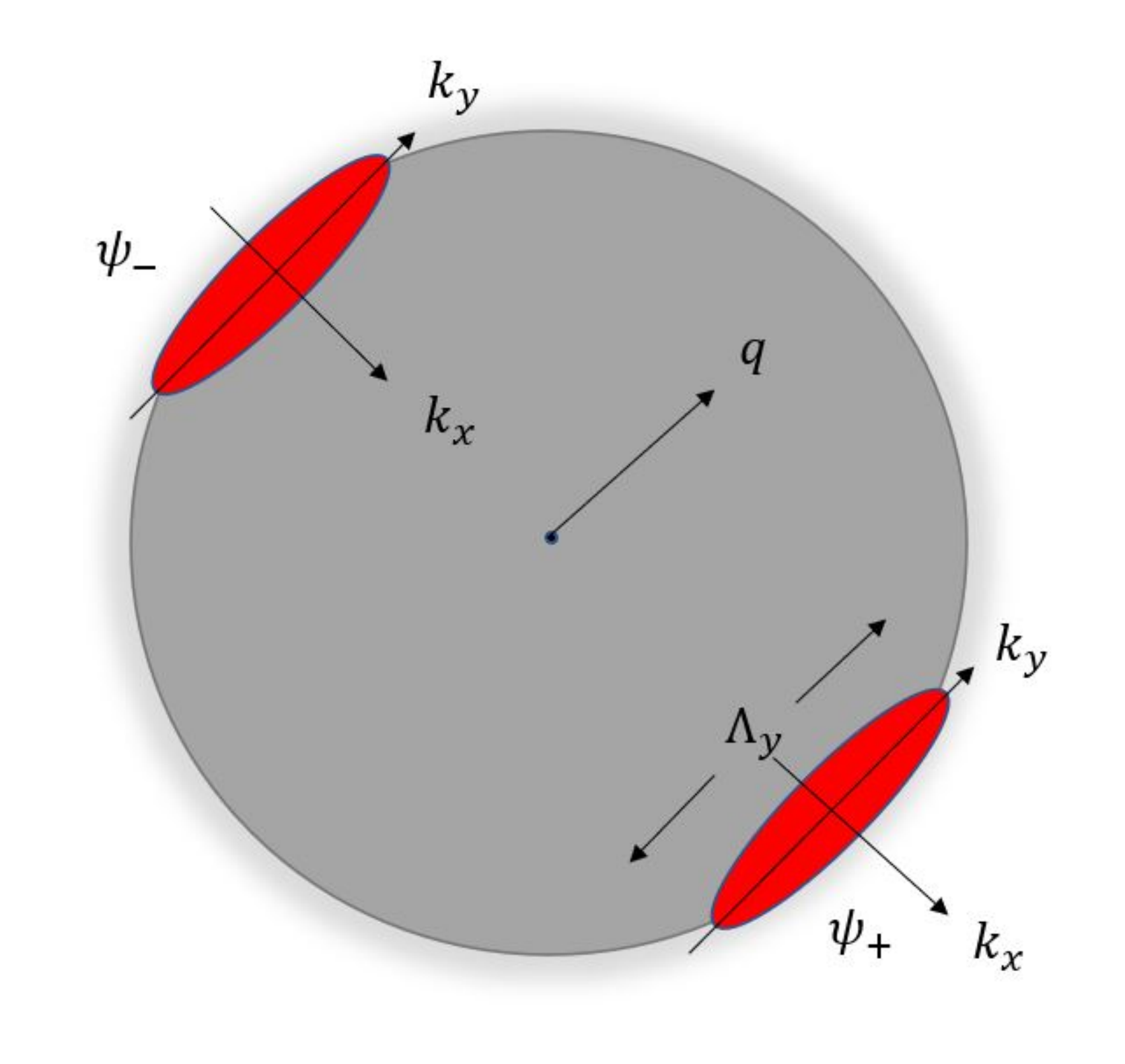}
\caption{
In the patch theory, a FS is partitioned into multiple patches ($\sim k_F/\Lambda_y$ of them, with $k_F$ the Fermi momentum and $\Lambda_y$ the patch size). Ignoring the short-range four-fermion interactions, 
couplings between modes from different patches are weak unless they have almost parallel Fermi velocities.
In this case,
one can focus on a pair of antipodal patches 
that have nearly collinear Fermi velocities.
}
\label{fig: FSpatch}
\end{figure}

In nFLs, fermionic quasiparticles are destroyed by scatterings that are singularly enhanced at small momenta. If large-momentum scatterings are suppressed strongly enough, 
one can understand physical observables that are local in momentum space (\eg the single-particle spectral function) within local patches of FS that include the momentum point of interest (see Fig. \ref{fig: FSpatch}). 
Although this patch description becomes ultimately invalid in the presence of a pairing instability driven by short-range four-fermion couplings,
which is {\it non-local} in momentum space, 
one may hope that the dominant physics in the normal state can be captured within the patch theory without invoking the entire FS. 
So the patch theory \cite{Lee2008, Lee2009, Metlitski2010, Mross2010, Metlitski2014} has been widely used to describe a large class of nFLs (see Refs. \cite{HLR,Senthil2008,Lee1992,Lee2005} for some prominent examples) . 

In MFLs, however, the validity of the patch theory is questionable even before taking into account the short-range four-fermion couplings, because large-momentum scatterings are only marginally suppressed.
If large-momentum scatterings create significant inter-patch couplings, the patch theory fails even for the purpose of describing observables local in momentum space.
In this case, the size of the FS, a UV paramter, qualitatively modifies the IR scaling behavior, showcasing UV/IR mixing.

While such UV/IR mixing does not show up at low orders in the perturbative expansion \cite{Mross2010, Metlitski2014}, 
a systematic understanding of higher-order effects is still lacking.
This issue is also pertinent to multiple experimentally relevant problems.
First, in the context of quantum Hall physics, it is debated whether the Halperin-Lee-Read theory \cite{HLR} and the recently proposed Son's theory \cite{Son2015}
describe the same 
universal physics 
of the composite Fermi liquid (CFL).
Second, in the continuous Mott transitions reported in {\ensuremath{\kappa}}-(ET)$_{2}$Cu$_{2}$(CN)$_{3}$ \cite{Kurosaki2005, Furukawa2015} and moir\'e materials \cite{Li2021, Ghiotto2021},
the observed phenomena appear to be compatible with the predictions of the patch theory \cite{Senthil2008}, but some specific critical properties seem to disagree \cite{Li2021}. 
To resolve these issues, it is crucial to understand the behaviors of the corresponding MFL theories.
Moreover, understanding higher-order effects in nFLs in general may provide new insight on
the nature of quantum phase transitions associated with sudden jumps of the FS size \cite{Onuki2005,Fang2020}, 
and deconfined metallic quantum criticality
\cite{Zou2020, Zou2020a, Zhang2020, Zhang2020a}.

\begin{figure}
     \centering
    \begin{subfigure}{\linewidth}
         \includegraphics[width=0.8\linewidth]{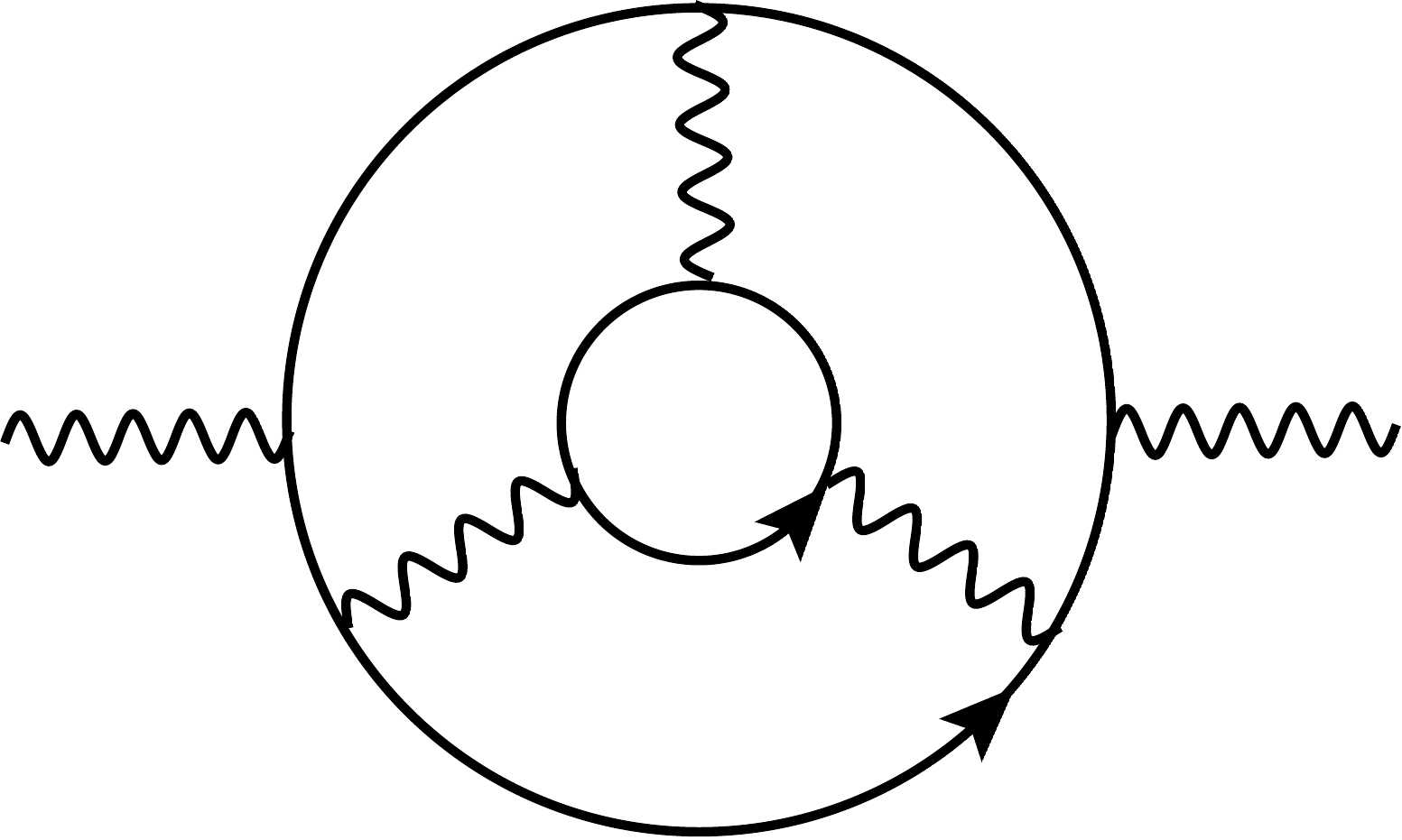}
          \caption{
          }
         \label{FD: Benz}
    \end{subfigure}
    \begin{subfigure}{\linewidth}
         \includegraphics[width=\linewidth]{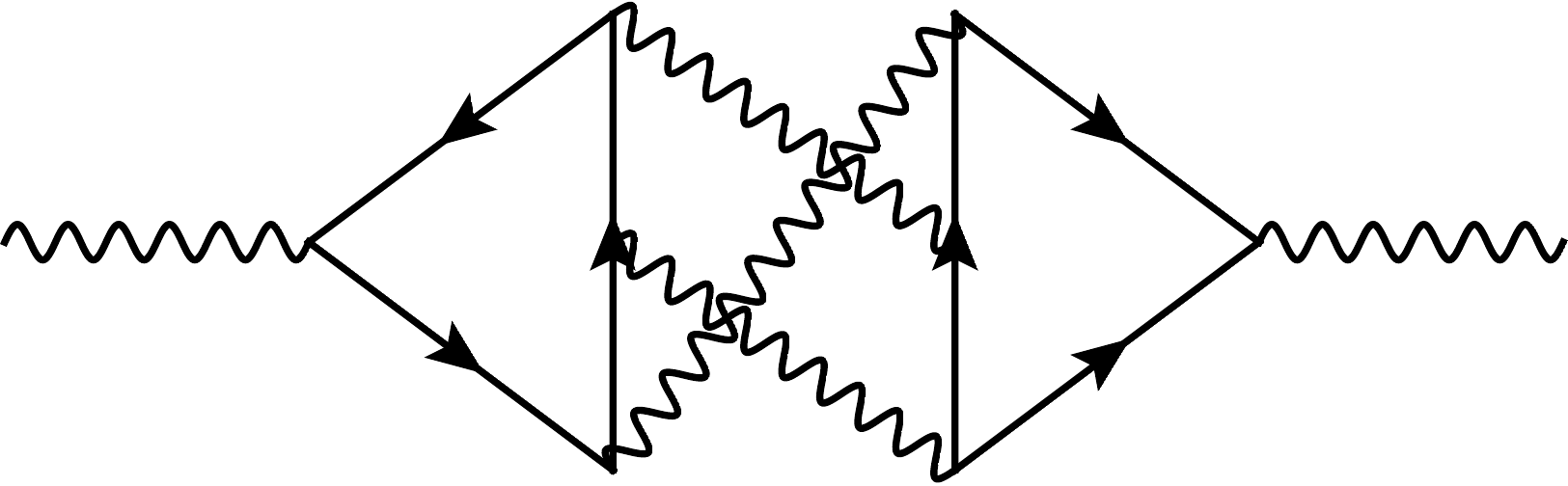}
        %  \hfill
        %  \hfill
          \caption{
          }\label{FD: 3Stringe}
         
    \end{subfigure}
        \caption{Eq. \eqref{eq: double log} comes from these two diagrams, together with their cousins with both internal fermion loops flipped in direction, which are not shown here. Note that only when the two fermion loops in each diagram run in the same direction do they contribute  to double-log divergence. See Sec. III.C.2 in \cite{supp} for all diagrams at this order.}
\label{FD: double log diagrams}
\end{figure}

In this paper, we study the higher-order behaviors of a theory of
$N$ flavors of two-dimensional FS coupled to a dynamical $U(1)$ gauge field, whose kinetic energy scales as $k_y^{1+\epsilon}$.
For the marginal exponent ($\epsilon=0$),
we indeed find potential UV/IR mixing in four-loop processes
that renormalize the gauge coupling (see Fig. \ref{FD: double log diagrams}), with a strength logarithmically singular
in the FS size.
This is caused by gapless {\it virtual} Cooper pairs that manage to explore the entire FS, assisted by large-momentum scatterings that are only marginally suppressed.

{\it Model and regularization scheme.} We denote the low-energy fermion fields near  a pair of antipodal patches by $\psi_{ip}$, 
with $p=\pm$ the patch index and $i=1,...,N$ the flavor index. 
$a$ represents the gauge field  (see Fig.~\ref{fig: FSpatch}). 
Due to the kinematic constraints, the most important interactions occur between fermionic modes and the gauge bosons 
with momenta nearly perpendicular 
to their Fermi velocity \cite{Polchinski1992, Lee2008}. 
The Euclidean action for the patch theory is \cite{ Mross2010}
\beq \label{eq: patch theory}
S = S_\psi + S_{\text{int}} + S_a,
\eeq
where 
\beq
\label{eq: Lagrangian of patch theory}
\begin{split}
S_\psi &= \int[dk] \sum_{i,p}\psi^\dagger_{ip}(k)(-i k_\tau + p k_x + k_y^2)\psi_{ip}(k),\\
S_{\text{int}} &= \int[dk_1][dk_2] \sum_{i,p}\lambda_p a(k_1) \psi^\dagger_{ip}(k_1+k_2) \psi_{ip}(k_2),\\
S_a &= \int[dk] \frac{N}{2e^2} |k_y|^{1+\epsilon} a(-k)a(k),\\
\end{split}
\eeq
with $[dk]=\frac{dk_\tau dk_x dk_y}{(2\pi)^3}$ and $\lambda_\pm=\pm 1$. The reason for the opposite signs of $\lambda_\pm$ is because $a$ couples to the currents of the fermions, and fermions from the two patches have opposite Fermi velocities. By power counting, the coupling $e^2$ is marginal (relevant) if $\epsilon=0$ ($\epsilon>0$). Physically, with decreasing $\epsilon$, 
the fermion-boson coupling gets weaker at small momenta, 
but large-momentum scatterings become stronger, 
which increases the ``risk" of UV/IR mixing.
%}
Below we primarily focus on the marginal case with $\epsilon=0$. 
Note that 
a large $N$ is still useful in organizing the calculations 
when $\epsilon=0$.

One can introduce two cutoffs.
$\Lambda$ denotes the energy cutoff\footnote{
UV cutoff associated to high-energy modes can be imposed on the 
$x$-momentum relative to the FS as well, yet this cutoff violates an emergent gauge invariance \cite{Mross2010}.
}, and $\Lambda_y$ is the cutoff of $y$-momentum.
The former is the usual UV cutoff,
while the latter represents the size of patch.
We take $\Lambda\rightarrow\infty$ for simplicity, \ie the theory is regularized by $\Lambda_y$ only. Crucially, $\Lambda_y$ also serves as IR data that measures the number of gapless modes near the FS,
and there is a priori no guarantee that low-energy observables are insensitive to
$\Lambda_y$.
If the $\Lambda_y$-dependence cannot be removed 
in low-energy observables 
by renormalization, the theory has UV/IR mixing,
and the patch description
fails.

{\it UV/IR mixing.} We consider the photon self-energy, $\Pi(k)$, which is $O(N^1)$ to the leading order.
To order $N^0$, $\Pi(k)$ is finite as $\Lambda_y\rightarrow\infty$ \cite{Metlitski2010, Mross2010}, due to a kinematic constraint that is nevertheless absent at higher orders (see Eq. (30) of the Supplemental Material \cite{supp} or \cite{Holder2015b}).
At order $N^{-1}$, by exact calculations we find a {\it double}-log divergence in the diagram in Fig. \ref{FD: Benz}:
$\Pi_0(k_\tau=0, k_y)= \frac{1}{N}\frac{|k_y|}{e^2}\frac{2\alpha^4}{3\pi^2}\left(\ln\left(\frac{\Lambda_y}{k_y}\right)\right)^2$, with $\alpha\equiv e^2/(4\pi)$ (see Sec. III.C.3 in \cite{supp}). Other diagrams are harder to compute explicitly.
However, under reasonable assumptions, we argue that the only other net contribution to the double-log divergence is from Fig. \ref{FD: 3Stringe}, whose contribution is also $\Pi_0$ at $k_\tau=0$ (see Sec. III.C.4 in \cite{supp}). Double-log divergences are usually from divergences in sub-diagrams, which can then be cancelled by  diagrams with counter terms. 
But the present double-log divergences are not due to this, since the only divergent sub-diagram in Fig.~\ref{FD: double log diagrams} 
is the three-loop vertex correction, but the corresponding counter term does not contribute to the renormalization of the boson kinetic term to order $N^{-1}$. Taken all diagrams together (including the ones with counter terms), the total double-log divergence to order $N^{-1}$ is
\beq
\label{eq: double log}
\Pi(k_\tau=0, k_y)\sim \frac{1}{N}\frac{|k_y|}{e^2}\frac{4\alpha^4}{3\pi^2}\left(\ln\left(\frac{\Lambda_y}{k_y}\right)\right)^2.
\eeq

\begin{figure}[h]
     \centering
     \includegraphics[width=\linewidth]{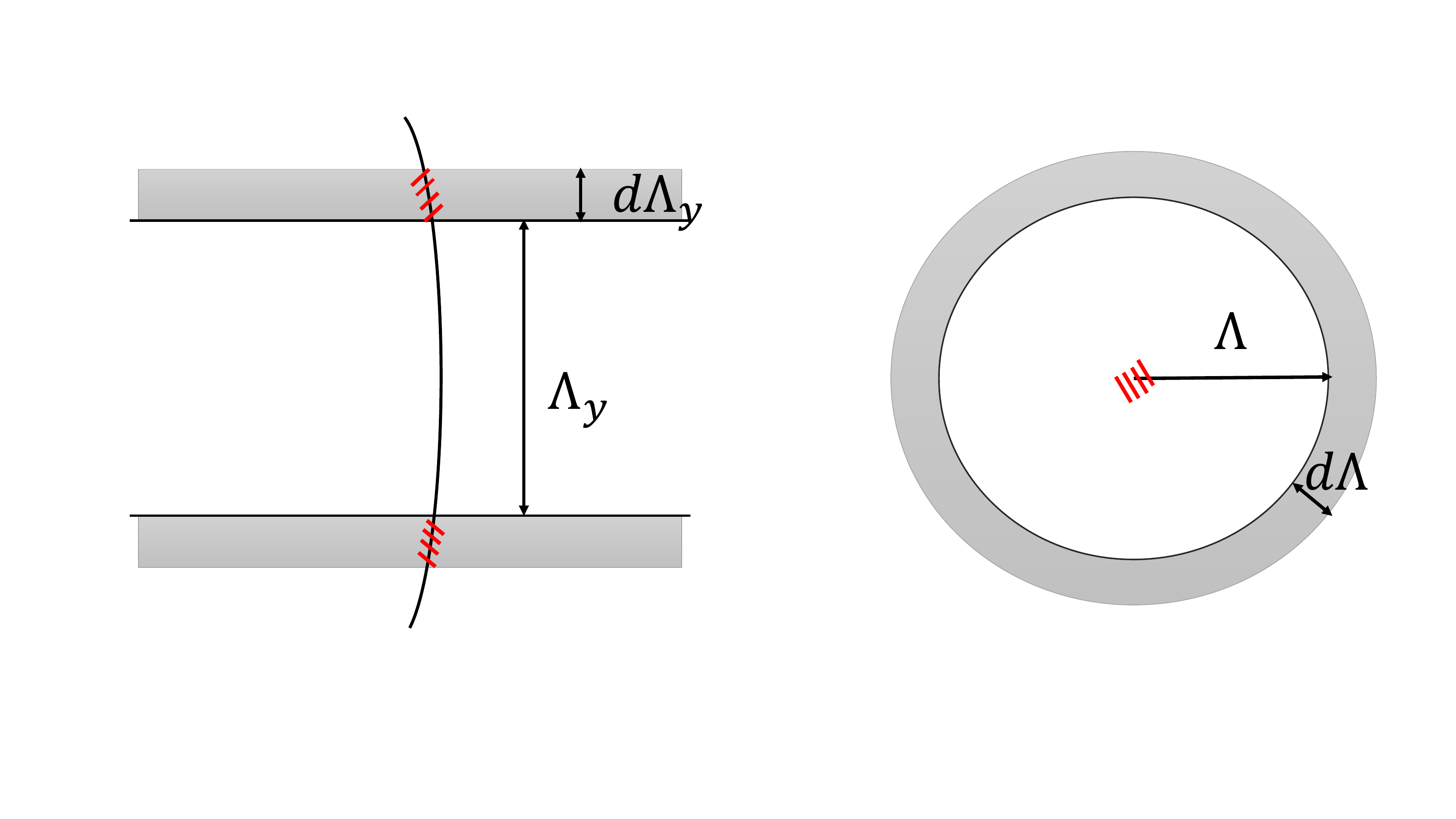}
         \caption{
The gray regions illustrate modes that are integrated out {\it if} we tune $\Lambda_y$ in MFL (left) or $\Lambda$ in a usual field theory without FS (right). The latter has gapless modes only at a single point in the momentum space (shown in red), while the former has gapless modes overlapping with the gray regions. 
In MFL, $\Pi(k)$ calculated at a {\it fixed} $\Lambda_y$ can have
IR singularities
stronger 
than $\ln(\Lambda_y/k)$.
}
         \label{fig: Integration Region}
\end{figure}

To better understand this result, first consider the usual renormalizable field theories without a FS, \eg $3+1$ dimensional $\phi^4$ theory.
In such theories, given a UV cutoff $\Lambda$, quantities like
$d\Pi/d\ln\Lambda$ 
are analytic in the external momentum $k\ll\Lambda$, since this derivative measures the contribution of {\it high-energy}
modes in the energy window $[\Lambda, \Lambda+d\Lambda)$ (see Fig. \ref{fig: Integration Region}). 
Consequently, the non-analyticity in $\Pi$ can at most take the form of $k^2 \ln\frac{\Lambda}{k}$, and the $\Lambda$-dependence of the results can then be eliminated by local counter terms, allowing any observable at a scale $k_1 \ll\Lambda$ to be expressed solely in terms of renormalized quantities measured at another scale $k_2 \ll\Lambda$, and the IR physics is insensitive to the UV physics.

However, the present theory has another short-distance scale, $\Lambda_y$, which measures the number of {\it gapless} modes near the FS.
Low-energy observables in general can depend on $\Lambda_y$ in a sensitive manner. Especially, $d\Pi/d\ln\Lambda_y$ does not have to be analytic in $k$
(see Fig. \ref{fig: Integration Region}). Gapless modes can not only renormalize the existing non-local term through
$|k_y| \ln(\Lambda_y/|k_y|)$,
but also generate stronger non-analyticity
in the quantum effective action, such as 
$|k_y| \ln^n(\Lambda_y/|k_y|)$ with $n>1$ ($n=2$ in Eq. \eqref{eq: double log}).
In this case, the $\Lambda_y$-dependence cannot be removed in low-energy observables through renormalization of the existing terms (local or not) in the action, signaling UV/IR mixing.

UV/IR mixing is known to arise 
in metals {\footnote{We note that UV/IR mixing with different origins is also proposed in other setups \cite{Ghamari2014, Shao2020, You2021, Shackleton2021, Zhou2021}.}. 
First, the FS size $k_F$, a UV parameter,
becomes relevant at low energies when a critical boson is coupled with FS
whose dimension is greater than one, as a boson can decay into particle-hole pairs along the ``great circle'' of FS whose tangent space includes the boson momentum \cite{Mandal2015,Mandal2016}.
Second, the FS size is important in the presence of pairing instabilities driven by short-range four-fermion interactions, via which Cooper pairs residing on the FS with zero total momentum can be scattered throughout the entire FS without violating momentum or energy conservation \cite{Cooper1956, Metlitski2014,Wang2016}.

The origin of the UV/IR mixing we find here is related to the second one but different. The contribution in Eq. \eqref{eq: double log} comes from {\it virtual} Cooper pairs (VCP), represented by the two fermion loops that come from opposite patches and run in the same direction in Fig. \ref{FD: double log diagrams}. Via the {\it marginal} long-range interactions mediated by the gauge field, these VCP spread over the entire FS, which enjoys a large phase space for scattering and can have singular contributions (see Eq. (86) in \cite{supp}).
Indeed, the double-log divergence disappears if either the VCP are absent (\eg by taking the fermion loops in Fig. \ref{FD: double log diagrams} to be in the same patch 
and/or run in opposite directions), or large-momentum scatterings are further suppressed (\eg by taking $\epsilon>0$)\footnote{
Our double-log divergence appears similar to the Sudakov double pole \cite{Peskin1995, Stewart2013} in QED and effective theories of QCD.
Both of them originate from gapless degrees of freedom.
However, the Sudakov double-log originates from small momentum modes
while our double-log divergence is from modes with large $y$-momentum.
}. 
This is reminiscent to the enhanced quasiparticle decay rate due to VCP in Fermi liquids \cite{Pimenov2021}.

{\it Consequences of UV/IR mixing.} Eq. \eqref{eq: double log} forces us to view $\Lambda_y$ as another 
{\it coupling constant} of the theory \cite{Mandal2015}. 
In particular, $\tilde \Lambda_y \equiv \Lambda_y/\mu$ plays the role of a relevant coupling as the size of FS blows up relative to the decreasing scale $\mu$.
The beta functions of the theory are (see Sec. I and II in \cite{supp} for details)
\beq\label{eq: Confusing RG term}
\frac{d \tilde \Lambda_y}{d \ln\mu} = - \tilde \Lambda_y,  ~~ 
\frac{d\alpha}{d \ln\mu} = 
\frac{2\alpha^2}{\pi N} -
\frac{8\alpha^5}{3\pi^2N^2} \ln\tilde \Lambda_y.
\eeq
Let us analyze these beta functions in the weak-coupling regime
with 
$\alpha \lesssim 1$ and low-energy limit
with $\mu\ll \Lambda_y$, together with
a large but finite $N$.
The first term 
in $d\alpha/d\ln \mu$
is the lowest-order term in $1/N$ and $\tilde \Lambda_y$-independent.
As the scale $\mu$ is lowered, 
it makes the 
gauge coupling decrease logarithmically through screening.
If the initial coupling $\alpha_0$ defined at energy scale $\mu_0$ satisfies $(\alpha_0^3/N)\cdot\ln  (\Lambda_y/\mu_0) \lesssim 1$, this term dominates and
the gauge coupling flows to zero at low energies.
On the other hand, 
for $(\alpha_0^3/N)\cdot\ln (\Lambda_y/\mu_0) \gtrsim 1$,
the second term dominates, 
which tends to enhance the coupling at low energies.
In this case,
we can ignore the first term to the leading order.
Then
the gauge coupling grows as $\alpha = \alpha_0\left[1-\frac{16}{3\pi^2 N^2}\alpha_0^4\left( \ln^2\left(\frac{\mu}{\Lambda_y}\right)-\ln^2\left(\frac{\mu_0}{\Lambda_y}\right)\right)\right]^{-1/4}$.
This solution shows a divergence of the gauge coupling with decreasing $\mu$, although it cannot be trusted in the strong coupling regime.
For theories defined at scale $\mu_0$,
the basin of attraction for the $\alpha=0$ fixed point  
is given by
$
{\cal B}_{\mu_0}
\equiv
\{
\alpha_0 | \alpha_0^3 < c N/ \ln (\Lambda_y/\mu_0) 
\}$,
with $c$ an $O(1)$ constant (see the shaded region in Fig. \ref{fig: Flow}).
The salient feature is that 
${\cal B}_{\mu_0}$
shrinks to a measure-zero set   in the low-energy limit 
(\ie $\mu_0\ll\Lambda_y$), due to the scale dependence in the beta function.
The fact that the beta function explicitly depends on $\Lambda_y$ is a hallmark of UV/IR mixing.

\begin{figure}[h]
\centering
\includegraphics[width=0.5\textwidth]{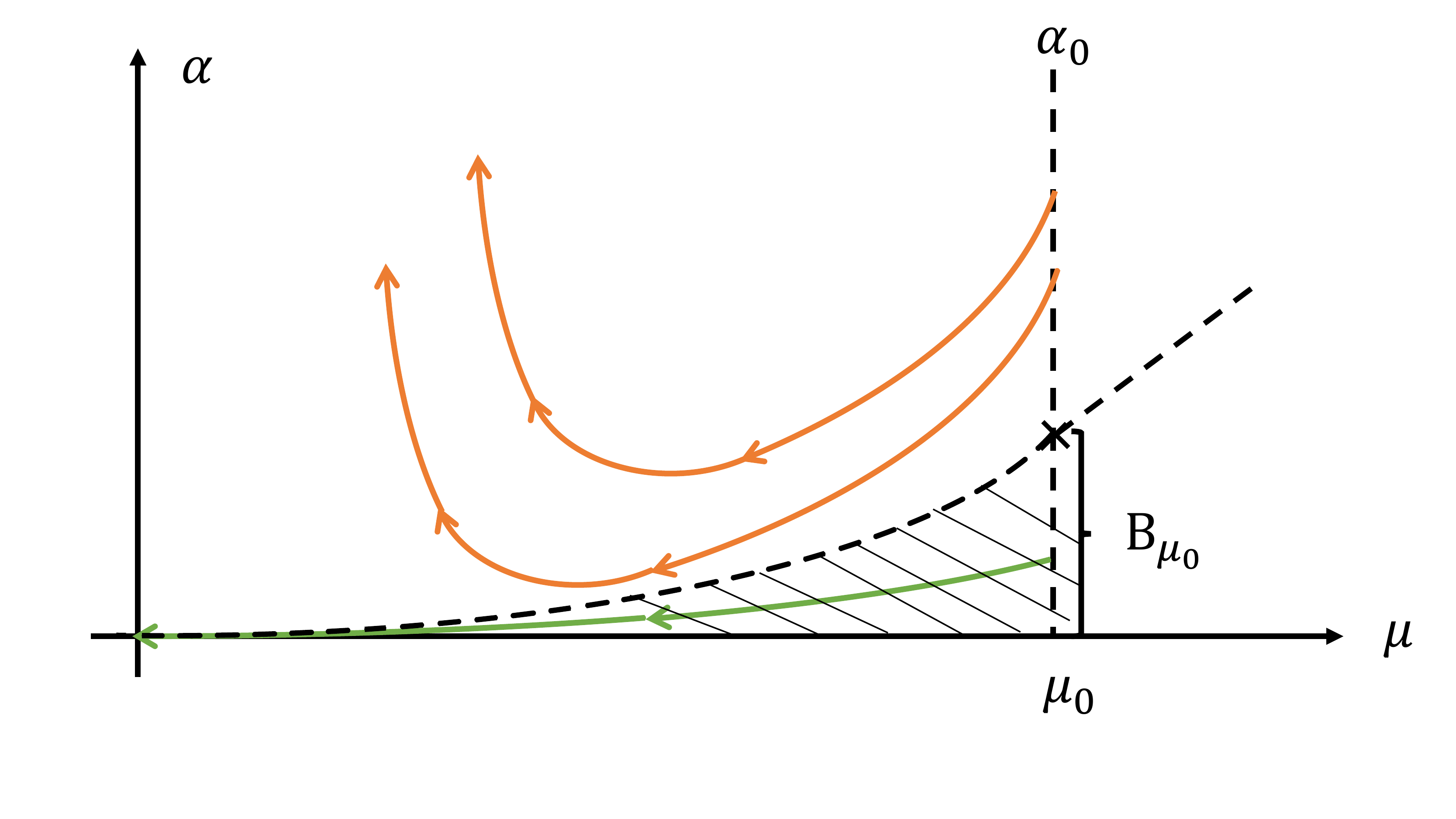}
\caption{
The flow of $\alpha=e^2/(4\pi)$ with initial condition $\alpha=\alpha_0$ at $\mu=\mu_0$. For each $\mu_0$, there is a critical value $\alpha^*\approx N/ \ln (\Lambda_y/\mu_0)$ (the dashed curve) :
 when $\alpha_0<\alpha^*$ the gauge coupling flows to zero at low energies (green), while when $\alpha>\alpha^*$ it flows to infinity (orange).}
\label{fig: Flow}
\end{figure}

In the presence of the UV/IR mixing, the FS size cannot be dropped in low-energy physical observables.
For example,
the single-fermion spectral function takes the form of
$\mc{A}(\omega, \vec k, T)
=
\omega^\Delta f\left(\frac{\omega}{k^z_\parallel},  \frac{k_\parallel}{k_F^{z'}}, \frac{\omega}{T}\right)$, 
where $k_F$ is the FS size, $k_{\parallel}$ is the distance of $\vec k$ away from the FS, $T$ is the temperature, $\Delta$, $z$ and $z'$ are critical exponents ($z'=2$ from Eq. \eqref{eq: double log}), and $f$ is a universal function  \cite{Mandal2015}.
It is interesting to test this in CFLs at various filling factors that can be realized in Chern bands \cite{Zou2020a}. 

The UV/IR mixing in MFLs also has implications for the $\epsilon$-expansion  scheme \cite{Nayak1994,Mross2010}, which has been used to approach nFLs 
with $\epsilon =1$ from MFLs with $\epsilon=0$ perturbatively in $\epsilon$.
To see it, we examine
how the UV/IR mixing in the base theory 
with $\epsilon=0$ affects the perturbative $\epsilon$-expansion.
In theories with $\epsilon > 0$, the UV/IR mixing disappears since the diagrams
in Fig. \ref{FD: double log diagrams}
are no longer divergent in $\Lambda_y$, as large-momentum scatterings are  further suppressed.
Instead, the double-log  
in Eq. \eqref{eq: double log}
is translated to a double pole in $\epsilon$ as{\footnote{See Ref. \cite{Holder2015} for a different but related calculation at $\epsilon=1$.}}
\beq\label{eq: double log epsilon}
\begin{split}
\Pi(k_\tau=0, &k_y)\sim \frac{1}{N}\frac{|k_y|}{\tilde e^2}\frac{8\tilde\alpha^4}{\pi^2}\times\\
&\left( \frac{1}{27\epsilon^2}+\frac{1}{9\epsilon}\ln(\mu/|k_y|)+\frac{1}{6}\left(\ln(\mu/|k_y|)\right)^2\right),
\end{split}
\eeq
where
$\tilde e^2 = e^2\mu^{-\epsilon}$ is  the dimensionless coupling and $\tilde\alpha=\tilde e^2/(4\pi)$. 
Since $1/\epsilon$-poles cannot be absorbed by terms already present in the action,
the naive perturbative expansion appears ill-defined. Moreover, this singular self-energy suggests that $\epsilon$ is renormalized to a larger value, further indicating that the $\epsilon$-expansion may break down.
This calls for alternative control schemes for nFLs.
See  Refs. \cite{Dalidovich2013,
Sur2014,
Lunts2017}
for the dimensional regularization scheme that has no UV/IR mixing,
and Refs. \cite{Damia2019, Aldape2020, Kim2020, Esterlis2021}
for other proposals.

{\it Summary and discussion.} 
We provide strong evidence that a $(2+1)$-dimensional MFL exhibits UV/IR mixing, caused by virtual Cooper pairs that spread over the entire FS due to large-momentum scatterings.
Our finding suggests the breakdown of the patch theory for MFLs
and a potential issue
in the $\epsilon$-expansion that uses MFLs as the base theory for nFLs.

We conclude with a few final remarks.
First, the UV/IR mixing identified in $(2+1)$-dimensional MFLs can be extended to more general cases.
Consider metals with $m$-dimensional FS (\eg a spherical or cylindrical FS has $m=2$ and a Weyl nodal line has $m=1$) coupled to a critical boson whose kinetic energy goes as $k_y^{1+\epsilon}$. The contribution of virtual Cooper pairs to loop corrections scales as $\sim\int\frac{d^mk_y}{k_y^{1+\epsilon}}$, which suggests that for
$m \geqslant 1+\epsilon$,
there exist UV divergences associated with the extended size of FS,
and UV/IR mixing can arise.
So we expect virtual-Cooper-pair-induced UV/IR mixing in $(3+1)$-dimensional gauge theories with $m=2$ and $\epsilon=1$ 
(on top of the UV/IR mixing identified in Ref. \cite{Mandal2015}).
This is relevant to quantum spin liquids \cite{Zhou2008, Lawler2008} and mixed-valence insulators \cite{Chowdhury2017}\footnote{
Since $\epsilon=1$ corresponds to local kinetic term for gauge boson, we expect that UV/IR mixing identified here does not originate from the non-analyticity of the kinetic term of the gauge boson.}.

Second, our result is obtained within the standard patch theory. 
To understand the  full consequences of
the UV/IR mixing 
caused by large-momentum scatterings, 
one should consider a general theory that keeps 
track of how the boson-fermion coupling is renormalized at {\it large} boson momenta.
For this, instead of a coupling constant, one should take into account a {\it momentum-dependent coupling function}, reminiscent of the familiar form factors in the interaction vertices in various settings \cite{Parameswaran2013}. Moreover, the four-fermion couplings, which should also be described by a coupling function, are not considered here, but they should in principle be studied on equal footing as the gauge coupling. Whether there is UV/IR mixing can depend on the microscopic details of the physical system.
What we have shown is the presence of a UV/IR mixing in systems where the two effects above are negligible. 
In the future, it will be of great interest to understand whether such UV/IR mixing exists in systems where these effects are significant and should be incorporated into the theory.
In any case, our results suggest that the physics of MFLs is richer than originally expected,
and mandates a qualitative 
improvement of the current theoretical understanding.

{\it Acknowledgement.} 
We thank Tobias Holder, Subir Sachdev and T. Senthil for helpful discussion. W.Y. would like to especially thank Timothy Hsieh for introducing the topic of non-Fermi liquid, Haoran Jiang for explaining double-log behavior in phenomenology and Minyong Guo for hospitality during his stay in Beijing where the bulk calculation was carried out. 
Research at Perimeter Institute is supported in part by the Government of Canada through the Department of Innovation, Science and Economic Development Canada and by the Province of Ontario through the Ministry of Colleges and Universities.
S.L. acknowledges the support of the Natural Sciences and Engineering Research Council of
Canada.

\bibliography{main.bib} 

\end{document}

\title{Supplemental Material for ``UV/IR Mixing in Marginal Fermi Liquids''}
	
\author{Weicheng Ye}
\affiliation{Perimeter Institute for Theoretical Physics, Waterloo, Ontario, Canada N2L 2Y5}
\affiliation{Department of Physics and Astronomy, University of Waterloo, Waterloo, Ontario, Canada N2L 3G1}

\author{Sung-Sik Lee}
\affiliation{Perimeter Institute for Theoretical Physics, Waterloo, Ontario, Canada N2L 2Y5}
\affiliation{Department of Physics and Astronomy, McMaster University, Hamilton, Ontario, Canada L8S 4M1}

\author{Liujun Zou}
\affiliation{Perimeter Institute for Theoretical Physics, Waterloo, Ontario, Canada N2L 2Y5}

%\begin{abstract}
%\end{abstract}

\maketitle

\tableofcontents\

This Supplemental Material contains the framework for the renormalization group (RG) analysis (Section \ref{app: framework}), a summary of the RG results (Section \ref{app: summary}), and the details of the Feynman diagram claculations (Section \ref{app: Diagram Calculation}).

\section{Framework for Renormalization Group Analysis}\label{app: framework}

In this section, we present the framework for our field-theoretic RG analysis. 
In terms of the renormalized fields and couplings, 
the action of the patch theory for $N$ Fermi surfaces (FS) coupled to a $U(1)$ gauge field 
is written as
\beq \label{eq: patch theory}
S = S_\psi + S_{\text{int}} + S_a,
\eeq
where 
\beq\label{eq: Lagrangian of patch theory}
\begin{split}
	S_\psi &= \int[dk] \sum_{i,p}\psi^\dagger_{ip}(k)\left[-i Z_1 k_\tau + Z_2(p k_x + k_y^2)\right]\psi_{ip}(k),\\
	S_{\text{int}} &= \int[dk_1][dk_2] \sum_{i,p}Z_2\lambda_p a(-k_1) \psi^\dagger_{ip}(k_2) \psi_{ip}(k_1+k_2),\\
	S_a &= \int[dk]Z_3\frac{N}{2e^2\mu^{\epsilon}} |k_y|^{1+\epsilon} a(-k)a(k)\\
\end{split}
\eeq
with $[dk]\equiv\frac{dk_\tau dk_x dk_y}{(2\pi)^3}$, $\lambda_\pm=\pm 1$, $\mu$ a renormalization scale with the same dimension as $k_y$, and $Z_{1,2,3}$ some renormalization factors. In the above, the rotational symmetry of the patch theory is invoked for the absence of the relative renormalization between $k_x$ and $k_y^2$, and the Ward identity is used to guarantee that 
the multiplicative renormalization of $S_{\rm int}$ is the same as
that of $(pk_x+k_y^2)$ part of the fermion kinetic term \cite{Metlitski2010}. 
For latter convenience, define $\alpha\equiv\gamma e^2$ with $\gamma\equiv 1/(4\pi)$ as a proxy for the coupling constant $e^2$. 
The renormalized fields, couplings and momentum are related to the bare ones via
\beq \label{eq: bare-renormalized relation}
\begin{split}
&k_{\tau B}=\frac{Z_1}{Z_2}k_\tau\equiv Z_\tau k_\tau,
\quad
k_{xB}=k_x,
\quad
k_{yB}=k_y,\\
&\psi_{ipB}(k_B)=\sqrt{\frac{Z_2}{Z_\tau}}\psi_{ip}(k)\equiv\sqrt{Z_\psi}\psi_{ip}(k),
\quad
a_B(k_B)=\frac{1}{Z_\tau}a(k)\equiv\sqrt{Z_a}a(k),\\
&\alpha_B\equiv\gamma e_B^2=\frac{1}{Z_3Z_\tau}\gamma e^2\mu^\epsilon\equiv Z_{\alpha}\alpha\mu^{\epsilon}
\end{split}
\eeq
where the subscript $B$ stands for bare. 
$Z_\tau$ is introduced to account for a nontrivial dynamical exponent, $Z_\psi$ and $Z_a$ are introduced to account for anomalous dimensions of the fermion and gauge fields, respectively, and $Z_{\alpha}$ is introduced to account for the flow of $\alpha$.
At the tree level,
the fields and couplings have the following scaling dimensions in momentum space:
\beq\label{eq: engineering dimensions}
\begin{split}
	&[k_\tau]_0=2,
	\quad
	[k_x]_0=2,
	\quad
	[k_y]_0=1,\\
	&[\psi_{ip}(k)]_0=-\frac{7}{2},
	\quad
	[a(k)]_0=-3,
	\quad
	[e^2]_0=0.
\end{split}
\eeq

The $Z$ factors are chosen so that the following renormalization conditions are satisfied:
\beq \label{eq: renormalization conditions 1}
\begin{split}
\frac{\partial\Sigma(k_\tau=\mu^2, k_x=k_y=0)}{\partial k_\tau}&=0,\\
\Sigma(k_y=\mu, k_\tau=k_x=0)&=0,\\
\Pi(k_y=\mu, k_\tau=k_x=0)&=0,
\end{split}
\eeq
where $\Sigma$ and $\Pi$ are the 1-particle irreducible self-energy of the renormalized fermionic and gauge fields, respectively, and $\mu$ is the renormalization scale. Below we focus on the case with $\epsilon=0$. 
In this case, 
it turns out that physical observables have divergence as $\Lambda_y$, the cutoff of $k_y$, goes to infinity. Notice that $\Lambda_y$ measures the abundance of {\it gapless} degrees of freedom (DOF) on the FS, so it is not a cutoff as in a usual field theory, which separates high-energy DOF from low-energy ones. Accordingly, it is more appropriate to view $\Lambda_y$ as a parameter that is on equal footing as the couplings of the theory.

Denote a correlation function involving $n_\psi$ fermionic fields and $n_a$ gauge fields by $G^{(n_\psi, n_a)}$, which is generically written as a function of a set of momenta, $\alpha$, $\mu$ and $\Lambda_y$. Then $G^{(n_\psi, n_a)}_B\cdot \delta(\{k_B\})=Z_\psi^{\frac{n_\psi}{2}}Z_a^{\frac{n_a}{2}}G^{(n_\psi, n_a)}(\{k\})\cdot \delta(\{k\})$, where $\{k\}$ denotes the momenta this correlation function depends on, and the $\delta$-functions impose momentum conservation. 
Due to $k_{\tau B}=Z_\tau k_\tau$, $\delta(\{k\})=Z_\tau\delta(\{k_B\})$, $G_B^{(n_\psi, n_a)}=Z_\psi^{\frac{n_\psi}{2}}Z_a^{\frac{n_a}{2}}Z_\tau G^{(n_\psi, n_a)}(\{k\})$. Because all bare quantities and fields are independent of $\mu$, $\mu\frac{d}{d\mu}G^{(n_\psi, n_a)}_B(\{k_B\})=0$, which leads to a Callan-Symanzik equation for the renormalized correlation function:
\beq \label{eq: Callan-Symanzik 1}
\left[\mu\frac{\partial}{\partial\mu}+b_{\alpha}\alpha\frac{\partial}{\partial\alpha}+\gamma_zk_\tau\frac{\partial}{\partial k_\tau}+\frac{n_\psi\gamma_\psi}{2}+(n_a-1)\gamma_z\right]G^{(n_\psi, n_a)}=0
\eeq
with
\beq \label{eq: RG quantities}
b_{\alpha}\equiv\frac{\mu}{\alpha}\frac{d\alpha}{d\mu}=-\frac{\mu}{Z_{\alpha}}\frac{d Z_{\alpha}}{d\mu},
\quad
\gamma_z\equiv-\frac{\mu}{Z_\tau}\frac{dZ_\tau}{d\mu},
\quad
\gamma_\psi\equiv\frac{\mu}{Z_\psi}\frac{dZ_\psi}{d\mu}.
\eeq
Note that the beta function of $\alpha$ is $\beta(\alpha)=b_{\alpha}\alpha$.

On the other hand, suppose the engineering dimension of $G^{(n_\psi, n_a)}$ is $\Delta_0$. By dimensional analysis, it must have the form $G^{(n_\psi, n_a)}=\mu^{\Delta_0}g\left(\frac{\{k_\tau\}}{\mu^2}, \frac{\{k_x\}}{\mu^2}, \frac{\{k_y\}}{\mu}, \alpha, \frac{\Lambda_y}{\mu}\right)$, where $g$ is some function. Thus,
\beq
\left(\mu\frac{\partial}{\partial\mu}+2k_\tau\frac{\partial}{\partial k_\tau}+2k_x\frac{\partial}{\partial k_x}+k_y\frac{\partial}{\partial k_y}+\Lambda_y\frac{\partial}{\partial\Lambda_y}-\Delta_0\right)G^{(n_\psi, n_a)}=0.
\eeq
Combining this equation with Eq. \eqref{eq: Callan-Symanzik 1} yields another form of the Callan-Symanzik equation:
\beq \label{eq: Callan-Symanzik useful 1}
\left[(2-\gamma_z)k_\tau\frac{\partial}{\partial k_\tau}+2k_x\frac{\partial}{\partial k_x}+k_y\frac{\partial}{\partial k_y}+\Lambda_y\frac{\partial}{\partial\Lambda_y}-b_{\alpha}\alpha\frac{\partial}{\partial\alpha}-\Delta_0-\frac{n_\psi\gamma_\psi}{2}-(n_a-1)\gamma_z\right]G^{(n_\psi, n_a)}=0.
\eeq
At the fixed point, $b_{\alpha}=0$, and the above equation implies the following scaling structure of the correlation function
\beq
G^{(n_\psi, n_a)}(s^z\{k_\tau\}, s^2\{k_x\}, s\{k_y\}, s\Lambda_y)=s^{\Delta}G^{(n_\psi, n_a)}(\{k_\tau\}, \{k_x\}, \{k_y\}, \Lambda_y)
\eeq
with $z\equiv 2-\gamma_z$ the dynamical exponent, and $\Delta\equiv \Delta_0+\frac{n_\psi\gamma_\psi}{2}+(n_a-1)\gamma_z$ the total scaling dimension of this correlation function. It is clear that $\gamma_\psi$ and $2\gamma_z$ can be interpreted as the anomalous dimensions of the fermionic and gauge fields, respectively. Note that the anomalous dimension of the gauge field $a(k)$ is entirely from a rescaling of time (by setting $k_{\tau B}=Z_\tau k_\tau$), while its real-space counterpart $a(x)$ has no anomalous dimension due to the Ward identity. 
Also note that the renormalized correlation function would not depend on $\Lambda_y$ in the absence of any UV/IR mixing, but one of the central results of this paper is that there is UV/IR mixing.

Before finishing this section, we describe an equivalent way to analyze the theory, which is also often used in the literature and differs from the above one by some convention \cite{Lee2008, Lee2009, Metlitski2010, Mross2010, Zou2020, Zou2020a}. 
We will mainly use the previous convention, and readers not interested in this part can skip it.
In the alternative convention,
the bare action is written as
\beq\label{eq: Lagrangian of patch theory 2}
\begin{split}
	S_\psi &= \int[dk] \sum_{i,p}\psi^\dagger_{ip}(k)\left(-i Z_fZ_\eta\eta k_\tau + Z_f(p k_x + k_y^2)\right)\psi_{ip}(k),\\
	S_{\text{int}} &= \int[dk_1][dk_2] \sum_{i,p}Z_f\lambda_p a(-k_1) \psi^\dagger_{ip}(k_2) \psi_{ip}(k_1+k_2),\\
	S_a &= \int[dk]\frac{1}{Z_e} \frac{N}{2e^2\mu^\epsilon} |k_y|^{1+\epsilon} a(-k)a(k),\\
\end{split}
\eeq
and $[dk]=\frac{dk_\tau dk_x dk_y}{(2\pi)^3}$, $\lambda_\pm=\pm 1$. Again, the Ward identity is used to guarantee that the renormalization factor in $S_{\rm int}$ is $Z_f$, the same as that of the $pk_x+k_y^2$ part of the fermionic kinetic term. The difference between this convention and our previous one is that in this case $k_\tau$ has the same dimension as  $k_x$.
Instead, another dimensionless parameter $\eta$ is introduced to keep track of a nontrivial dynamical exponent. 
Below we will see that, not surprisingly, these two conventions are equivalent.
The bare quantities and fields are related to the renormalized ones via
\beq
(k_{\tau B}, k_{xB}, k_{yB})=(k_\tau, k_x, k_y),~\psi_{ip B}=\sqrt{Z_f} \psi_{ip},~a_{B}=a,~\eta_B=Z_{\eta}\eta=1,~e_{B}^2\equiv Z_{e} e^2.
\eeq
The engineering dimensions are also given by \eqref{eq: engineering dimensions} with the extra $[\eta]_0=0$. We also define $\tilde\alpha \equiv \frac{\gamma e^2}{\eta}$ with $\gamma=1/(4\pi)$.
Similar to the previous convention, the $Z$ factors are determined by demanding the renormalization conditions given in Eq. \eqref{eq: renormalization conditions 1}, but with $\mu^2$ changed to $\mu^2/\eta$ in the first renormalization condition:
\beq
\frac{\partial\Sigma(k_\tau=\mu^2/\eta, k_x=k_y=0)}{\partial k_\tau}&=0.
\eeq
Now consider a correlation function involving $n_f$ fermionic fields and $n_c$ gauge fields, denoted by $G^{(n_f, n_c)}$. One can show that $G^{(n_f, n_c)}=\eta^{\frac{n_f}{2}+n_c-1}\mu^{\Delta_0}g\left(\frac{\eta\{k_\tau\}}{\mu^2}, \frac{\{k_x\}}{\mu^2}, \frac{\{k_y\}}{\mu}, \tilde \alpha, \frac{\Lambda_y}{\mu}\right)$, where $\Delta_0$ is the engineering dimension of the correlation function, and $g$ is the same function as in the previous convention. This implies that $\eta\frac{\partial}{\partial\eta}G^{(n_f, n_c)}=\left(\frac{n_f}{2}+n_c-1+k_\tau\frac{\partial}{\partial k_\tau} - e^2\frac{\partial}{\partial e^2} \right)G^{(n_f, n_c)}$. Combining this result and analogous derivations as before, similar Callan-Symanzik equations can be obtained
\beq
\begin{split}
	\left(\mu\frac{\partial}{\partial\mu}+b_e e^2\frac{\partial}{\partial e^2}+b_\eta\eta\frac{\partial}{\partial\eta}+\frac{n_f}{2}\gamma_f\right)G^{(n_f, n_c)}&=0,\\
	\left[\mu\frac{\partial}{\partial\mu}+(b_e-b_\eta)e^2\frac{\partial}{\partial e^2}+b_\eta k_\tau\frac{\partial}{\partial k_\tau}+\frac{n_f}{2}(\gamma_f+b_\eta)+(n_c-1)b_\eta\right]G^{(n_f, n_c)}&=0,\\
	\left[(2-b_\eta)k_\tau\frac{\partial}{\partial k_\tau}+2k_x\frac{\partial}{\partial k_x}+k_y\frac{\partial}{\partial k_y}+\Lambda_y\frac{\partial}{\partial\Lambda_y}-(b_e-b_\eta)e^2\frac{\partial}{\partial e^2}-\Delta_0-\frac{n_f}{2}(\gamma_f+b_\eta)-(n_c-1)b_\eta\right]G^{(n_f, n_c)}&=0,
\end{split}
\eeq
where
\beq
b_\eta\equiv\frac{\mu}{\eta}\frac{d\eta}{d\mu},~b_e\equiv\frac{\mu}{e^2}\frac{d e^2}{d\mu},~\gamma_f\equiv\frac{\mu}{Z_f}\frac{dZ_f}{d\mu}.
\eeq
It is not hard to see that $b_\eta$, $b_e-b_\eta$ and $\gamma_f+b_\eta$ should be identified as $\gamma_z$, $b_\alpha$ and $\gamma_\psi$ in the previous convention, respectively, and the two conventions ultimately lead to identical results. Note that in this convention, the beta function for $\tilde\alpha$ is $\beta(\tilde\alpha)=\left(b_e-b_\eta\right)\tilde\alpha$. In this paper, we will mainly use the previous convention.

\section{Summary of the RG Results}\label{app: summary}

The $Z$ factors introduced in Eq. \eqref{eq: Lagrangian of patch theory} are calculated in Sec. \ref{app: Diagram Calculation}. 
Here we summarize the results. 
We expand $Z$ in powers of $1/N$, and at each order of $1/N$, we keep results relevant to the calculations of $b_\alpha$, $\gamma_z$ and $\gamma_\psi$ to the leading order in $\Lambda_y/\mu$:
\beq
\begin{split}
	&Z_1=1-\frac{1}{N}\cdot\frac{2\alpha}{\pi}\ln\frac{\Lambda_y}{\sqrt{\alpha}\mu}+\frac{1}{N^2}\cdot\frac{2\alpha^3}{\pi^2}\left(S^a(\alpha)+T^a(\alpha)+2G^a(\alpha)-S^b(\alpha)-T^b(\alpha)-2G^b(\alpha)\right)\ln\frac{\Lambda_y}{\mu},\\
	&Z_2=1+\frac{1}{N^2}\cdot\frac{2\alpha^3}{\pi^2}\left(S^a(\alpha)+T^a(\alpha)-S^b(\alpha)-T^b(\alpha)\right)\ln\frac{\Lambda_y}{\mu},\\
	&Z_3\approx1-\frac{1}{N}\cdot\frac{2\alpha^3}{\pi}F(\alpha)+\frac{1}{N^2}\cdot\frac{4\alpha^4}{3\pi^2}\left(\ln\frac{\Lambda_y}{\mu}\right)^2,
\end{split}
\eeq
where the functions $F(\alpha),G^{a}(\alpha),S^{a}(\alpha),T^{a}(\alpha),G^{b}(\alpha),S^{b}(\alpha),T^{b}(\alpha)$ can be found in Eqs. \eqref{eq: def of F}, \eqref{eq: def of Ga}, \eqref{eq: def of Sa}, \eqref{eq: def of Ta}, \eqref{eq: def of Gb}, \eqref{eq: def of Sb}, \eqref{eq: def of Tb}, respectively.
Using these expressions,
Eqs. \eqref{eq: bare-renormalized relation} and \eqref{eq: RG quantities}, $b_\alpha$, $\gamma_z$ and $\gamma_\psi$ are obrained to be
\beq
\begin{split}
	&b_\alpha\approx\frac{1}{N}\cdot\frac{2\alpha}{\pi} -\frac{1}{N^2}\cdot\frac{8\alpha^4}{3\pi^2}\ln\frac{\Lambda_y}{\mu},\\
	&\gamma_z=-\frac{1}{N}\cdot\frac{2\alpha}{\pi}+\frac{1}{N^2}\cdot\left(\frac{4\alpha^3}{\pi^2}\left(G^a(\alpha)-G^b(\alpha)\right)-\frac{2\alpha^2}{\pi^2}\right),\\
	&\gamma_\psi=-\frac{1}{N}\cdot\frac{2\alpha}{\pi}+\frac{1}{N^2}\cdot\left(\frac{2\alpha^3}{\pi^2}\left(2G^a(\alpha)-S^a(\alpha)-T^a(\alpha)-2G^b(\alpha)+S^b(\alpha)+T^b(\alpha)\right)-\frac{2\alpha^2}{\pi^2}\right).
\end{split}
\eeq
Here we ignore $\frac{1}{N^2}\ln\frac{\Lambda_y}{\mu}$ term in $Z_3$ and keep only terms proportional to $\ln\frac{\Lambda_y}{\mu}$ at order $1/N^2$ in $b_\alpha$. Note that almost all terms in $b_\alpha$, $\gamma_z$ and $\gamma_\psi$ come from explicit $\mu$ dependence in $Z$ except for the $\frac{1}{N^2}\alpha^2$ term, which comes from the $\ln\sqrt{\alpha}$ term in $Z_1$.

\section{Feynman Diagram Calculations}\label{app: Diagram Calculation}

In this section,
we present the details of the Feynman diagram calculations leading to the results summarized in Section \ref{app: summary}. 
Parts of the relevant calculations can be found 
in Refs.~\cite{Lee2009,Metlitski2010,Mross2010,Zou2020,Holder2015,Holder2015b}. 
Here we focus on the case where $\epsilon=0$.\footnote{As discussed in Ref. \cite{Lee2009}, if we use the second convention, we need to set $\eta\sim\mc{O}(1)$ such that the large-$N$ expansion has no extra factors of $N$ from $\eta$.} 
We will also assume that external $k_y$ to be positive for self-energies unless mentioned otherwise.

First, we collect results of all the Feynman diagrams contributing to the self-energy of boson and fermion to the third leading order in the large-$N$ expansion. We will see that all of the divergent diagrams give the traditional logarithmic divergence except for the Benz diagram and the 3-String diagram, which give the unconventional double logarithmic divergence. 
In any case, $Z_{1,2,3}$  can be expanded in $1/N$ as
\beq
Z_i = 1 + \frac{1}{N}\delta_i^{(1)} + \frac{1}{N^2}\delta_i^{(2)}\quad i=1,2,3
\eeq
such that renormalization conditions Eq. \eqref{eq: renormalization conditions 1} are satisfied. The propagator of $a$, denoted as $\mc{D}(k)$, and the propagator of $\psi$,  denoted as $\mc{G}(k)$, to the third leading order in $1/N$ are written as
\beq
\begin{split}
\mc{D}^{-1} &= N D^{-1} - \Pi^{(1)} -\frac{1}{N}\Pi^{(2)} + ...\\
\mc{G}^{-1} &= G^{-1} - \frac{1}{N}\Sigma^{(1)} - \frac{1}{N^2}\Sigma^{(2)} + ... ,
\end{split}
\eeq
where $D$ is the leading-order propagator for $a$ as in e.g. Ref.~\cite{Metlitski2010}
\beq\label{eq: bare photon propagator 1}
D(k_\tau, k_y)=\frac{1}{\gamma\frac{|k_\tau|}{|k_y|}+\frac{|k_y|}{e^2}}
\eeq
with $\gamma = 1/(4\pi)$, and $G$ is the bare fermion propagator,
\beq\label{eq: bare fermion propagator 1}
G_\pm(k_\tau, k_x, k_y) = \frac{1}{-ik_\tau \pm k_x + k_y^2}.
\eeq
We can also write down a large-$N$ expansion of $b_\alpha$, $\gamma_z$ and $\gamma_\psi$ to the order of $1/N^2$, 
\beq\label{eq: beta function expansion 1}
\begin{split}
	b_{\alpha} & = \frac{1}{N} b_{\alpha}^{(1)} + \frac{1}{N^2} b_{\alpha}^{(2)} + ...\\
	\gamma_z & = \frac{1}{N} \gamma_z^{(1)} + \frac{1}{N^2} \gamma_z^{(2)} + ...\\ 
	\gamma_\psi &= \frac{1}{N} \gamma_\psi^{(1)} +  \frac{1}{N^2} \gamma_\psi^{(2)} + ...\\
\end{split}
\eeq
which can be calculated through solving the Callan-Symanzik equation Eq. \eqref{eq: Callan-Symanzik 1} from the propagator $\mc{D}, \mc{G}$ at the third leading order of $1/N$, or through Eq. \eqref{eq: RG quantities}. 

\subsection{Second Leading Order Result of Self-Energy}\label{subapp: second leading order result}
\begin{figure}[h]
\begin{center}
\includegraphics[scale=0.38]{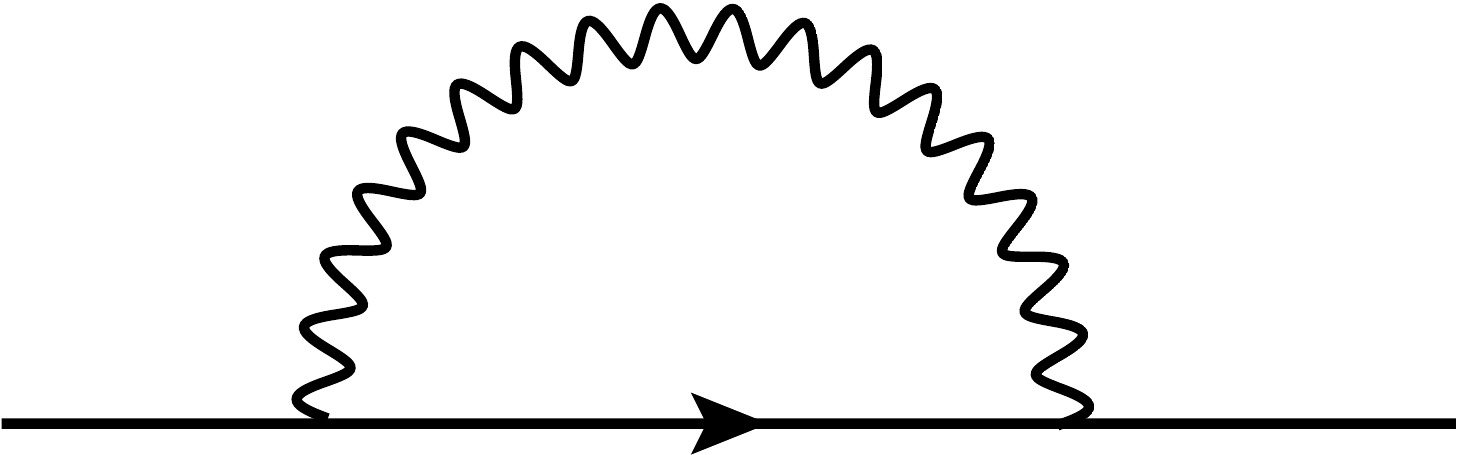}
\end{center}
\caption{One-loop diagram for fermion self-energy.} 
\label{FD: 1LF}
\end{figure}

The second leading order result of the fermion self-energy is given by the standard one-loop diagram shown in Fig. 
\ref{FD: 1LF}.
The diagram has been computed in \cite{Mross2010,Zou2020}. Here we write down the result of the diagram at $\epsilon=0$ with $y$-momentum cutoff $\Lambda_y$,
\beq
\Sigma_{1LF}(k) = \int [dl]~G_f(k-l)D(l)=i k_\tau \frac{e^2}{2\pi^2}\left(\frac{1}{2}+\ln\left(\frac{\Lambda_y}{\sqrt{\gamma e^2|k_\tau|}}\right)\right).
\eeq
To impose the renormalization conditions in Eq. \eqref{eq: renormalization conditions 1}, we take
\beq
\delta_1^{(1)} = -\frac{2\alpha}{\pi}\ln\left(\frac{\Lambda_y}{\mu\sqrt{\alpha}}\right),
\quad 
\delta_2^{(1)}=0
\eeq
with $\alpha\equiv e^2/(4\pi)$.
\footnote{
We note that in the simple minimal subtraction scheme, one would take $\delta^{(1)}=-\frac{2\alpha}{\pi}\ln\frac{\Lambda_y}{\mu}$, which amounts  to imposing a different RG condition.
}

\begin{figure}[h]
\centering
\includegraphics[width=0.4\textwidth]{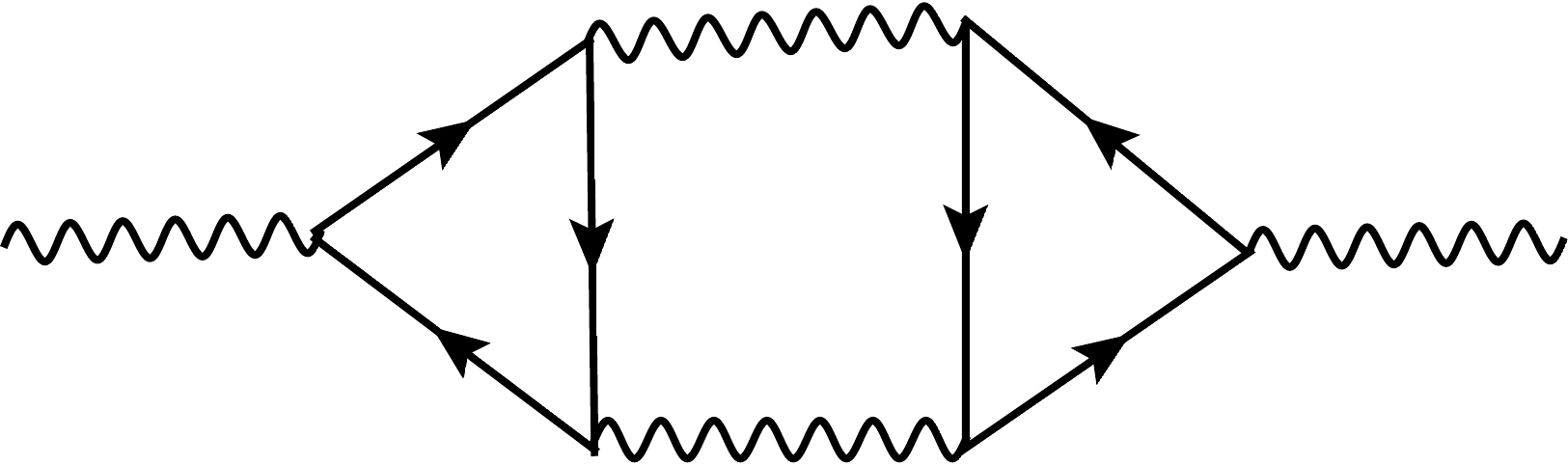}
\caption{Aslamazov-Larkin (AL) diagram for boson self-energy.}
\label{FD: AL}
\end{figure}

Let us then turn to the second leading order result of the self-energy of boson, $\Pi^{(1)}$. At this order only the so-called Aslamazov-Larkin (AL) diagram \cite{Aslamazov1968} in Fig.~\ref{FD: AL} gives non-zero but finite contribution as argued in Ref. \cite{Mross2010}.
The AL diagram reads \cite{Metlitski2010}
\beq\label{eq: AL pre}
\Pi_{AL}(k) = \frac{1}{2}\int\frac{dl_\tau dl_x dl_y}{(2\pi)^3}\Gamma^3(k, l, -(k+l))\Gamma^3(-k, -l, k+l)D(l)D(k+l)
\eeq
where $\Gamma^3$ is the fermion-induced three-boson vertex and receives contributions from 
both fermion patches as well as both directions in which fermions can travel in the loop,
\beq
\Gamma^3 = \Gamma^3_+ + \Gamma^3_-,
\eeq
\beq
\Gamma^3_s(l_1, l_2, -(l_1+l_2)) = \lambda_s^3\left(f_s(l_1, l_2, -(l_1+l_2)) + f_s(l_2, l_1, -(l_1+l_2))\right),
\eeq
\beq
f_s(l_1, l_2, -(l_1+l_2)) = -\int \frac{dp_\tau dp_x dp_y}{(2\pi)^3}G_s(p) G_s(p - l_1) G_s(p+l_2).
\eeq
First let us compute $f_\pm(k, l, -(k+l))$. We perform the integral in the sequence of $p_x$, $p_y$ and $p_\tau$ 
to obtain
\beq
\begin{split}
f_\pm(k, l, -(k+l)) 
&=\frac{1}{4\pi\left( k_y(il_\tau\mp l_x-l_y^2)-l_y(ik_\tau \mp k_x+k_y^2)\right)}\\
&\quad\quad\quad\quad\times\left(k_\tau  \theta\left(\frac{k_\tau }{k_y}\right) + l_\tau \theta\left(\frac{l_\tau}{l_y}\right) - (k+l)_\tau \theta\left(\frac{(k+l)_\tau}{(k+l)_y}\right)\right).
\end{split}
\eeq
Here $\theta(x)$ is the unit step function. An important property is that if we take the static limit $k_\tau\rightarrow0$, the fermion-induced three-boson vertex is non-zero only when the internal boson satisfies,
\beq\label{eq: 3-boson kinetic constraint}
-k_y<l_y<0,~~~\text{when}~k_y>0~~~~~~\text{or}~~~~~~0<l_y<-k_y,~~~\text{when}~k_y<0,
\eeq
i.e. the integration of the internal boson momentum only has nonzero contribution from a restricted finite region in the static limit. \footnote{This is the reason why we do not see the double log divergence in lower order diagrams. In $U(1)\times U(1)$ or $U(2)$ gauge theory \cite{Zou2020}, this also leads to the fact that there is a fixed line instead of several isolated fixed points at the leading order of the RG equation. It is straightforward to see that the fixed line will in principle be lifted because of the third leading order results of (boson) self-energy.} 
Note that the imaginary part of the pole of $l_x$ for both $f_s(k, l, - (k+l))$ and $f_s(l, k, -(k+l))$ with a fixed $s=\pm$ are at $s(l_\tau - k_\tau l_y/k_y)$. Considering the integration of $l_x$ in Eq. \eqref{eq: AL pre}, we see that $\Pi_{AL}$ is non-zero only when the two fermion loops come from different patches. 
After the integration of $l_x$, we see that the $k_x$ dependence drops out and we have
\beq
\begin{split}
\Pi_{\text{AL}}(k_\tau, k_y) 
&= -\frac{1}{64\pi^4 k_y^2} \int d l_\tau d l_y D(l) D(k+l) \left(k_\tau  \theta\left(\frac{k_\tau }{k_y}\right) + l_\tau \theta\left(\frac{l_\tau}{l_y}\right) - (k+l)_\tau \theta\left(\frac{(k+l)_\tau}{(k+l)_y}\right)\right)^2\\ &  \quad\quad\quad\quad\quad\quad\quad\quad\quad\quad\quad\quad 
\times \frac{\left(l_y^2+k_y l_y\right)^2}{|l_\tau-k_\tau l_y/k_y|\left(\left(l_\tau-k_\tau l_y/k_y\right)^2 + \left(l_y^2+k_y l_y\right)^2\right)}.\\
\end{split}
\eeq
It is straightforward to see that this integral is convergent from the kinematic constraint in Eq. \eqref{eq: 3-boson kinetic constraint}.
Here, we focus on 
$\Pi_{AL}(k_\tau=0, k_y)$. 
Also we temporarily assume $k_y>0$, which indicates that $-k_y<l_y<0$. 
After a change of variable with $l_\tau = k_y^2\Delta$ ($-\infty<\Delta<\infty$) and $l_y = k_y \delta$ ($-1<\delta<0$), the final result is given as follows,
\beq\label{eq: N^0 photon propagator}
\Pi_{AL}(k_\tau = 0, k_y) = -\frac{|k_y|}{e^2}\frac{2\alpha^3}{\pi}F(\alpha), 
\eeq
where
\beq\label{eq: def of F}
F(\alpha) \equiv \int_0^1 d \delta \int_0^\infty d\Delta \frac{\delta}{\delta^2+\alpha\Delta}\frac{1-\delta}{(1-\delta)^2+\alpha\Delta}\frac{\Delta\left(\delta^2-\delta\right)^2}{\Delta^2+\left(\delta^2-\delta\right)^2}.
\eeq
To impose the renormalization conditions in Eq. \eqref{eq: renormalization conditions 1}, 
we take the counterterm to be
\beq
\delta_3^{(1)}=-\frac{2\alpha^3}{\pi}F(\alpha)
\eeq

\subsection{Third Leading Order Result of Fermion Self-Energy}\label{subapp: fermion}

\begin{figure}[!htbp]
     \centering
     \begin{subfigure}[h]{0.34\textwidth}
     	\centering
     	\includegraphics[scale = 0.4]{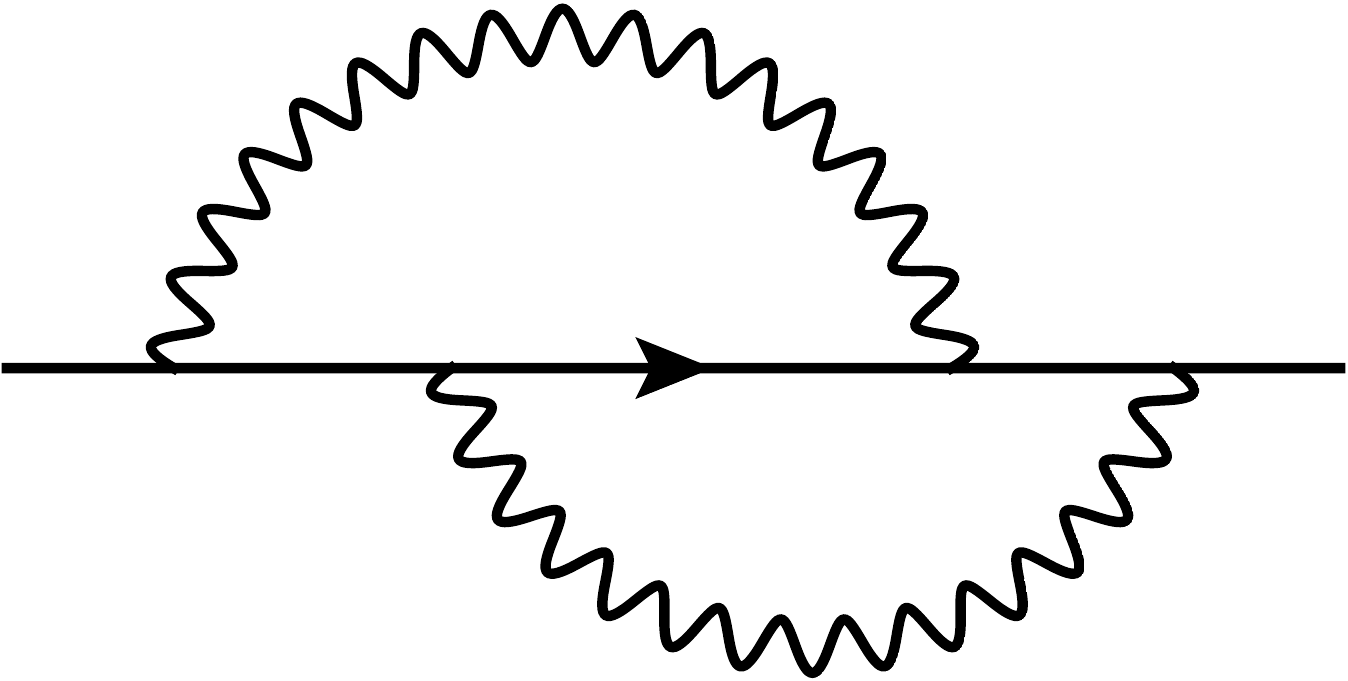}
     	\caption{Two-Loop Fermion Diagram.}
     	\label{FD: 2LF}
     \end{subfigure}
     \begin{subfigure}[h]{0.65\textwidth}
         \centering
         \includegraphics[scale = 0.64]{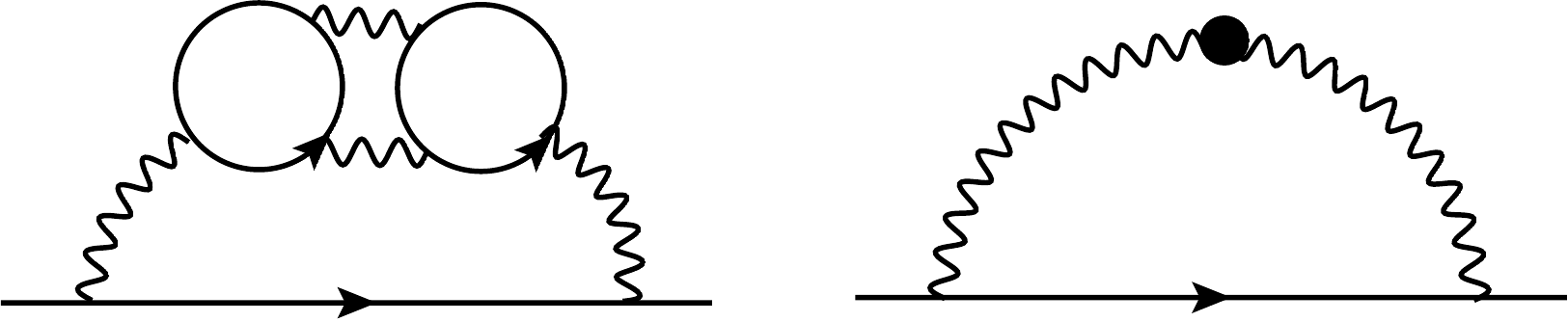}
         \caption{Nested Diagrams of One-Loop Fermion Diagram with AL diagram and its corresponding counterterm.}
         \label{FD: Nested}
     \end{subfigure}
     \hfill
     \begin{subfigure}[h]{0.5\textwidth}
         \centering
         \includegraphics[scale = 0.5]{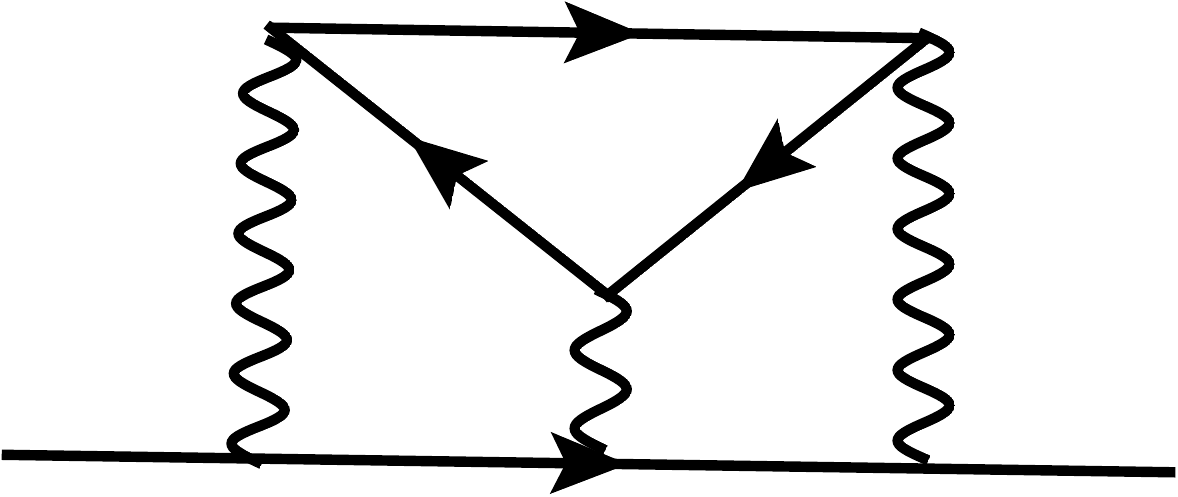}
         \caption{3-Loop Fermion Diagram.}
         \label{FD: 3LF}
     \end{subfigure}
        \caption{Diagrams contributing to the third leading order of fermion self-energy.  }
        \label{FD: all fermion 1/N2 diagrams}
\end{figure}

Next, we move to the self-energies at the third leading order of $1/N$. 
Because the goal is to compute the beta functions to the order of $1/N^2$, it is sufficient to keep only the divergent parts in the self-energies.
We start from the fermion self-energy. There are four non-zero diagrams as shown in Fig.~\ref{FD: all fermion 1/N2 diagrams}. The two-loop fermion diagram in Fig.~\ref{FD: 2LF} is convergent, while the divergence of the two nested diagrams shown in Fig.~\ref{FD: Nested} cancel with each other. Thus, the divergent part only comes from the three-loop diagram, shown in Fig.~\ref{FD: 3LF}. 

\begin{figure}
     \centering
     \begin{subfigure}[h]{0.35\textwidth}
         \centering
         \includegraphics[width=\textwidth]{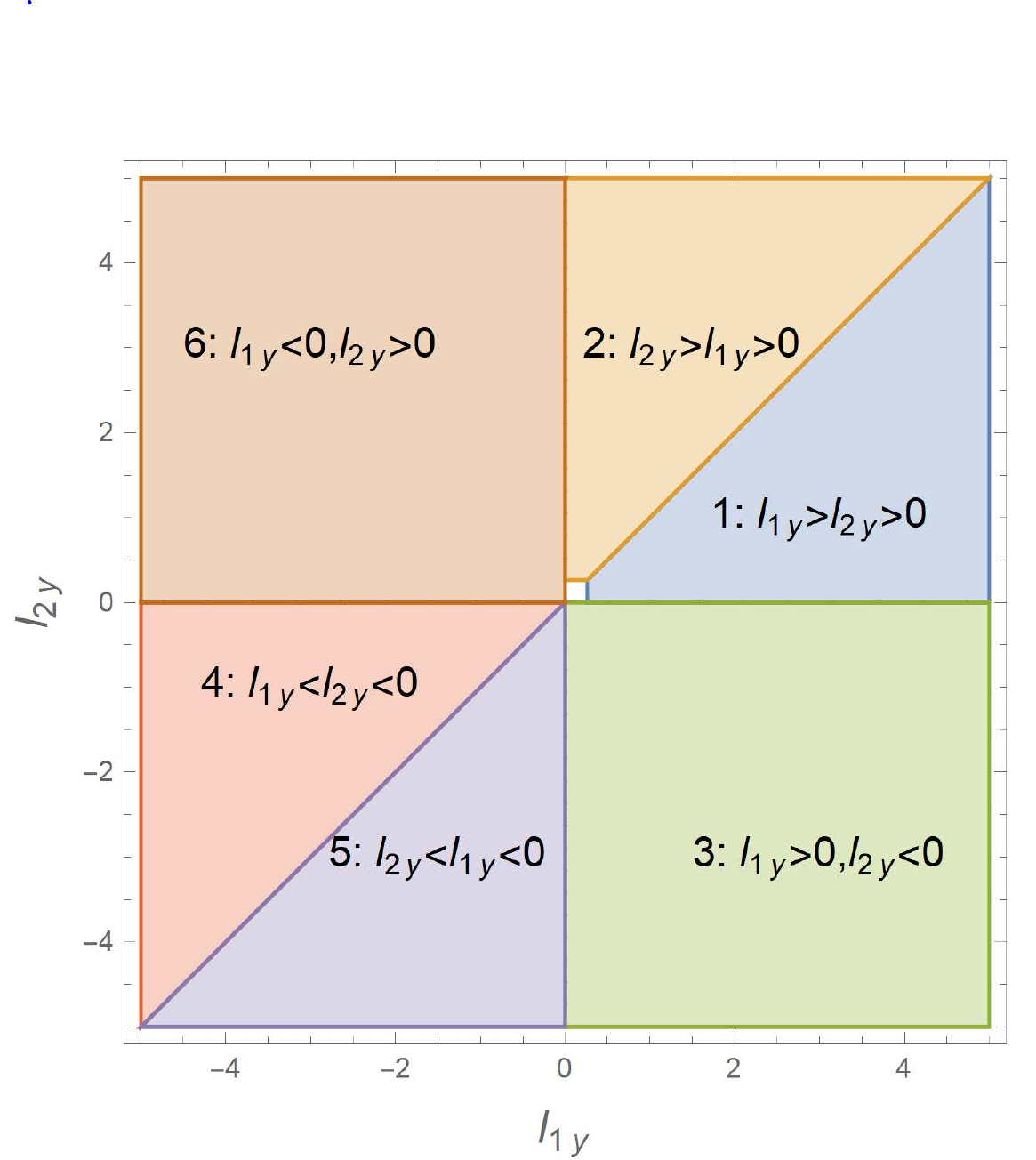}
         \caption{Our separtation of $l_{1y}$-$l_{2y}$ plane.}
         \label{fig: yregion}
     \end{subfigure}
     \hfill
     \begin{subfigure}[h]{0.63\textwidth}
         \centering
         \includegraphics[width=\textwidth]{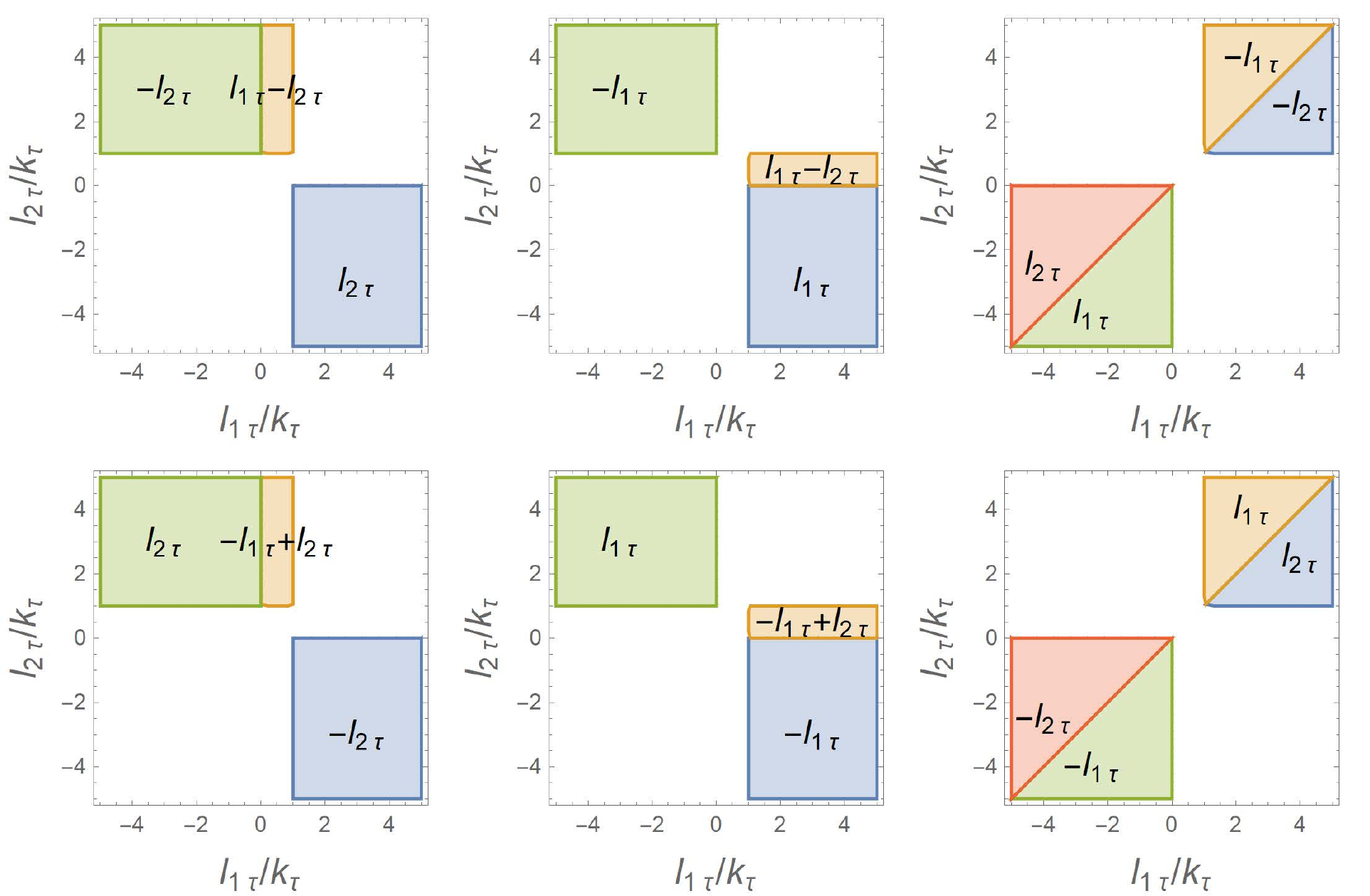}
         \caption{Result of $\mc{T}$ in each region. The units in the $l_{1\tau}$ and $l_{2\tau}$ axes are both $k_\tau$.}
         \label{fig: 3LFtregion}
     \end{subfigure}
        \caption{Result of $\mc{T}$ given different sections on the $l_{1y}-l_{2y}$ plane. Different sections of $l_{1y}-l_{2y}$ plane are labeled in Fig.~\ref{fig: yregion} and the result of $\mc{T}$ in each region is given in Fig.~\ref{fig: 3LFtregion} where each graph from left to right then from top down corresponds to the result in section 1-6 respectively.
        }
        \label{fig: 3LF graphs}
\end{figure}

In the three-loop diagram, the fermion in the top loop can run in the clockwise or counterclockwise direction. 
Moreover, only when the fermion running in the loop lies in a different $\pm$ patch than the external fermion \cite{Metlitski2010,Sur2014} will the diagram give logarithmic divergence. 
Let us assume that the external fermion lies in the $+$ patch. 
First, consider the case in which the fermion in the loop travels in counterclockwise direction. 
Its contribution reads
\beq
\begin{split}
\Sigma_{\text{3LF}}^a(k) = -\lambda_+^3 \lambda_-^3 \int\frac{[dl_1][dl_2][dp]}{(2\pi)^9}G_+(k-l_1) G_+(k-l_2)&G_-(p)G_-(p+l_1) G_-(p+l_2)\\
\times D(l_1) D(l_1-l_2) D(l_2).
\end{split}
\eeq
Integrating over $l_{1x}$, $l_{2x}$ 
results in
\beq
\begin{split}
\Sigma_{3LF}^a&(k) = -\int\frac{dp_\tau dp_x dp_y dl_{1\tau} dl_{1y} dl_{2\tau} dl_{2y}}{(2\pi)^7}\frac{\theta( l_{1\tau}-k_\tau)-\theta(- l_{1\tau}-p_\tau)}{i(k_\tau-p_\tau-2l_{1\tau})+2(k_y+p_y) l_{1y}-p_x+p_y^2-k_x-k_y^2}\\
&\times\frac{\theta(l_{2\tau}-k_\tau)-\theta(-l_{2\tau}-p_\tau)}{i(k_\tau-p_\tau-2l_{2\tau})+2(k_y+p_y) l_{2y}-p_x+p_y^2-k_x-k_y^2}\cdot\frac{1}{-ip_\tau - p_x +p_y^2}D(l_1) D(l_1-l_2) D(l_2).
\end{split}
\eeq
Integrating over $p_x$  and $p_y$, we obtain
\beq\label{eq: main form of 3LF}
\Sigma_{3LF}^a(k) = \int&\frac{dp_\tau  dl_{1\tau} dl_{1y} dl_{2\tau} dl_{2y}}{(2\pi)^5}\frac{D(l_1) D(l_1-l_2) D(l_2)}{2\left((l_2-l_1)_y G_+(k)^{-1} + 2i(l_{2y} l_{1\tau}-l_{1y} l_{2\tau})\right)}T(p_\tau; l_{1\tau}, l_{2\tau}; l_{1y}, l_{2y}; k_\tau),
\eeq
where $G_+(k)^{-1}$ is just the inverse of bare $+$ patch fermion propagator and 
\beq\label{eq: def of mathsfLF}
\begin{split}
T(p_\tau; l_{1\tau}, l_{2\tau}; l_{1y}, l_{2y}; k_\tau) =\left(\theta( l_{1\tau}-k_\tau)-\theta(- l_{1\tau}-p_\tau)\right)\left(\theta( l_{2\tau}-k_\tau)-\theta(- l_{2\tau}-p_\tau)\right) \\
\times\Bigg[\left(\theta\left(\frac{2l_{2\tau}-k_\tau}{l_{2y}}\right)-\theta\left(\frac{2l_{1\tau}-k_\tau}{l_{1y}}\right)\right)\theta(-p_\tau) + \left(\theta\left(\frac{2l_{1\tau}-k_\tau}{l_{1y}}\right)-\theta\left(\frac{l_{2\tau}-l_{1\tau}}{l_{2y} - l_{1y}}\right)\right)\theta\left(k_\tau-p_\tau-2l_{1\tau}\right)\\ 
-\left(\theta\left(\frac{2l_{2\tau}-k_\tau}{l_{2y}}\right)-\theta\left(\frac{l_{2\tau}-l_{1\tau}}{l_{2y} - l_{1y}}\right)\right)\theta\left(k_\tau-p_\tau-2l_{2\tau}\right)\Bigg].\\
\end{split}
\eeq
Suppose $k_\tau>0$. Separating $l_{1y}, l_{2y}$ plane into six regions as in Fig.~\ref{fig: yregion} and doing the integration of $p_\tau$, we obtain
\beq
\mathcal{T}(l_{1\tau}, l_{2\tau}; l_{1y}, l_{2y}; k_\tau) \equiv \int d p_\tau T(p_\tau; l_{1\tau}, l_{2\tau}; l_{1y}, l_{2y}; k_\tau) 
\eeq
in each region shown in Fig.~\ref{fig: 3LFtregion} ($k_\tau$ is set to be 1 in these graphs).

Note that
\beq
\mathcal{T}(l_{1\tau}, l_{2\tau}; -l_{1y}, -l_{2y}; k_\tau) = -\mathcal{T}(l_{1\tau}, l_{2\tau}; l_{1y}, l_{2y}; k_\tau)
\eeq
and therefore regions connected by $l_{1y}, l_{2y} \Longleftrightarrow -l_{1y}, -l_{2y}$ gives the same result. 
We can thus restrict to regions with $l_{1y} > 0$. First consider region 1. In order to isolate the divergent terms, let us do the following change of parameters: 
\beq
l_{2y} = l_{1y}\delta,~~~~~~l_{1\tau} = l_{1y}^{2}\Delta_1,~~~~~~l_{2\tau} = l_{1y}^{2}\Delta_2,~~~~~~k_\tau = l_{1y}^{2}\Delta_0.
\eeq
Moreover, we can write down $\mathcal{T}$ as
\beq
\mathcal{T}(l_{2\tau}, l_{1\tau}; l_{2y}, l_{1y}; k_\tau) = l_{1y}^{2}\mathcal{T}_0(\Delta_1, \Delta_2;\delta; \Delta_0).
\eeq
Then we have
\beq
\begin{split}
\Sigma_{3LF}^a(k) = &\frac{i \alpha^3}{2\pi^2}\int dl_{1y} d\delta d\Delta_1 d\Delta_2 
\frac{l_{1y}}{(\Delta_2-\delta \Delta_1) -\frac{1}{2}i(1-\delta)\frac{G_+(k)^{-1}}{l_{1y}^2}}\mathcal{T}_0\\
&\quad\quad\quad\quad\times\frac{1}{1 + \alpha|\Delta_1|}\frac{|\delta|}{|\delta|^{2} + \alpha|\Delta_2|}\frac{|1-\delta|}{|1-\delta|^{2} + \alpha|\Delta_1-\Delta_2|}.
\end{split}
\eeq
The divergent terms come from the integration region $l_{1y}\rightarrow\infty$, so let us do the integration of $l_{1y}$ from $k_y$, the dynamically generated IR cutoff, to infinity instead. We can thus first do a large $l_{1y}$ expansion in this region,
\beq
\Sigma_{3LF}^a(k) = \frac{i \alpha^3}{2\pi^2}\left(\sigma_1\int l_{1y} d l_{1y} + i\sigma_2~G_+(k)^{-1} \int l_{1y}^{-1} d l_{1y} + \sigma_3~ k_\tau \int l_{1y}^{-1} d l_{1y}+...\right),
\eeq
where
\beq
\begin{split}
\sigma_1 &= \int d\delta d\Delta_1 d\Delta_2\frac{1}{\Delta_2-\delta \Delta_1} \frac{1}{1 + \alpha|\Delta_1|}\frac{|\delta|}{|\delta|^{2} + \alpha|\Delta_2|}\frac{|1-\delta|}{|1-\delta|^{2} + \alpha|\Delta_1-\Delta_2|}\mathcal{T}_0(\Delta_1, \Delta_2;\delta; 0), \\
\sigma_2 &= \int d\delta d\Delta_1 d\Delta_2\frac{1-\delta}{2(\Delta_2-\delta \Delta_1)^2}\frac{1}{1 + \alpha|\Delta_1|}\frac{|\delta|}{|\delta|^{2} + \alpha|\Delta_2|}\frac{|1-\delta|}{|1-\delta|^{2} + \alpha|\Delta_1-\Delta_2|} \mathcal{T}_0(\Delta_1, \Delta_2;\delta; 0), \\
\sigma_3 &= \frac{d}{d\Delta_0}\left(\int d\delta d\Delta_1 d\Delta_2\frac{1}{\Delta_2-\delta \Delta_1}\frac{1}{1 + \alpha|\Delta_1|}\frac{|\delta|}{|\delta|^{2} + \alpha|\Delta_2|}\frac{|1-\delta|}{|1-\delta|^{2} + \alpha|\Delta_1-\Delta_2|} \mathcal{T}_0(\Delta_1, \Delta_2;\delta; \Delta_0)\right)\Bigg|_{\Delta_0=0}.\\
\end{split}
\eeq
The integration of the ... terms is convergent. The term proportional to $\sigma_1$ is polynomially divergent and does not contribute to the logarithmic divergence, so it should give an unimportant shift of chemical potential. The term proportional to $\sigma_2$ and $\sigma_3$ give the following logarithmically divergent contribution to $\Sigma^a_{3LF}(k)$:
\beq\label{eq: 3LF region 1}
\left(\frac{ \alpha^3}{2\pi^2} S^a(\alpha)G_+(k)^{-1}-\frac{ \alpha^3}{\pi^2} G^a(\alpha)i k_\tau\right)\ln\left(\frac{\Lambda_y}{k_y}\right),
\eeq
where $G^a(\alpha)$ and $S^a(\alpha)$ are defined as follows
\beq\label{eq: def of Ga}
G^a(\alpha) \equiv \int_0^1 d \delta \int_0^\infty d\Delta \frac{\delta}{\delta^2+\alpha\Delta}\frac{1-\delta}{(1-\delta)^2+\alpha\Delta},
\eeq
\beq\label{eq: def of Sa}
S^a(\alpha) \equiv \int_0^1 d \delta \int_0^\infty d \Delta_1 \int_0^\infty d \Delta_2 \frac{(1-\delta)\Delta_2}{(\Delta_2+\delta \Delta_1)^2}\frac{1}{1 + \alpha\Delta_1}\frac{\delta}{\delta^{2} + \alpha\Delta_2}\frac{1-\delta}{(1-\delta)^2 + \alpha(\Delta_1+\Delta_2)} .
\eeq
The calculation in the region 2 is completely analogous to the calculation in the region 1. 
If 1 and 2 are exchanged in Eqs. \eqref{eq: main form of 3LF} and \eqref{eq: def of mathsfLF}, one can see that the result in region 2 is identical to the result in region 1.
For region 3, it is clear that $\sigma_3 = 0$ while from $\sigma_2$ we obtain the divergent part
\beq
\frac{\alpha^3}{\pi^2}G_+(k)^{-1} T^a(\alpha)\ln\left(\frac{\Lambda_y}{k_y}\right),
\eeq
where
\beq\label{eq: def of Ta}
T^a(\alpha) \equiv \int_0^\infty d \delta \int_0^\infty d \Delta_1 \int_0^{\Delta_1} d \Delta_2 \frac{(1+\delta)\Delta_2}{(\Delta_2+\delta \Delta_1)^2}\frac{1}{1 + \alpha\Delta_1}\frac{\delta}{\delta^{2} + \alpha\Delta_2}\frac{1+\delta}{(1+\delta)^2 + \alpha(\Delta_1-\Delta_2)}.
\eeq
Adding the results of all regions up, we get the result for the case in which the fermion travels in counterclockwise direction:
\beq\label{eq: 3LFa div part}
\Sigma_{3LF}^a(k) \rightarrow \frac{ 2\alpha^3}{\pi^2}\left(S^a(\alpha) + T^a(\alpha)\right)G_+(k)^{-1}\ln\left(\frac{\Lambda_y}{k_y}\right)-\frac{ 4\alpha^3}{\pi^2}G^a(\alpha)i k_\tau\ln\left(\frac{\Lambda_y}{k_y}\right).
\eeq

The calculation of the case in which the fermion travels clockwise is completely analogous with similar expressions. The diagram reads
\beq
\begin{split}
\Sigma_{\text{3LF}}^b(k) = -\lambda_+^3 \lambda_-^3 \int\frac{[dl_1][dl_2][dp]}{(2\pi)^9}G_+(k-l_1) G_+(k-l_2)&G_-(p)G_-(p-l_1) G_-(p-l_2)\\
\times D(l_1) D(l_1-l_2) D(l_2).
\end{split}
\eeq
Integrate over $l_{1x}$, $l_{2x}$, $p_x$ and $p_y$ in order, we have
\beq
\Sigma_{3LF}^b(k) = -\int&\frac{dp_\tau dl_{1\tau} dl_{1y} dl_{2\tau} dl_{2y}}{(2\pi)^5}\frac{D(l_1) D(l_1-l_2) D(l_2)}{2\left[ (l_2-l_1)_y G_+(k)^{-1} + 2i \left(l_{2y} l_{1\tau}-l_{1y} l_{2\tau}-i(l_{1y}^2 l_{2y}-l_{1y} l_{2y}^2)\right)\right]
}T(p_\tau; l_{1\tau}, l_{2\tau}; l_{1y}, l_{2y}; k_\tau),
\eeq
where $T(p_\tau; l_{1\tau}, l_{2\tau}; l_{1y}, l_{2y}; k_\tau)$ is precisely the same as what was defined in Eq. \eqref{eq: def of mathsfLF}. Following closely the procedure outlined, we have
\beq\label{eq: 3LFb div part}
\Sigma_{3LF}^b(k) \rightarrow -\frac{ 2\alpha^3}{\pi^2}\left(S^b(\alpha) + T^b(\alpha)\right)G_+(k)^{-1}\ln\left(\frac{\Lambda_y}{k_y}\right)+\frac{ 4\alpha^3}{\pi^2}G^b(\alpha)i k_\tau\ln\left(\frac{\Lambda_y}{k_y}\right),
\eeq
where $G^b(\alpha)$ is defined as
\beq\label{eq: def of Gb}
G^b(\alpha)\equiv \int_0^1 d \delta \int_0^\infty d\Delta \frac{\delta}{\delta^2+\alpha\Delta}\frac{1-\delta}{(1-\delta)^2+\alpha\Delta} \frac{\Delta^2}{\Delta^2+\left(\delta^2-\delta\right)^2},
\eeq
and $S^b(\alpha)$ and $T^b(\alpha)$ are defined as
\beq\label{eq: def of Sb}
\begin{split}
S^b(\alpha) &\equiv \int_0^1 d \delta \int_0^\infty d \Delta_1 \int_0^\infty d \Delta_2 \frac{(1-\delta)\Delta_2\left((\Delta_2+\delta\Delta_1)^2-(\delta^2-\delta)^2 \right)}{\left((\Delta_2+\delta\Delta_1)^2+(\delta-\delta^2)^2 \right)^2}\\
&\quad\quad\quad\quad\quad\quad\quad\quad\quad\quad\quad\quad\times\frac{1}{1 + \alpha\Delta_1}\frac{\delta}{\delta^{2} + \alpha\Delta_2}\frac{1-\delta}{(1-\delta)^2 + \alpha(\Delta_1+\Delta_2)},
\end{split}
\eeq
\beq\label{eq: def of Tb}
\begin{split}
T^b(\alpha) &\equiv \int_0^\infty d \delta \int_0^\infty d \Delta_1 \int_0^{\Delta_1} d \Delta_2 \frac{(1+\delta)\Delta_2\left((\Delta_2+\delta\Delta_1)^2-(\delta^2+\delta)^2 \right)}{\left((\Delta_2+\delta\Delta_1)^2+(\delta^2+\delta)^2 \right)^2}\\
&\quad\quad\quad\quad\quad\quad\quad\quad\quad\quad\quad\quad\times\frac{1}{1 + \alpha\Delta_1}\frac{\delta}{\delta^{2} + \alpha\Delta_2}\frac{1+\delta}{(1+\delta)^2 + \alpha(\Delta_1-\Delta_2)}.
\end{split}
\eeq

Collecting the results for fermion self-energy at this order, we obtain the counter terms:
\beq
\begin{split}
&\delta_1^{(2)}=\frac{2\alpha^3}{\pi^2}\left(S^a(\alpha)+T^a(\alpha)+2G^a(\alpha)-S^b(\alpha)-T^b(\alpha)-2G^b(\alpha)\right)\ln\frac{\Lambda_y}{\mu}\\
&\delta_2^{(2)}=\frac{2\alpha^3}{\pi^2}\left(S^a(\alpha)+T^a(\alpha)-S^b(\alpha)-T^b(\alpha)\right)\ln\frac{\Lambda_y}{\mu}
\end{split}
\eeq
where $G^a(\alpha), S^a(\alpha), T^a(\alpha), G^b(\alpha), S^b(\alpha), T^b(\alpha)$ are defined in Eqs. \eqref{eq: def of Ga}, \eqref{eq: def of Sa}, \eqref{eq: def of Ta}, \eqref{eq: def of Gb}, \eqref{eq: def of Sb}, \eqref{eq: def of Tb} respectively.

\begin{figure}[h]
\begin{center}
\includegraphics[scale=0.32]{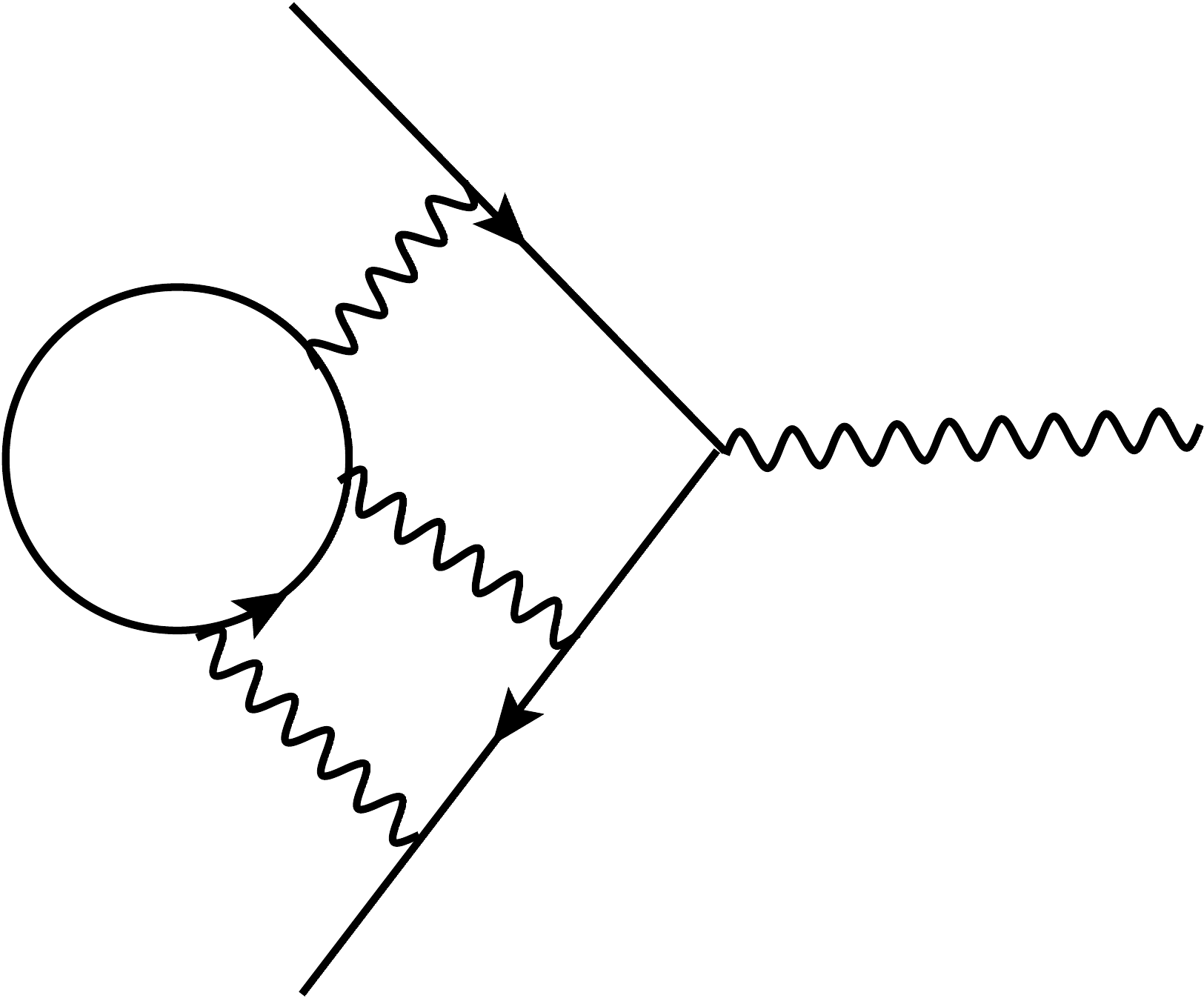}
\end{center}
\caption{Diagram contributing to the divergence of fermion-fermion-boson vertex at order $1/N^2$.}
\label{FD: vertex}
\end{figure}

Before we continue, let us mention that from the Ward identity \cite{Metlitski2010}, the fermion field strength divergence is the same as the divergence of fermion-gauge vertex, whose contribution to order $1/N^2$ comes solely from Fig.~\ref{FD: vertex} \cite{Holder2015b}.

\subsection{Third leading Order Result of Boson Self-Energy}

In this subsection, we evaluate the boson self energy at the third leading order and calculate $\delta_3^{(2)}$, the contribution to $Z_3$ at the order of $1/N^2$. Again, it is sufficient to keep only the divergent parts. It turns out that $\delta_3^{(2)}$ is double-logarithmic (double-log) divergent, \ie $\delta_3^{(2)}\propto\left(\ln\frac{\Lambda_y}{\mu}\right)^2$, with $\mu$ the renormalization scale. The leading contribution in the limit of $\mu\ll\Lambda_y$ is
\beq \label{eq: b e 2}
\delta_3^{(2)}=\frac{4\alpha^4}{3\pi^2}\left(\ln\frac{\Lambda_y}{\mu}\right)^2
\eeq
This result can be obtained by combining Eqs. \eqref{eq: double log Benz}, \eqref{eq: double log 3String} and the renormalization condition in Eq. \eqref{eq: renormalization conditions 1}. The rest of this subsection is devoted to the computational details.

\subsubsection{General Strategies of Calculation and Assumptions}\label{subsubapp: general strategies}

\begin{figure}[!htbp]
     \centering
     \begin{subfigure}[h]{0.45\textwidth}
     	\centering
     	\includegraphics[width=0.7\textwidth]{Benz.pdf}
     	\caption{Benz Diagram.}
     	\label{FD: Benz-app}
     \end{subfigure}
    \begin{subfigure}[h]{0.45\textwidth}
 	\centering
 	\includegraphics[width=\textwidth]{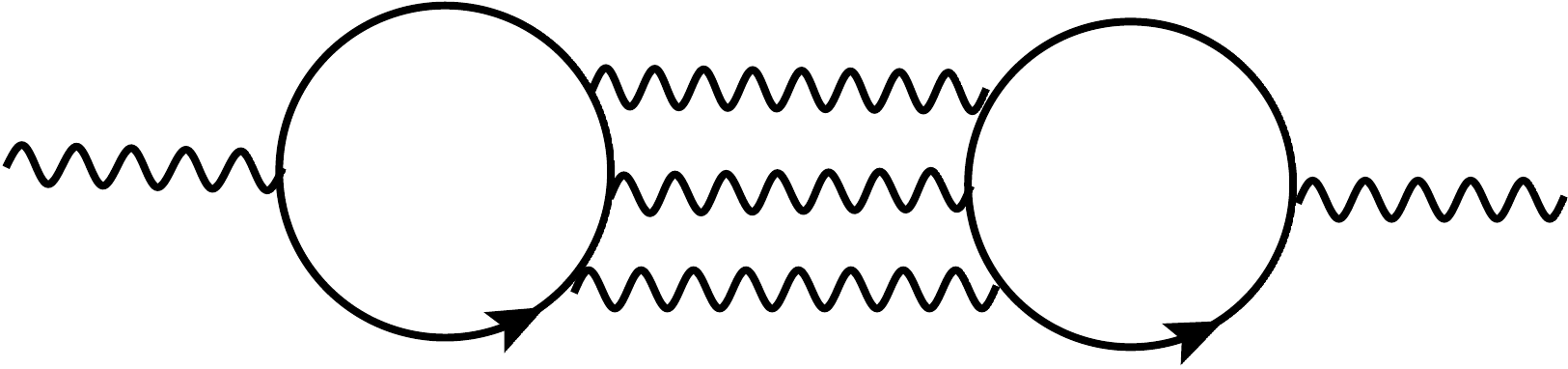}
 	\caption{3-String Diagram.}
 	\label{FD: 3String}
   \end{subfigure}
 
     \begin{subfigure}[h]{\textwidth}
         \centering
         \includegraphics[width=\textwidth]{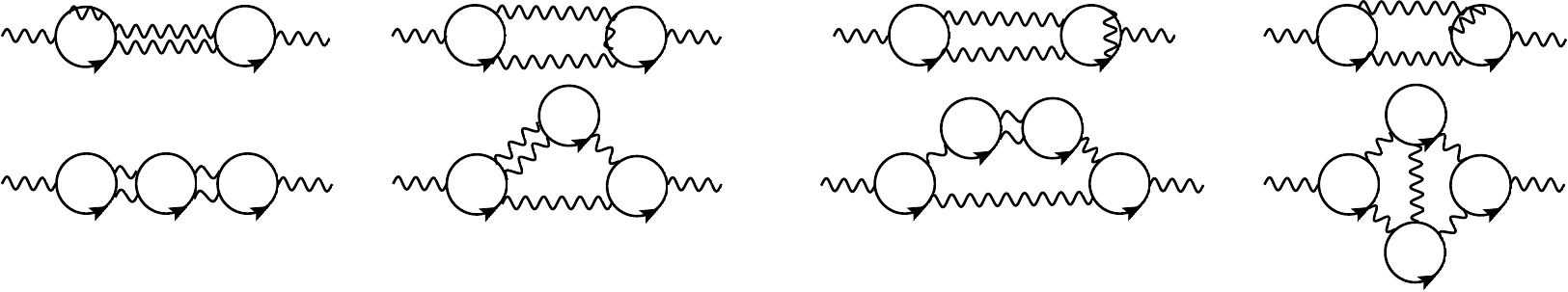}
         \caption{Feynman diagrams with one fermion loop attached to three boson lines.}
         \label{FD: SEFf}
     \end{subfigure}   
     \begin{subfigure}[h]{\textwidth}
         \centering
         \includegraphics[width=\textwidth]{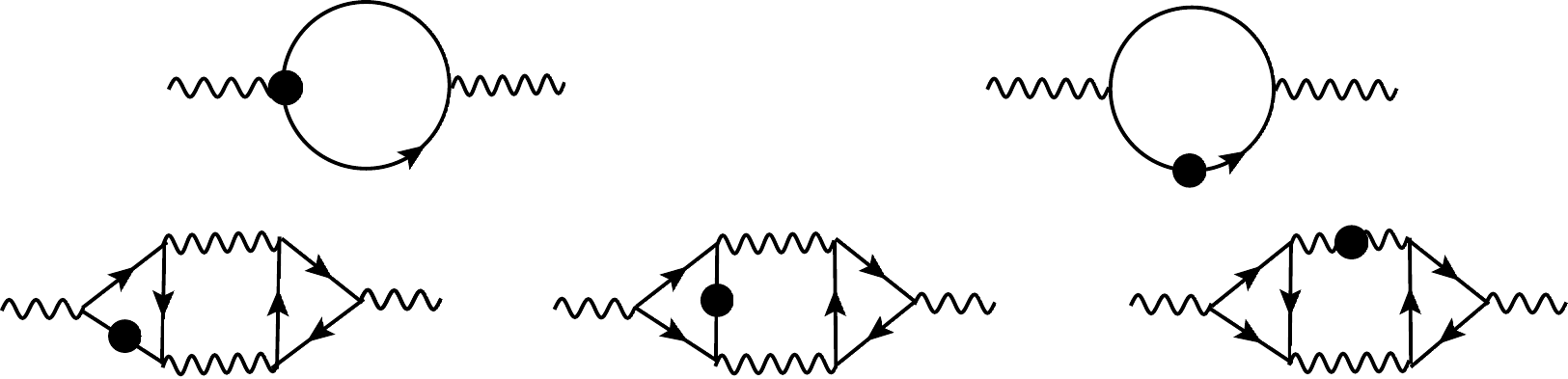}
         \caption{Feynman diagrams involving counterterms.}
         \label{FD: SEFC}
     \end{subfigure}
        \caption{Diagrams contributing to the third leading order of boson self-energy. We do not repeat all the possible reshuffling of boson lines when attached to a fermion loop. We also do not include Feynman diagrams with only one fermion loop, which are deemed to be zero in the static limit $k_\tau\rightarrow 0$ \cite{Lee2009,Metlitski2010}.}
        \label{FD: all 1/N boson self-energy diagrams}
\end{figure}

Since the higher-loop diagrams are much more difficult to compute explicitly, we begin by outlining  the general strategies we use and spell out assumptions we make to simplify our calculations. 

We draw all diagrams contributing to the third leading order, i.e. order $1/N$, of boson self-energy (see Fig. \ref{FD: all 1/N boson self-energy diagrams}). It turns out that some diagrams have internal boson momenta restricted as in Eq. \eqref{eq: 3-boson kinetic constraint}. 
Using the calculation of the AL diagram and the results in quantum field theory (QFT), we make the following {\it local divergence} assumption : {\it integration with restricted $y$-momentum is convergent when the Feynman diagram contains no divergent subdiagrams}. 
Hence, only two sets of diagrams can contribute to double-log divergence, which we call Benz diagram and 3-String diagram, and denote the results of these two sets of diagrams by $\Pi_{\rm Benz}(k_y)$ and $\Pi_{\rm 3String}(k_y)$, respectively.

For these two sets of diagrams, we can first use the residue theorem to integrate the $x$-momentum of all internal propagators, as well as frequency and $y$-momentum of the two fermion loops. Then we are left with the integration of internal boson frequency and $y$-momentum, denoted as $l_{i\tau}$ and $l_{iy}$, with $i=1, 2$ for two internal boson lines. It turns out that at $k_\tau=0$ these two contributions to the self energy take the form of $\Pi_{\rm Benz}(k_y)=\frac{1}{|k_y|}\int dl_{1\tau}dl_{1y}dl_{2\tau}dl_{2y}\mathsf{B}(k_y, l_{1\tau}, l_{1y}, l_{2\tau}, l_{2y})$ and $\Pi_{\rm 3String}(k_y)=\frac{1}{|k_y|}\int dl_{1\tau}dl_{1y}dl_{2\tau}dl_{2y}\mathsf{S}(k_y, l_{1\tau}, l_{1y}, l_{2\tau}, l_{2y})$, 
where $\mathsf{B}$ and $\mathsf{S}$ can be written as
\beq
\begin{split}
&\mathsf{B}(k_y, l_{1\tau}, l_{1y}, l_{2\tau}, l_{2y}) = \mathsf{B}(0, l_{1\tau}, l_{1y}, l_{2\tau}, l_{2y}) +a_{B}(l_{1\tau}, l_{1y}, l_{2\tau}, l_{2y})\cdot k_y+f_B(k_y, l_{1\tau}, l_{1y}, l_{2\tau}, l_{2y})\cdot k_y^2,\\
&\mathsf{S}(k_y, l_{1\tau}, l_{1y}, l_{2\tau}, l_{2y}) = \mathsf{S}(0, l_{1\tau}, l_{1y}, l_{2\tau}, l_{2y}) +a_S(l_{1\tau}, l_{1y}, l_{2\tau}, l_{2y})\cdot k_y+f_S(k_y, l_{1\tau}, l_{1y}, l_{2\tau}, l_{2y})\cdot k_y^2,
\end{split}
\eeq
where $a_{B,S}$ is the coefficient of the $k_y$-term in the Taylor expansion of $\mathsf{B}, \mathsf{S}(k_y, l_{1\tau}, l_{1y}, l_{2\tau}, l_{2y})$, and $f_{B,S}(k_y, l_{1\tau}, l_{1y}, l_{2\tau}, l_{2y})$ are some analytic functions of $k_y$ {\it defined} by the above equations. It turns out that the integrations of both $\mathsf{B}(0, l_{1\tau}, l_{1y}, l_{2\tau}, l_{2y})$ and $\mathsf{S}(0, l_{1\tau}, l_{1y}, l_{2\tau}, l_{2y})$ are polynomially divergent for the individual diagrams. 
In the end, they are canceled among all diagrams related to each other by rearranging the order in which photon lines attach to the fermion loops, e.g. between the Benz diagram in Fig.~\ref{FD: Benz-app} and its twin diagram in Fig.~\ref{FD: BenzTwin}, and among the six cases of 3-String diagrams shown in Fig.~\ref{FD: 3Stringextended} \cite{Holder2015}. Furthermore, $\Pi(k)$ is an even function of $k_y$ due to the inversion symmetry of the system, so the second terms (\ie the terms proportional to $k_y$) in the above equations do not contribute. Hence we just need to integrate $f_{B,S}(k_y, l_{1\tau}, l_{1y}, l_{2\tau}, l_{2y})$, with a cutoff $\Lambda_y$ for $l_{1y}$ and $l_{2y}$.
 
To extract potential double-log divergence in a simpler way, we apply the following {\it expansion trick} similar to the calculation of fermion self-energy. 
Since we are interested in the divergent part of the integration as $\Lambda_y\rightarrow\infty$, it is enough to consider the region with sufficiently large $l_{iy}$ and $l_{i\tau}$. 
Therefore, we can simply integrate $f(k_y, l_{1\tau}, l_{1y}, l_{2\tau}, l_{2y})$ 
at $k_y=0$ here, and impose suitable {\it dynamically generated IR  cutoff} for the integrals, \ie the scale above which $f(0, l_{1\tau}, l_{1y}, l_{2\tau}, l_{2y})$ is comparable with $f(k_y, l_{1\tau}, l_{1y}, l_{2\tau}, l_{2y})$. In this way, the coefficient of the double-log divergent term can be extracted. To check the validity of this strategy,
we compute the Benz diagram in Sec. \ref{subsubapp: Benz}
explicitly and confirm that it gives the same answer as the one obtained from the method  explained above.

\subsubsection{All Diagrams at the Third Leading Order}

All diagrams contributing to the boson self-energy
 to the third leading order in $1/N$ are listed in Fig.~\ref{FD: all 1/N boson self-energy diagrams}. 
For brevity, we do not repeat diagrams related to the ones listed through rearranging the order in which boson lines are attached to a fermion loop. 
These diagrams are related to each other by the fact that their summation vanishes when external momentum vanishes \cite{Holder2015}. Moreover, since we are mostly concerned with divergence in front of the term $|k_y|$ of boson self-energy, we do not include Feynman diagrams  which have only one fermion loop because they vanish in the static limit $k_\tau\rightarrow 0$ \cite{Lee2009,Metlitski2010}.

Because of the kinetic constraint mentioned in the calculation of AL diagram in \eqref{eq: 3-boson kinetic constraint} \cite{Holder2015b}, at the static limit $k_\tau=0$, at least one boson's $y$-momentum in Fig.~\ref{FD: SEFf} is bounded by external boson $y$-momentum $k_y$. 
Hence, {\it we conjecture that diagrams in Fig.~\ref{FD: SEFf} with no divergent subdiagrams are convergent.} 
In fact, the first two diagrams have divergent subdiagrams corresponding to one-loop fermion self-energy diagram as in Fig.~\ref{FD: 1LF}, whose divergence is cancelled by the third and fourth diagrams in Fig.~\ref{FD: SEFC} involving counterterms.\footnote{When we apply the Callan-Symanzik equation \eqref{eq: Callan-Symanzik 1} to the third leading order of $1/N$, $\mu \frac{\partial}{\partial \mu}$ acted on these two diagrams involving counterterms is nonzero, yet it is cancelled by $\gamma_z k_\tau\frac{\partial}{\partial k_\tau}$ acted on AL diagram.} 
We conjecture that the rest six diagrams are simply finite like the AL diagram. 
The first and second diagram in Fig.~\ref{FD: SEFC} only give divergence to the Landau damping term, \ie divergence in the term $|k_\tau|/|k_y|$, and cancel with each other due to the Ward identity. The fifth diagram in Fig.~\ref{FD: SEFC} is convergent.

Finally, the Benz diagram as in Fig.~\ref{FD: Benz-app} is explicitly calculated in Section \ref{subsubapp: Benz} while the calculation of the 3-String diagram in Fig.~\ref{FD: 3String} is   outlined in Section \ref{subsubapp: 3string} \footnote{See also Ref. \cite{Holder2015} for the calculation of 3-String diagram with quadratic boson kinetic term, i.e. at $\epsilon=1$, where some $(\ln\frac{\Lambda_y}{k_y})^5$ divergence is identified.}, with their double-log divergence given in Eqs
. \eqref{eq: double log Benz} and \eqref{eq: double log 3String} (note that they are identically the same).

\subsubsection{Benz Diagram}\label{subsubapp: Benz}

\begin{figure}[h]
\begin{center}
\includegraphics[scale=0.4]{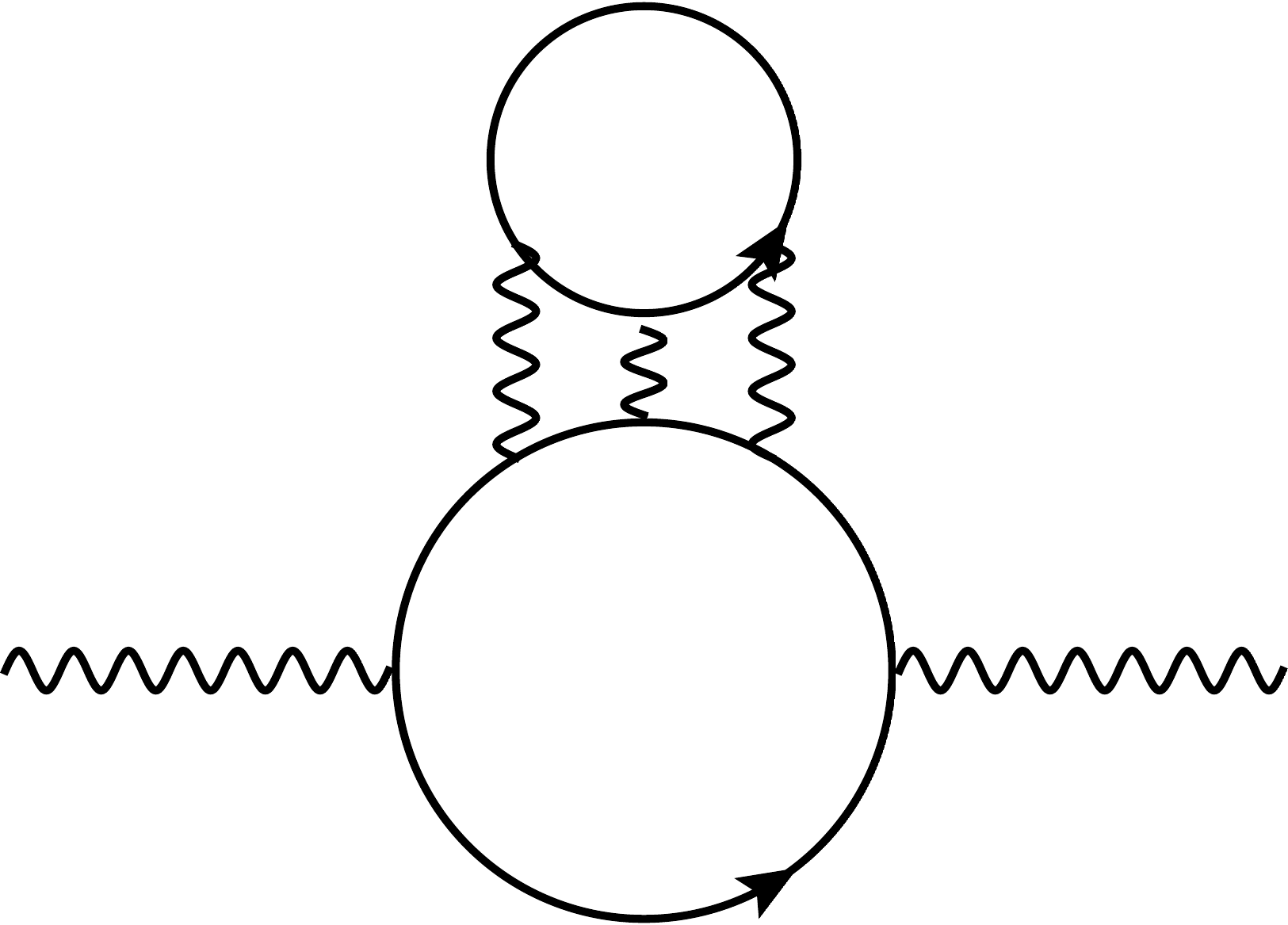}
\end{center}
\caption{The twin diagram of the Benz diagram after rearranging the order photon propagators attach to the fermion loop.} 
\label{FD: BenzTwin}
\end{figure}

Let us briefly sketch the calculation of the Benz diagram. 
Before the calculation, we mention that rearanging the order in which photon propagators attach to fermion loops results in a ``twin'' diagram of the Benz diagram (see Fig.~\ref{FD: BenzTwin}), which is simply the nested diagram of the 3-loop fermion self-energy diagram in Fig.~\ref{FD: 3LF} together with a single fermion loop. 
It only gives divergence to the Landau damping term $|k_\tau|/|k_y|$ that is canceled by the second diagram in Fig.~\ref{FD: SEFC} involving counterterms corresponding to fermion field strength renormalization. Similarly, the $|k_\tau|/|k_y|$ divergence of the Benz diagram is canceled by the first diagram in Fig.~\ref{FD: SEFC} involving counterterms corresponding to vertex renormalization, which can be explicity checked.

Considering different directions fermions in the loops of Benz diagram can travel as well as different patches the fermions can belong to, there are altogether $16$ different contributions from the Benz diagram, which we temporarily denote as $\Pi_\text{Benz} = \sum (\Pi_{\pm,C/A; \pm,C/A})$, 
where $\Pi_{+C, -A}$ represents the Benz diagram whose outer fermion loop lies on the $+$ patch and runs in the clockwise (C) direction, while the inner fermion loop lies on the $-$ patch and runs in the anti-clockwise (A) direction, etc. 
Unless the two fermion loops belong to the opposite patches, the diagram vanishes \cite{Sur2014,Lee2009}. 
Since the photon propagator is independent of the $x$-component momentum $k_x$, it is convenient to take $k_x=0$, and it is then straightforward to see that simultaneously flipping the patch labels of the two fermion loops does not change the result of the diagram, \ie $\Pi_{+C, -A}=\Pi_{-C, +A}$, etc. Furthermore, simultaneously flipping the directions in which fermions run in the two loops will complex conjugate the result, \ie $\Pi_{+A, -C}=\Pi^*_{+C, -A}$. 
Therefore, the sum of all Benz diagrams can be written as 
\beq\label{eq: Benz case assembling}
4{\rm Re}(\Pi_{+C, -C}+\Pi_{+C,-A}).
\eeq
Therefore, only two cases need to be considered : we can first set the outer loop fermion to be in the $+$ patch travelling clockwise, while the inner loop fermion is set to be in the $-$ patch travelling either clockwise (case $a$) or anti-clockwise (case $b$), corresponding to $\Pi_{+C, -C}$ and $\Pi_{+C, -A}$ respectively.

The Benz diagram in the two cases read
\beq
\begin{split}
\Pi_{\text{Benz}}^a(k) &= \lambda_+^5 \lambda_-^3 \int[d l_1][d l_2][d p][d q]~G_+(q)G_+(q+k)G_+(q+k-l_1)G_+(q-l_1)G_+(q-l_2)\\
&\quad\quad\quad\quad
\times G_-(p)G_-(p+l_1)G_-(p+l_2)D(l_1)D(l_2)D(l_1-l_2),
\end{split}
\eeq
\beq
\begin{split}
\Pi_{\text{Benz}}^b(k) &= \lambda_+^5 \lambda_-^3 \int[d l_1][d l_2][d p][d q]~G_+(q)G_+(q+k)G_+(q+k-l_1)G_+(q-l_1)G_+(q-l_2)\\
&\quad\quad\quad\quad
\times G_-(p)G_-(p-l_1)G_-(p-l_2)D(l_1)D(l_2)D(l_1-l_2).
\end{split}
\eeq
Using the residue theorem to calculate the integration of $l_{1x}$, $l_{2x}$, $p_x$, $q_x$, $p_y$ and $q_y$ in order, and taking the real part of the expression, we obtain the following expression, 
\beq\label{eq: Benz results overview}
\Pi_{\text{Benz}}'(k) \equiv{\rm Re}(\Pi_{\text{Benz}}^a + \Pi_{\text{Benz}}^b) = \frac{1}{16(2\pi)^6|k_y|}\int d l_{1\tau}d l_{1y}d l_{2\tau}d l_{2y} d p_\tau d q_\tau 
\left(\sum_{i=1}^6 \mathsf{B}_{i}\right)D(l_1)D(l_2)D(l_1-l_2),
\eeq
where $\mathsf{B}_{i}$ are defined as follows
\beq
\begin{split}
\mathsf{B}_{1} \equiv & \left(\theta(l_{2\tau}-q_\tau)-\theta(-l_{2\tau}-p_\tau)\right)\left(\theta(l_{1\tau}-k_\tau-q_\tau)-\theta(-l_{1\tau}-p_\tau)\right)\left(\theta(k_\tau-2l_{1\tau}-p_\tau+q_\tau)-\theta(-p_\tau)\right)\\
&\times\left(\theta(k_\tau-2l_{1\tau}+q_\tau)-\theta(k_{\tau}+q_\tau)\right)\left(\theta\left(\frac{l_{1\tau}}{l_{1y}}\right)-\theta\left(\frac{-k_{\tau}+2l_{1\tau}-2l_{2\tau}}{2l_{1y}-2l_{2y}}\right)\right) \\
&\times\left(\theta\left(\frac{k_\tau l_{1y}+2l_{1y}l_{2\tau}-2l_{1\tau}l_{2y}}{2l_{1y}}\right)-\theta(k_\tau)\right)\mc{B}_{1},
\end{split}
\eeq
\beq
\begin{split}
\mathsf{B}_{2} \equiv & \left(\theta(l_{2\tau}-q_\tau)-\theta(-l_{2\tau}-p_\tau)\right)\left(\theta(l_{1\tau}-k_\tau-q_\tau)-\theta(-l_{1\tau}-p_\tau)\right)\left(\theta(-2l_{2\tau}-p_\tau+q_\tau)-\theta(-p_\tau)\right)\\
&\times\left(\theta(-2l_{2\tau}+q_\tau)-\theta(k_{\tau}+q_\tau)\right)\left(\theta\left(\frac{-k_{\tau}+2l_{1\tau}-2l_{2\tau}}{2l_{1y}-2l_{2y}}\right)-\theta\left(\frac{k_{\tau}+2l_{2\tau}}{2l_{2y}}\right)\right) \\
&\times\left(\theta\left(\frac{k_\tau l_{1y}+2l_{1y}l_{2\tau}-2l_{1\tau}l_{2y}}{2l_{1y}}\right)-\theta(k_\tau)\right)\mc{B}_{1},
\end{split}
\eeq
\beq
\begin{split}
\mathsf{B}_{3} \equiv & \left(\theta(l_{2\tau}-q_\tau)-\theta(-l_{2\tau}-p_\tau)\right)\left(\theta(l_{1\tau}-k_\tau-q_\tau)-\theta(-l_{1\tau}-p_\tau)\right)\left(\theta(k_\tau-2l_{1\tau}-p_\tau+q_\tau)-\theta(-p_\tau)\right)\\
&\times\left(\theta(k_\tau-2l_{1\tau}+q_\tau)-\theta(q_\tau)\right)\left(\theta\left(\frac{-k_{\tau}+2l_{1\tau}-2l_{2\tau}}{2l_{1y}-2l_{2y}}\right)-\theta\left(\frac{-k_{\tau}+2l_{1\tau}}{2l_{1y}}\right)\right) \\
&\times\left(\theta\left(\frac{k_\tau l_{2y}+2l_{1y}l_{2\tau}-2l_{1\tau}l_{2y}}{2l_{2y}}\right)-\theta(k_\tau)\right)\mc{B}_{2},
\end{split}
\eeq
\beq
\begin{split}
\mathsf{B}_{4} \equiv & \left(\theta(l_{2\tau}-q_\tau)-\theta(-l_{2\tau}-p_\tau)\right)\left(\theta(l_{1\tau}-k_\tau-q_\tau)-\theta(-l_{1\tau}-p_\tau)\right)\left(\theta(-2l_{2\tau}-p_\tau+q_\tau)-\theta(-p_\tau)\right)\\
&\times\left(\theta(-2l_{2\tau}+q_\tau)-\theta(q_\tau)\right)\left(\theta\left(\frac{l_{2\tau}}{l_{2y}}\right)-\theta\left(\frac{-k_{\tau}+2l_{1\tau}-2l_{2\tau}}{2l_{1y}-2l_{2y}}\right)\right) \\
&\times\left(\theta\left(\frac{k_\tau l_{2y}+2l_{1y}l_{2\tau}-2l_{1\tau}l_{2y}}{2l_{2y}}\right)-\theta(k_\tau)\right)\mc{B}_{2},
\end{split}
\eeq
\beq
\begin{split}
\mathsf{B}_{5} \equiv & \left(\theta(l_{2\tau}-q_\tau)-\theta(-l_{2\tau}-p_\tau)\right)\left(\theta(l_{1\tau}-q_\tau)-\theta(-l_{1\tau}-p_\tau)\right)\left(\theta(-2l_{1\tau}-p_\tau+q_\tau)-\theta(-p_\tau)\right)\\
&\times\left(\theta(-2l_{1\tau}+q_\tau)-\theta(k_{\tau}+q_\tau)\right)\left(\theta\left(\frac{l_{1\tau}-l_{2\tau}}{l_{1y}-l_{2y}}\right)-\theta\left(\frac{k_{\tau}+2l_{1\tau}}{2l_{1y}}\right)\right) \\
&\times\left(\theta\left(\frac{k_\tau(l_{1y}-l_{2y})+2l_{1y}l_{2\tau}-2l_{1\tau}l_{2y}}{2(l_{1y}-l_{2y})}\right)-\theta(k_\tau)\right)\mc{B}_{3},
\end{split}
\eeq
\beq
\begin{split}
\mathsf{B}_{6} \equiv & \left(\theta(l_{2\tau}-q_\tau)-\theta(-l_{2\tau}-p_\tau)\right)\left(\theta(l_{1\tau}-q_\tau)-\theta(-l_{1\tau}-p_\tau)\right)\left(\theta(-2l_{2\tau}-p_\tau+q_\tau)-\theta(-p_\tau)\right)\\
&\times\left(\theta(-2l_{2\tau}+q_\tau)-\theta(k_{\tau}+q_\tau)\right)\left(\theta\left(\frac{k_{\tau}+2l_{2\tau}}{2l_{2y}}\right)-\theta\left(\frac{l_{1\tau}-l_{2\tau}}{l_{1y}-l_{2y}}\right)\right) \\
&\times\left(\theta\left(\frac{k_\tau(l_{1y}-l_{2y})+2l_{1y}l_{2\tau}-2l_{1\tau}l_{2y}}{2(l_{1y}-l_{2y})}\right)-\theta(k_\tau)\right)\mc{B}_{3},
\end{split}
\eeq
and $\mc{B}_i = \mc{B}^a_i - \mc{B}^b_i$ are defined as follows
\beq
\begin{split}\label{eq: mcBa}
\mc{B}^a_1&\equiv \frac{l_{1y} \left((l_{1y} l_{2\tau}-l_{1\tau} l_{2y})^2 + k_{y}^2 l_{1y}^2 l_{2y}(l_{1y}-l_{2y})\right)}{\left((l_{1y} l_{2\tau}-l_{1\tau} l_{2y})^2 + k_{y}^2 l_{1y}^2 l_{2y}^2\right)\left((l_{1y} l_{2\tau}-l_{1\tau} l_{2y})^2 + k_{y}^2 l_{1y}^2 (l_{1y}-l_{2y})^2\right)},\\
\mc{B}^a_2&\equiv \frac{l_{2y}}{(l_{1y} l_{2\tau}-l_{1\tau} l_{2y})^2 + k_{y}^2 l_{1y}^2 l_{2y}^2},\\
\mc{B}^a_3&\equiv \frac{l_{1y}-l_{2y}}{(l_{1y} l_{2\tau}-l_{1\tau} l_{2y})^2 + k_{y}^2 l_{1y}^2 (l_{1y}-l_{2y})^2},
\end{split}
\eeq
\beq\scriptsize
\begin{split}
\mc{B}^b_1&\equiv \frac{l_{1y} \left((l_{1y} l_{2\tau}-l_{1\tau} l_{2y})^2 -\left(l_{1y} l_{2y}^2-l_{1y}^2 l_{2y}+k_y l_{1y} l_{2y}\right)\left(l_{1y} l_{2y}^2-l_{1y}^2 l_{2y}-k_y l_{1y} (l_{1y} -l_{2y})\right)\right)}{\left((l_{1y} l_{2\tau}-l_{1\tau} l_{2y})^2+(l_{1y} l_{2y}^2-l_{1y}^2 l_{2y}+k_y l_{1y} l_{2y})^2\right)\left((l_{1y} l_{2\tau}-l_{1\tau} l_{2y})^2+(l_{1y} l_{2y}^2-l_{1y}^2 l_{2y}-k_y l_{1y}(l_{1y}- l_{2y}))^2\right)},\\
\mc{B}^b_2&\equiv \frac{l_{2y}\left((l_{1y} l_{2\tau}-l_{1\tau} l_{2y})^2-(l_{1y} l_{2y}^2-l_{1y}^2 l_{2y})(l_{1y} l_{2y}^2-l_{1y}^2 l_{2y}+k_y l_{1y} l_{2y})\right)}{\left((l_{1y} l_{2\tau}-l_{1\tau} l_{2y})^2+(l_{1y} l_{2y}^2-l_{1y}^2 l_{2y})^2\right)\left((l_{1y} l_{2\tau}-l_{1\tau} l_{2y})^2+(l_{1y} l_{2y}^2-l_{1y}^2 l_{2y}+k_y l_{1y} l_{2y})^2\right)},\\
\mc{B}^b_3&\equiv \frac{(l_{1y}-l_{2y})\left((l_{1y} l_{2\tau}-l_{1\tau} l_{2y})^2-(l_{1y} l_{2y}^2-l_{1y}^2 l_{2y})(l_{1y} l_{2y}^2-l_{1y}^2 l_{2y}-k_y l_{1y} (l_{1y} -l_{2y}))\right)}{\left((l_{1y} l_{2\tau}-l_{1\tau} l_{2y})^2+(l_{1y} l_{2y}^2-l_{1y}^2 l_{2y})^2\right)\left((l_{1y} l_{2\tau}-l_{1\tau} l_{2y})^2+\left(l_{1y} l_{2y}^2-l_{1y}^2 l_{2y}-k_y l_{1y}(l_{1y}- l_{2y})\right)^2\right)}.\\
\end{split}
\eeq
Note that $\mc{B}^a_1=\mc{B}^a_2+\mc{B}^a_3$ and $\mc{B}^b_1=\mc{B}^b_2+\mc{B}^b_3$. 

Setting $k_\tau=0$, we obtain
\beq\label{eq: Benz master Equation}
\begin{split}
\sum_{i=1}^6 \mathsf{B}_{i} =& \left(\hat{T_1}\theta\left(\frac{l_{1\tau}}{l_{1y}}\right) - \hat{T_2}\theta\left(\frac{l_{2\tau}}{l_{2y}}\right) + (\hat{T_2}-\hat{T_1})\theta\left(\frac{l_{1\tau}-l_{2\tau}}{l_{1y}-l_{2y}}\right)\right)\left(\theta(l_{1y})-\theta(l_{2y})\right)\text{sgn}\left(l_{1y} l_{2\tau}-l_{1\tau} l_{2y}\right)\mc{B}_{2}\\
&+ \left(l_1\rightarrow - l_1,~~~l_2\rightarrow l_2-l_1\right),
\end{split}
\eeq
where the second term denotes the expression coming from transforming the first expression accordingly and the simplification comes from the left-right flipping symmetry of the Benz diagram. Here $\hat{T}_{1,2}$ are defined as follows
\beq\label{eq: def of T1}
\begin{split}
\hat{T}_1\equiv &\left(\theta(l_{2\tau}-q_\tau)-\theta(-l_{2\tau}-p_\tau)\right)\left(\theta(l_{1\tau}-q_\tau)-\theta(-l_{1\tau}-p_\tau)\right)\\
&\left(\theta(-2l_{1\tau}-p_\tau+q_\tau)-\theta(-p_\tau)\right)\left(\theta(-2l_{1\tau}+q_\tau)-\theta(q_\tau)\right),
\end{split}
\eeq
\beq\label{eq: def of T2}
\begin{split}
\hat{T}_2\equiv &\left(\theta(l_{2\tau}-q_\tau)-\theta(-l_{2\tau}-p_\tau)\right)\left(\theta(l_{1\tau}-q_\tau)-\theta(-l_{1\tau}-p_\tau)\right)\\
&\left(\theta(-2l_{2\tau}-p_\tau+q_\tau)-\theta(-p_\tau)\right)\left(\theta(-2l_{2\tau}+q_\tau)-\theta(q_\tau)\right).
\end{split}
\eeq
Then the first term and the second term should give the same contribution to $\Pi^{a,b}$, and therefore we focus on the first term. We can now explicitly see that $\Pi^{a,b}$ are even functions of $k_y$, as expected.

Following the procedure outlined in Section \ref{subsubapp: general strategies}, we expand $\mc{B}_{2}^a$ at small $k_y$ and focus on terms proportional to $k_y^2$,
\beq\label{eq: ky2 term for B2a}
\mc{B}^a_2\rightarrow-\frac{l_{1y}^2l_{2y}^3}{(l_{1y}l_{2\tau}-l_{1\tau}l_{2y})^2\left((l_{1y}l_{2\tau}-l_{1\tau}l_{2y})^2+k_y^2 l_{1y}^2l_{2y}^2\right)}k_y^2\xrightarrow{f(k_y=0)\cdot k_y^2} -\frac{l_{1y}^2l_{2y}^3}{(l_{1y}l_{2\tau}-l_{1\tau}l_{2y})^4}k_y^2.
\eeq
We will just use the expression $f(0)k_y^2$ for most of our calculation, 
yet the full expression is important in identifying the dynamically generated IR cutoff for calculating the coefficient of the double-log. 
For $\mc{B}_2^b$, since only single-log is present, we just write down the expression of $f(0)k_y^2$ here for simplicity, 
\beq
\begin{split}
\mc{B}^b_2\rightarrow & l_{1y}^2l_{2y}^3\Bigg(-\frac{1}{\left((l_{1y}l_{2\tau}-l_{1\tau}l_{2y})^2+(l_{1y}l_{2y}^2-l_{1y}^2l_{2y})^2\right)^2} + \frac{8(l_{1y}l_{2y}^2-l_{1y}^2 l_{2y})^2}{\left((l_{1y}l_{2\tau}-l_{1\tau}l_{2y})^2+(l_{1y}l_{2y}^2-l_{1y}^2l_{2y})^2\right)^3}\\
&-\frac{8(l_{1y}l_{2y}^2-l_{1y}^2l_{2y})^4}{\left((l_{1y}l_{2\tau}-l_{1\tau}l_{2y})^2+(l_{1y}l_{2y}^2-l_{1y}^2l_{2y})^2\right)^4}\Bigg)k_y^2.
\end{split}
\eeq
Note that the relevant terms in $\sum_{i=1}^6\mathsf{B}_i$ are invariant if we simultaneously flip the sign of $l_{1y}$, $l_{2y}$ or $l_{1\tau}$, $l_{2\tau}$. 
Therefore, let us restrict to the integration region where $l_{1y}>0$ as well as $l_{1\tau}>0$. 
In this regime,
the first term of $\sum_{i=1}^6\mathsf{B}_i$ is nonzero only when $l_{2y}<0$ and $l_{2\tau}>0$, and after integrating over $p_\tau$ and $q_\tau$ it becomes $\mc{B}_2\times\min(l_{1\tau}, l_{2\tau})^2$.
Now we are left with the integral over $l_{1\tau}$ $l_{1y}$, and $l_{2\tau}$ and $l_{2y}$. 
To perform this integral, first consider the region where $l_{1\tau}>l_{2\tau}$. 
In order to isolate the divergence, we change the integration variables as
\beq
l_{2y} = -l_{1y}\delta,~~~~~~l_{1\tau} = l_{1y}^2\Delta_1,~~~~~~l_{2\tau} = l_{1y}^2\Delta_2.   
\eeq
After the integration of $\delta$, $\Delta_{1,2}$ as well as $l_{1y}$, we have the following logarithmic divergent piece of $\Sigma^{a,b}$ coming from the first term in $\sum_{i=1}^6\mathsf{B}_i$ in the region $l_{1y}>0$, $l_{2y}<0$, $l_{1\tau}>l_{2\tau}>0$, 
\beq\label{eq: Benz pre log one}
\frac{|k_y|}{e^2}\frac{\alpha^4}{4\pi^2}\left( \int\frac{d l_{1y}}{l_{1y}}\int_0^\infty d\delta\int_0^\infty d\Delta_1 \int_0^{\Delta_1} d\Delta_2~\Delta_2^2\delta^3 (\omega^a-\omega^b)\frac{1}{1+\alpha\Delta_1}\frac{\delta}{\delta^2+\alpha\Delta_2}\frac{1+\delta}{(1+\delta)^2+\alpha(\Delta_1-\Delta_2)}\right)
\eeq
where $\omega^a$ and $\omega^b$ comes from $\mc{B}^a_2$ and $\mc{B}^b_2$ respectively,
\beq\label{eq: def of omegaa}
\omega^a = \frac{1}{(\Delta_2+\delta  \Delta_1)^2\left((\Delta_2+\delta  \Delta_1)^2 + (k_y/l_{1y})^2\delta^2\right)}\xrightarrow{k_y=0} \frac{1}{(\Delta_2+\delta  \Delta_1)^4},
\eeq
\beq\label{eq: def of omegab}
\omega^b = \frac{1}{\left((\delta^2+\delta)^2+(\Delta_2+\delta  \Delta_1)^2\right)^2} -\frac{8 (\delta^2+\delta)^2}{\left((\delta^2+\delta)^2+(\Delta_2+\delta  \Delta_1)^2\right)^3} +\frac{8 (\delta^2+\delta)^4}{\left((\delta^2+\delta)^2+(\Delta_2+\delta  \Delta_1)^2\right)^4}.
\eeq
Next consider the region where $l_{2\tau}>l_{1\tau}$. To compare two regions let us do the following change of parameters in this region:
\beq
l_{2y} = -|l_{1y}|,~~~~~~l_{1y}=|l_{1y}|\delta,~~~~~~l_{1\tau} = l_{1y}^2\Delta_2,~~~~~~l_{2\tau} =l_{1y}^2\Delta_1.
\eeq
The logarithmic divergent piece in this region is then written as follows,
\beq\label{eq: Benz pre log two}
\frac{|k_y|}{e^2}\frac{\alpha^4}{4\pi^2} \left(\int\frac{d l_{1y}}{l_{1y}}\int_0^\infty d\delta\int_0^\infty d\Delta_1 \int_0^{\Delta_1} d\Delta_2~\Delta_2^2\delta^2 (\omega^a-\omega^b)\frac{1}{1+\alpha\Delta_1}\frac{\delta}{\delta^2+\alpha\Delta_2}\frac{1+\delta}{(1+\delta)^2+\alpha(\Delta_1-\Delta_2)}\right)
\eeq

Adding the contribution of all regions up is now straightforward, which is simply adding Eqs. \eqref{eq: Benz pre log one} and \eqref{eq: Benz pre log two} before multiplied by 4. The second term in $\sum_{i=1}^6\mathsf{B}_i$ gives an extra factor of 2 while different directions in which fermions travel as well as different fermion patch numbers give an extra factor of 4. 
Carefully considering all these factors, the Benz diagram gives the following log divergent contribution to boson self-energy
\beq
4\Pi_{\text{Benz}}'(k)\rightarrow \frac{|k_y|}{e^2}\frac{8\alpha^4}{\pi^2}\int\frac{d l_{1y}}{l_{1y}}(M^a(\alpha)-M^b(\alpha)),
\eeq
where $M_{a,b}(\alpha)$ are defined as
\beq
M^{a,b}(\alpha) \equiv \int_0^\infty d \delta \int_0^\infty d \Delta_1 \int_0^{\Delta_1} d \Delta_2~\Delta_2^2\delta^2(1+\delta)\omega^{a,b}\frac{1}{1 + \alpha\Delta_1}\frac{\delta}{\delta^{2} + \alpha\Delta_2}\frac{1+\delta}{(1+\delta)^2 +\alpha(\Delta_1-\Delta_2)}
\eeq
with $\omega^{a,b}$ defined in Eqs. \eqref{eq: def of omegaa} and \eqref{eq: def of omegab}.
At $k_y=0$, even though $M^b(\alpha)$ is just a regular function of $\alpha$, $M^a(\alpha)$ is not, because of the divergent integration coming from the region where $\Delta_1$ and $\Delta_2$ go to zero simultaneously, which can be translated to the dynamical region where 
\beq\label{eq: double log region}
|l_{1y}|, |l_{2y}| \gg |k_y|,~~~~~~|l_{1\tau}|, |l_{2\tau}| \approx |k_yl_{1y}|.
\eeq
Note that $|k_yl_{1y}|$ is approximately the on-shell energy of a fermion with $x$-momentum $l_{1y}^2$ and $y$-momentum $k_y+l_{1y}$. Therefore the integral is actually double-log divergent instead of single-log divergent.
Denoting the integrand of $M^a(\alpha)$ as $m^a(\alpha)$, we separate $M^a(\alpha)$ in the following way
\beq\label{eq: extracting double log Benz}
\begin{split}
M^a(\alpha)&\equiv\int_0^\infty d \delta \int_0^\infty d \Delta_1 \int_0^{\Delta_1} d \Delta_2~m^a(\alpha)\\
&= \left(\int_0^\infty d \delta \int_1^\infty d \Delta_1 \int_0^{\Delta_1} d \Delta_2~m^a(\alpha)\right) \\
&+ \left(\int_0^\infty d \delta \int_0^1 d \Delta_1 \int_0^{\Delta_1} d \Delta_2 \left(m^a(\alpha)-\frac{\Delta_2^2\delta}{(\Delta_2+\delta  \Delta_1)^2\left((\Delta_2+\delta  \Delta_1)^2 + (k_y/l_{1y})^2\delta^2\right)}\right)\right) \\
&+\Bigg(\int_0^\infty d \delta \int_0^1 d \Delta_1 \int_0^{\Delta_1} d\Delta_2 \frac{\Delta_2^2\delta}{(\Delta_2+\delta  \Delta_1)^2\left((\Delta_2+\delta  \Delta_1)^2 + (k_y/l_{1y})^2\delta^2\right)}\Bigg).
\\
\end{split}
\eeq
Then only the third term is IR divergent at $k_y=0$, which will be related to the double-log divergence we are after. 

In order to compute the coefficient of the double-log through the {\it expansion trick} sketched in Section \ref{subsubapp: general strategies}, we need to first identify the IR cutoff for $l_\tau$
and $l_{y}$. 
For this, we return to the full expression of the integrand,
\beq\label{eq: tem for extracting double log Benz}
\int_0^{\Lambda_y} d l_{1y} \int_0^{\Lambda_y} d l_{2y} \int_0^{l_{1y}^2} d l_{1\tau} \int_0^{l_{1\tau}} d l_{2\tau}\frac{l_{2\tau}^2 l_{1y} l_{2y}}{(l_{1y} l_{2\tau} - l_\tau l_{2y})^2\left((l_{1y} l_{2\tau} - l_\tau l_{2y})^2 + k_y^2 l_{1y}^2 l_{2y}^2\right)}.
\eeq
From Eq. \eqref{eq: ky2 term for B2a} or \eqref{eq: tem for extracting double log Benz}, the crossover scale appears when $(l_{1y} l_{2\tau} - l_\tau l_{2y})^2$ is of the same order as $k_y^2 l_{1y}^2 l_{2y}^2$, and therefore the dynamically generated IR cutoff for $l_\tau$ should be $k_y l_y$. Again, note that $|k_y l_{1y}|$ is approximately the on-shell energy of a fermion with $x$-momentum $l_{1y}^2$ and $y$-momentum $k_y+l_{1y}$. Accordingly, we can do the integration of $\Delta_1$ from $k_y/l_{1y}$ instead of 0 to 1, which gives us a log term. This log term is reminiscent of the BCS-log associated to the real Cooper pair \cite{Polchinski1992,Shankar1994}, which is also tied solely to the integration of frequency, and hence looks like a virtual BCS-log. Then the third term after the integration of $\delta, \Delta_1, \Delta_2$ as well as $l_{1y}$ is 
\beq
\int\frac{d l_{1y}}{l_{1y}}M^a(\alpha) \rightarrow\frac{1}{6}\int_{k_y}^{\Lambda_y}\frac{dl_{1y}}{l_{1y}}\int_{k_y/l_{1y}}^1\frac{d\Delta_1}{\Delta_1}=\frac{1}{12}\left(\ln\left(\frac{\Lambda_y}{k_y}\right)\right)^2.
\label{eq:doublelogexp}
\eeq

It is also possible to 
calculate the coefficient of the relevant double-log divergent terms explicitly.
Here, it is possible to do the explicit integration. 
After integrating over $\Delta_1$, $\Delta_2$ and $\delta$, the result is a function $f(k_y/l_{1y})$, explicitly written out as follows,
\beq
f(x) = \frac{x  \left(\left(x ^2+3\right) \ln \left(x ^2+1\right) -2x^2\ln (x )-4\right)+4 \arctan(x )}{12 x ^3}.
\eeq
and the third term now becomes $\int\frac{dl_{1y}}{l_{1y}}f\left(\frac{k_y}{l_{1y}}\right)$. From the limiting behavior of $f(x)$ at small and large $x$, 
\beq
f(x)=\Bigg\{
\begin{array}{rcl}
-\frac{1}{6}\ln(x) & & x\ll 1\\
\frac{1}{2}\frac{\ln(x)}{x^2} & & x\gg 1
\end{array} 
\eeq
we see that the integration $\int \frac{dx}{x} f(\frac{1}{x})$ is IR convergent and UV double-log divergent. Hence the integration of $l_{1y}$ is IR convergent and UV double-log divergent. This integration can be carried out explicitly as well, which at large $\Lambda_y/k_y$ is 
\beq 
\label{eq: Benz exact}
\frac{1}{12}\left(\ln\left(\frac{\Lambda_y}{k_y}\right)\right)^2 + \frac{5}{36}\ln\left(\frac{\Lambda_y}{k_y}\right) + \frac{3\pi^2+38}{432}.
\eeq
The double-log term 
in \eqref{eq: Benz exact}
agrees with Eq.  \eqref{eq:doublelogexp} that is obtained from the simplified scheme.
This yields the promised double-log divergent term
\beq\label{eq: double log Benz}
\Pi_{\text{Benz}}(k)\rightarrow \frac{|k_y|}{e^2}\frac{2\alpha^4}{3\pi^2}\left(\ln\left(\frac{\Lambda_y}{k_y}\right)\right)^2.
\eeq

\subsubsection{3-String Diagram}\label{subsubapp: 3string}

\begin{figure}[h]
\begin{center}
\includegraphics[width=\linewidth]{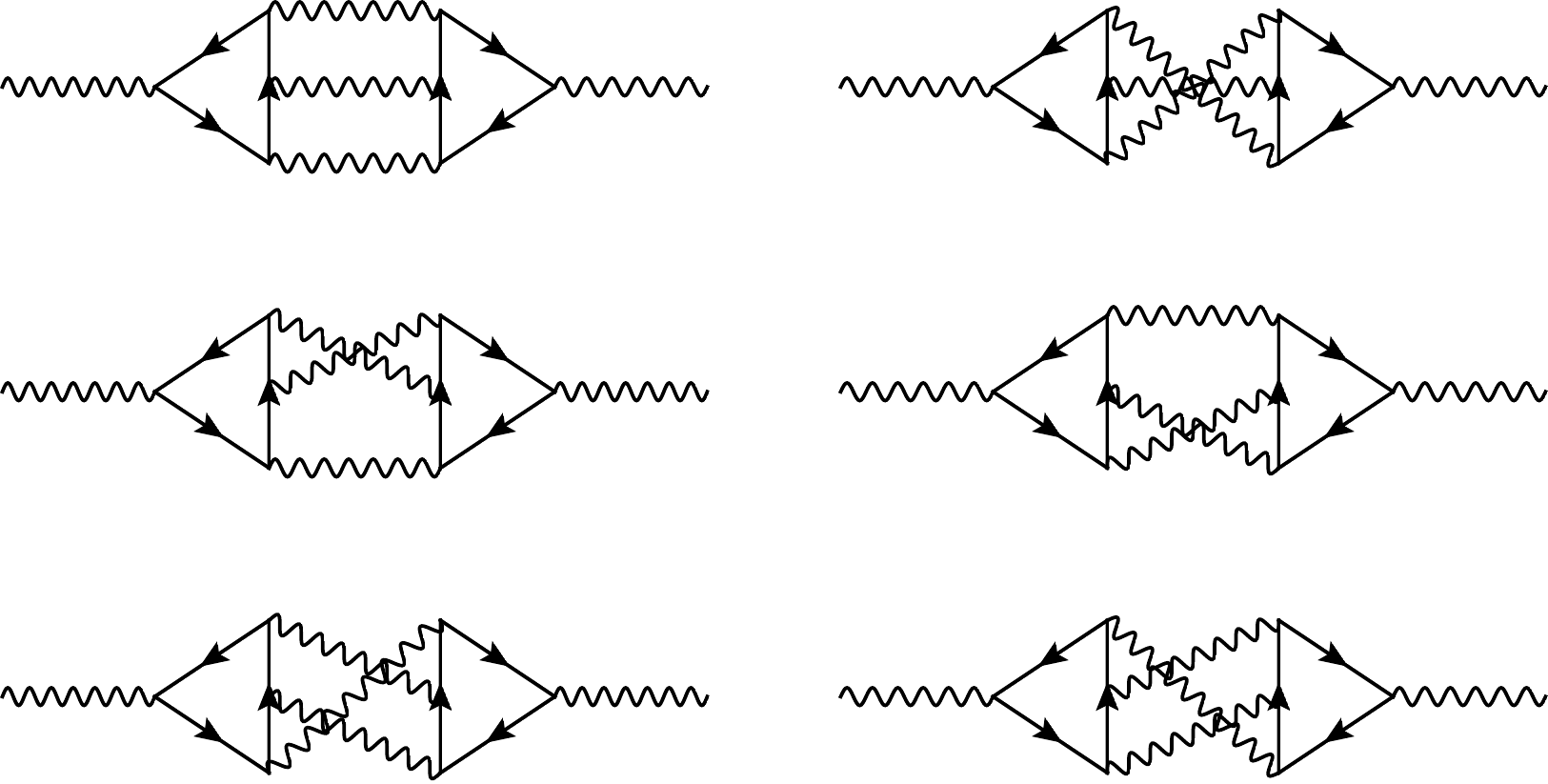}
\end{center}
\caption{All possible 3-String diagrams depending on how three internal photon propagators attach to the fermion loops. From left to right then from top down are case $a$, $b$, $c$, $d$, $e$ and $f$ respectively.} 
\label{FD: 3Stringextended}
\end{figure}

Finally, let us briefly sketch the calculation of the 3-String diagram. 
For simplicity, we focus on the case where $k_\tau=0$ whenever possible. 
As in previous cases, the two fermion loops must belong to different  patches to give rise to UV divergence. 
Since the photon propagator is independent of the $x$-component momentum $k_x$,
we can set the left fermion loop to be in the $-$ patch without loss of generality. 
We can also fix that fermion in the left loop runs in counterclockwise direction and fermion in the right loop runs in clockwise direction. 
Then there are 6 possible ways in which the three internal photon propagators attach to the two loops, shown in Fig.~\ref{FD: 3Stringextended}. 
After simultaneously flipping the directions in which fermions run in the two loops, 
diagram $a$ and $b$ do not change, while $c, d$ and $e, f$ transform into each other respectively. 
Therefore, the contribution of the 3-String diagram as a whole becomes
\beq\label{eq: 3String case assembling}
2(\Pi^a+\Pi^b)+4\text{Re}(\Pi^c+\Pi^e),
\eeq
where
\beq
\begin{split}
\Pi_{\text{3String}}^a(k) &= \lambda_+^4 \lambda_-^4 \int[d l_1][d l_2][d p][d q]~G_-(p)G_-(p+k)G_-(p+l_1+k/2)G_-(p+l_2+k/2)\\
&\quad\quad
\times G_+(q)G_+(q-k)G_+(q-l_1-k/2)G_+(q-l_2-k/2)D(k/2+l_1)D(k/2-l_2)D(l_1-l_2),
\end{split}
\eeq
\beq
\begin{split}
\Pi_{\text{3String}}^b(k) &= \lambda_+^4 \lambda_-^4 \int[d l_1][d l_2][d p][d q]~G_-(p)G_-(p+k)G_-(p+l_1+k/2)G_-(p+l_2+k/2)\\
&\quad\quad
\times G_+(q)G_+(q+k)G_+(q+l_1+k/2)G_+(q+l_2+k/2)D(k/2+l_1)D(k/2-l_2)D(l_1-l_2),
\end{split}
\eeq
\beq
\begin{split}
\Pi_{\text{3String}}^c(k) &= \lambda_+^4 \lambda_-^4 \int[d l_1][d l_2][d p][d q]~G_-(p)G_-(p+k)G_-(p+l_1+k/2)G_-(p+l_2+k/2)\\
&\quad\quad
\times G_+(q)G_+(q+l_1+k/2)G_+(q+l_2+k/2)G_+(q+l_2-k/2)D(k/2+l_1)D(k/2-l_2)D(l_1-l_2),
\end{split}
\eeq
\beq
\begin{split}
\Pi_{\text{3String}}^e(k) &= \lambda_+^4 \lambda_-^4 \int[d l_1][d l_2][d p][d q]~G_-(p)G_-(p+k)G_-(p+l_1+k/2)G_-(p+l_2+k/2)\\
&\quad\quad
\times G_+(q)G_+(q-l_1-k/2)G_+(q-l_2-k/2)G_+(q-l_2+k/2)D(k/2+l_1)D(k/2-l_2)D(l_1-l_2).
\end{split}
\eeq
Again, we use the residue theorem to calculate the integration of $l_{1x}$, $l_{2x}$, $p_x$, $q_x$, $p_y$ and $q_y$ in order. The expression at non-zero $k_\tau$ is extremely lengthy, but there is some simplification when $k_\tau=0$:
\beq\label{eq: 3String results overview}
\Pi_{\text{3String}}^{a,b,c,e}(k_y) = \frac{\sgn(k_y)}{(2\pi)^6|k_y|}\int d l_{1\tau}d l_{1y}d l_{2\tau}d l_{2y} d p_\tau d q_\tau \left(-\hat{T}_1\mathsf{S}_1^{a,b,c,e}+\hat{T}_2\mathsf{S}_2^{a,b,c,e}\right) D(k/2+l_1)D(k/2-l_2)D(l_2-l_1),
\eeq
where $\hat{T}_{1,2}$ are defined in \eqref{eq: def of T1} and \eqref{eq: def of T2} and
\beq
\begin{split}
\mathsf{S}_{i}^{a,b} \equiv & \left(\theta\left(\frac{2l_{i\tau}}{2l_{iy}-k_y}\right) - \theta\left(\frac{2l_{i\tau}}{2l_{iy}+k_y}\right)\right)
\left(\theta\left(\frac{2l_{i\tau}}{2l_{iy}-k_y}\right) \mc{S}^{a,b}_{i,-}- \theta\left(\frac{2l_{i\tau}}{2l_{iy}+k_y}\right)\mc{S}^{a,b}_{i,+}\right)\\
&+\Bigg(\left(\theta\left(\frac{2l_{i\tau}}{2l_{iy}-k_y}\right) - \theta\left(\frac{l_{1\tau}-l_{2\tau}}{l_{1y}-l_{2y}}\right)\right)\theta\left(\frac{k_y(l_{1\tau}-l_{2\tau})+2l_{1y}l_{2\tau}-2l_{1\tau}l_{2y}}{k_y(l_{1y}-l_{2y})}\right)\\
&-\left(\theta\left(\frac{2l_{i\tau}}{2l_{iy}+k_y}\right)
- \theta\left(\frac{l_{1\tau}-l_{2\tau}}{l_{1y}-l_{2y}}\right)\right)\theta\left(\frac{k_y(l_{1\tau}-l_{2\tau})-2l_{1y}l_{2\tau}+2l_{1\tau}l_{2y}}{k_y(l_{1y}-l_{2y})}\right)\Bigg)\mc{S}^{a,b}_0,
\end{split}
\eeq
\beq
\begin{split}
\mathsf{S}_1^{c,e} \equiv & \left(\theta\left(\frac{2l_{1\tau}}{2l_{1y} - k_y}\right) - \theta\left(\frac{2l_{1\tau}}{2l_{1y} + k_y}\right)\right)
\theta\left(\frac{2l_{1\tau}}{2l_{1y}+k_y}\right) \mc{S}^{c,e}_1\\
&+ \left(\theta\left(\frac{2l_{1\tau}}{2l_{1y}+k_y}\right) - \theta\left(\frac{l_{1\tau}-l_{2\tau}}{l_{1y}-l_{2y}}\right)\right)
\theta\left(\frac{k_y(l_{1\tau}-l_{2\tau})-2l_{1y}l_{2\tau}+2l_{1\tau}l_{2y}}{k_y(2l_{1y}+k_y)}\right) \mc{S}^{c,e}_1\\
&-\left(\theta\left(\frac{2l_{1\tau}}{2l_{1y}-k_y}\right)
- \theta\left(\frac{l_{1\tau}-l_{2\tau}}{l_{1y}-l_{2y}}\right)\right)\Bigg(\theta\left(\frac{k_y(l_{1\tau}-l_{2\tau})+2l_{1y}l_{2\tau}-2l_{1\tau}l_{2y}}{k_y(-2l_{2y}+k_y)}\right) \mc{S}^{c,e}_2\\
&+\theta\left(\frac{k_y(l_{1\tau}-l_{2\tau})+2l_{1y}l_{2\tau}-2l_{1\tau}l_{2y}}{k_y(l_{1y}-l_{2y})}\right)\mc{S}^{c,e}_0\Bigg),
\end{split}
\eeq
\beq
\begin{split}
\mathsf{S}_2^{c,e} \equiv & \left(\theta\left(\frac{2l_{2\tau}}{2l_{2y}+k_y}\right) - \theta\left(\frac{2l_{2\tau}}{2l_{2y}-k_y}\right)\right)
\left(\theta\left(\frac{2l_{2\tau}}{2l_{2y}-k_y}\right) \mc{S}^{c,e}_{2,-}-\theta\left(\frac{2l_{2\tau}}{2l_{2y}+k_y}\right) \mc{S}^{c,e}_{2,+}\right)\\
&+ \left(\theta\left(\frac{2l_{2\tau}}{2l_{2y}+k_y}\right) - \theta\left(\frac{l_{1\tau}-l_{2\tau}}{l_{1y}-l_{2y}}\right)\right)
\theta\left(\frac{k_y(l_{1\tau}-l_{2\tau})-2l_{1y}l_{2\tau}+2l_{1\tau}l_{2y}}{k_y(2l_{1y}+k_y)}\right) \mc{S}^{c,e}_1\\
&-\left(\theta\left(\frac{2l_{2\tau}}{2l_{2y}-k_y}\right)
- \theta\left(\frac{l_{1\tau}-l_{2\tau}}{l_{1y}-l_{2y}}\right)\right)\Bigg(\theta\left(\frac{k_y(l_{1\tau}-l_{2\tau})+2l_{1y}l_{2\tau}-2l_{1\tau}l_{2y}}{k_y(-2l_{2y}+k_y)}\right) \mc{S}^{c,e}_2\\
&+\theta\left(\frac{k_y(l_{1\tau}-l_{2\tau})+2l_{1y}l_{2\tau}-2l_{1\tau}l_{2y}}{k_y(l_{1y}-l_{2y})}\right)\mc{S}^{c,e}_0\Bigg),
\end{split}
\eeq
where the most important $\mc{S}$ factors for case $a$ are defined as follows,
\beq
\begin{split}
\mc{S}^a_{i,\pm} &\equiv \frac{k_y \pm 2l_{iy}}{16k_y l_{i\tau}\left(k_y(l_{1\tau}-l_{2\tau})\mp 2(l_{1y}l_{2\tau}-l_{1\tau}l_{2y})\right)},\\
\mc{S}^a_0&\equiv \frac{l_{1y}-l_{2y}}{4 \left(k_y(l_{1\tau}-l_{2\tau})+2l_{1y}l_{2\tau}-2l_{1\tau}l_{2y}\right)\left(k_y(l_{1\tau}-l_{2\tau})-2l_{1y}l_{2\tau}+2l_{1\tau}l_{2y}\right)},
\end{split}
\eeq
and for case $b$,
\beq
\begin{split}
\mc{S}^b_{i,\pm}  \equiv & (k_y\pm 2l_{iy})\Bigg/\Big\{2 k_y\left( 4 l_{i\tau}-i (k_y^2-4 l_{iy}^2)\right)\\
& \times\left(2\left(k_y(l_{1\tau}-l_{2\tau})\mp 2\left(l_{1y}l_{2\tau}-l_{1\tau}l_{2y}\right)\right)\pm i\left(k_y^2(l_{1y}-l_{2y})+4(l_{1y}^2l_{2y}-l_{2y}^2l_{1y})\right)+2ik_y(l_{1y}^2-l_{2y}^2)\right)\Big\},\\
\mc{S}^b_0  \equiv & (l_{1y}-l_{2y})\Bigg/\Big\{\left(2\left(k_y(l_{1\tau}-l_{2\tau})+ 2\left(l_{1y}l_{2\tau}-l_{1\tau}l_{2y}\right)\right)- i\left(k_y^2(l_{1y}-l_{2y})+4(l_{1y}^2l_{2y}-l_{2y}^2l_{1y})\right)+2ik_y(l_{1y}^2-l_{2y}^2)\right)\\
& \times\left(2\left(k_y(l_{1\tau}-l_{2\tau})- 2\left(l_{1y}l_{2\tau}-l_{1\tau}l_{2y}\right)\right)+ i\left(k_y^2(l_{1y}-l_{2y})+4(l_{1y}^2l_{2y}-l_{2y}^2l_{1y})\right)+2ik_y(l_{1y}^2-l_{2y}^2)\right)\Big\},\\
\end{split}
\eeq
and for case $c$,
\beq\scriptsize
\begin{split}
\mc{S}_1^c\equiv& (-k_y-2 l_{1y})\Bigg/\Big\{2 \left(k_y^2 l_{1y}-k_y^2 l_{2y}-2 i k_y l_{1\tau}+2 k_y l_{1y}^2+2 i
   k_y l_{2\tau}-2 k_y l_{2y}^2-4 i l_{1\tau} l_{2y}+4 l_{1y}^2 l_{2y}+4 i l_{1y} l_{2\tau}-4 l_{1y} l_{2y}^2\right)\\
   &\times\left(k_y^2 l_{1y}-k_y^2 l_{2y}-2 i k_y l_{1\tau}+2 k_y l_{1y}^2-4 k_y l_{1y} l_{2y}-2 i k_y l_{2\tau}+2 k_y
   l_{2y}^2+4 i l_{1\tau} l_{2y}-4 l_{1y}^2 l_{2y}-4 i l_{1y} l_{2\tau}+4 l_{1y} l_{2y}^2\right)\Big\},\\
\mc{S}_2^c\equiv& (-k_y+2 l_{2y})\Bigg/\Big\{2 \left(-k_{y}^2 l_{1y}+k_{y}^2 l_{2y}+2 i k_{y} l_{1\tau}-2 k_{y} l_{1y}^2+4 k_{y} l_{1y} l_{2y}-2 i
   k_{y} l_{2\tau}-2 k_{y} l_{2y}^2-4 i l_{1\tau} l_{2y}+4 l_{1y}^2 l_{2y}+4 i l_{1y} l_{2\tau}-4 l_{1y} l_{2y}^2\right)\\
   &\times\left(-k_{y}^2 l_{1y}+k_{y}^2 l_{2y}+2 ik_{y} l_{1\tau}-2 k_{y} l_{1y}^2+4
   k_{y} l_{1y} l_{2y}+2 ik_{y} l_{2\tau}-2 k_{y} l_{2y}^2-4 il_{1\tau} l_{2y}+4 l_{1y}^2 l_{2y}+4 il_{1y} l_{2\tau}-4 l_{1y} l_{2y}^2\right)\Big\},\\
\mc{S}_0^c=&\mc{S}_1^c-\mc{S}_2^c\equiv(-l_{1y} + l_{2y})\Bigg/\Big\{ \left(k_y^2 l_{1y}-k_y^2 l_{2y}-2 i k_y l_{1\tau}+2
   k_y l_{1y}^2+2 i k_y l_{2\tau}-2 k_y l_{2y}^2-4 i l_{1\tau}
   l_{2y}+4 l_{1y}^2 l_{2y}+4 i l_{1y} l_{2\tau}-4 l_{1y}
   l_{2y}^2\right) \\
   &\times\left(k_y^2
   l_{1y}-k_y^2 l_{2y}-2 i k_y l_{1\tau}+2 k_y l_{1y}^2-4
   k_y l_{1y} l_{2y}+2 i k_y l_{2\tau}+2 k_y l_{2y}^2+4 i
   l_{1\tau} l_{2y}-4 l_{1y}^2 l_{2y}-4 i l_{1y} l_{2\tau}+4 l_{1y}
   l_{2y}^2\right)\Big\},\\
\mc{S}^c_{2,-}\equiv& i(k_{y} -2 l_{2y})\Bigg/\Big\{8 k_y l_{2\tau}\left(-k_{y}^2 l_{1y}+k_{y}^2 l_{2y}+2 i
   k_{y} l_{1\tau}-2 k_{y} l_{1y}^2+4 k_{y} l_{1y} l_{2y}+2 i k_{y} l_{2\tau}-2 k_{y} l_{2y}^2-4 i l_{1\tau} l_{2y}+4 l_{1y}^2 l_{2y}+4 i l_{1y}
   l_{2\tau}-4 l_{1y} l_{2y}^2\right)\Big\},    \\
\mc{S}^c_{2,+}\equiv& i(k_{y} +2 l_{2y})\Bigg/\Big\{8 k_y l_{2\tau}\big(-k_{y}^2 l_{1y}+k_{y}^2 l_{2y}+2 i
   k_{y} l_{1\tau}-2 k_{y} l_{1y}^2-2 i k_{y} l_{2\tau}+2 k_{y} l_{2y}^2+4 i l_{1\tau} l_{2y}-4 l_{1y}^2 l_{2y}-4 i l_{1y} l_{2\tau}+4 l_{1y}
   l_{2y}^2\big)\Big\},\\    
\end{split}
\eeq
and, finally, for case $e$,
\beq
\begin{split}
\mc{S}_1^e\equiv& i(k_y+2 l_{1y})\Bigg/\Big\{4 (k_y l_{1\tau}-k_y l_{2\tau}+2 l_{1\tau} l_{2y}-2 l_{1y} l_{2\tau})\\
&\times\left(2 k_y^2 l_{1y}-2 k_y^2 l_{2y}+k_y^3+2 i k_y l_{1\tau}-4 k_y l_{1y} l_{2y}+2 i k_y l_{2\tau}-4 i l_{1\tau}
   l_{2y}+4 i l_{1y} l_{2\tau}\right)\Big\},\\
\mc{S}_2^e\equiv& (k_y-2 l_{2y})\Bigg/\Big\{4 \left(-k_y^2 l_{1y}+k_y^2
   l_{2y}-i k_y l_{1\tau}+2 k_y l_{1y} l_{2y}+i k_y l_{2\tau}-2 k_y l_{2y}^2+2 i l_{1\tau} l_{2y}-2 i l_{1y} l_{2\tau}\right)\\
   &\times\left(2 k_y^2 l_{1y}-2 k_y^2 l_{2y}+k_y^3+2 i k_y l_{1\tau}-4 k_y
   l_{1y} l_{2y}+2 i k_y l_{2\tau}-4 i l_{1\tau} l_{2y}+4 i l_{1y} l_{2\tau}\right)\Big\},\\
\mc{S}_0^e=& \mc{S}_1^e-\mc{S}_2^e\equiv -i(l_{1y}-l_{2y})\Bigg/\Big\{4 (k_y l_{1\tau}-k_y l_{2\tau}+2 l_{1\tau} l_{2y}-2 l_{1y} l_{2\tau})\\
&\times\left(-k_y^2 l_{1y}+k_y^2
   l_{2y}-i k_y l_{1\tau}+2 k_y l_{1y} l_{2y}+i k_y l_{2\tau}-2 k_y l_{2y}^2+2 i l_{1\tau} l_{2y}-2 i l_{1y} l_{2\tau}\right)\Big\},\\
\mc{S}^e_{2,-}\equiv& (-k_{y} + 2 l_{2y})\Bigg/\Big\{2 k_y \left(k_{y}^2+4 i l_{2\tau}-4 l_{2y}^2\right)\\
&\times\left(2 k_{y}^2 l_{1y}-2 k_{y}^2 l_{2y}+k_{y}^3+2 i k_{y} l_{1\tau}-4 k_{y} l_{1y} l_{2y}+2 ik_{y} l_{2\tau}-4 i l_{1\tau} l_{2y}+4 i l_{1y}
   l_{2\tau}\right)\Big\},    \\
\mc{S}^e_{2,+}\equiv& i(k_{y} +2 l_{2y})\Bigg/\Big\{4 k_y\left(k_y^2+4 i l_{2\tau}-4 l_{2y}^2\right)  \left(k_y l_{1\tau}-k_y l_{2\tau} -2 l_{1y}l_{2\tau}+2l_{1\tau}l_{2y}\right) \Big\}.    
\end{split}
\eeq

First consider case $a$ and $b$. Because of the theta pre-factors, only $\mc{S}^{a,b}_0$ contribute to UV divergence according to the {\it local divergence conjecture} in Section \ref{subsubapp: general strategies}. 
Setting $k_y=0$ everywhere in the theta factors except for an overall $\text{sgn}(k_y)$, 
we obtain
\beq\label{eq: 3String Master Equation ab}
\begin{split}
-\hat{T}_1\mathsf{S}_1^{a,b}+\hat{T}_2\mathsf{S}_2^{a,b}\approx
\left(-\hat{T_1}\theta\left(\frac{l_{1\tau}}{l_{1y}}\right) + \hat{T_2}\theta\left(\frac{l_{2\tau}}{l_{2y}}\right) + (\hat{T_1}-\hat{T_2})\theta\left(\frac{l_{1\tau}-l_{2\tau}}{l_{1y}-l_{2y}}\right)\right)\text{sgn}\left(\frac{l_{1y}l_{2\tau}-l_{1\tau}l_{2y}}{k_y(l_{1y}-l_{2y})}\right)\mc{S}^{a,b}_0.
\end{split}
\eeq
Again the real part of this expression is invariant if we simultaneously flip the sign of $k_y,l_{1y}$, $l_{2y}$ or $l_{1\tau}$, $l_{2\tau}$. Therefore let us assume $l_{1\tau}>0$ as well as $l_{1y}>0$. Then we have six regions where the expression is non-zero,
\beq
\begin{split}
\text{region 1:}~~~~~~l_{2\tau}>l_{1\tau}>0,~l_{1y}>l_{2y}>0,~~\int dp_\tau dq_\tau\left(-\hat{T}_1\mathsf{S}_1^{a,b}+\hat{T}_2\mathsf{S}_2^{a,b}\right)&=-(l_{2\tau}-l_{1\tau})^2\text{sgn}(k_y)\mc{S}_0^{a,b},\\
\text{region 2:}~~~~~~~~~~~~~~l_{2\tau}>l_{1\tau}>0,~l_{2y}<0,~~\int dp_\tau dq_\tau\left(-\hat{T}_1\mathsf{S}_1^{a,b}+\hat{T}_2\mathsf{S}_2^{a,b}\right)&=-l_{1\tau}^2\text{sgn}(k_y)\mc{S}_0^{a,b},\\
\text{region 3:}~~~~~~l_{1\tau}>l_{2\tau}>0,~l_{2y}>l_{1y}>0,~~\int dp_\tau dq_\tau\left(-\hat{T}_1\mathsf{S}_1^{a,b}+\hat{T}_2\mathsf{S}_2^{a,b}\right)&=(l_{2\tau}-l_{1\tau})^2\text{sgn}(k_y)\mc{S}_0^{a,b},\\
\text{region 4:}~~~~~~~~~~~~~~l_{1\tau}>l_{2\tau}>0,~l_{2y}<0,~~\int dp_\tau dq_\tau\left(-\hat{T}_1\mathsf{S}_1^{a,b}+\hat{T}_2\mathsf{S}_2^{a,b}\right)&=-l_{2\tau}^2\text{sgn}(k_y)\mc{S}_0^{a,b},\\
\text{region 5:}~~~~~~~~~~~~~~l_{2\tau}<0,~l_{2y}>l_{1y}>0,~~\int dp_\tau dq_\tau\left(-\hat{T}_1\mathsf{S}_1^{a,b}+\hat{T}_2\mathsf{S}_2^{a,b}\right)&=l_{1\tau}^2\text{sgn}(k_y)\mc{S}_0^{a,b},\\
\text{region 6:}~~~~~~~~~~~~~~l_{2\tau}<0,~l_{1y}>l_{2y}>0,~~\int dp_\tau dq_\tau\left(-\hat{T}_1\mathsf{S}_1^{a,b}+\hat{T}_2\mathsf{S}_2^{a,b}\right)&=-l_{2\tau}^2\text{sgn}(k_y)\mc{S}_0^{a,b}.\\
\end{split}
\eeq
After interchanging $l_{1\tau},l_{2\tau}$ as well as $l_{1y},l_{2y}$ simultaneously, we see that the expressions in region 1 and 3, 2 and 4, as well as 5 and 6 are identical to each other respectively, and therefore we can focus on region 1, 2 and 5. 
The analysis of case $c$ and $e$ are straightforward as well. Because of the theta pre-factors, $\mc{S}_{2,\pm}^{c,e}$ do not contribute to the divergence in the UV. 
Setting $k_y=0$ everywhere in the theta factors except for an overall $\text{sgn}(k_y)$, 
we obtain
\beq\label{eq: 3String Master Equation ce}
\begin{split}
-\hat{T}_1\mathsf{S}_1^{c,e}+\hat{T}_2\mathsf{S}_2^{c,e} \approx & \left(-\hat{T}_1\theta\left(\frac{l_{1\tau}}{l_{1y}}\right) + \hat{T}_2\theta\left(\frac{l_{2\tau}}{l_{2y}}\right) + (\hat{T}_1-\hat{T}_2)\theta\left(\frac{l_{1\tau}-l_{2\tau}}{l_{1y}-l_{2y}}\right)\right)\\
\times&\text{sgn}\left(l_{1y} l_{2\tau}-l_{1\tau} l_{2y}\right)\text{sgn}(k_y)\left(\theta(-l_{1y})\mc{S}_{1}^{c,e}-\theta(-l_{2y})\mc{S}_2^{c,e}-\theta(l_{1y}-l_{2y})\mc{S}_0^{c,e}\right).
\end{split}
\eeq
Restricting to the regime where $l_{1y}>0$ as well as $l_{1\tau}>0$, 
we have four different sections on the $l_{2y}-l_{2\tau}$ plane where the expression is non-zero,
\beq
\begin{split}
\text{region 1:}~~~~~~l_{2\tau}>l_{1\tau}>0,~l_{1y}>l_{2y}>0,~~\int dp_\tau dq_\tau\left(-\hat{T}_1\mathsf{S}_1^{c,e}+\hat{T}_2\mathsf{S}_2^{c,e}\right)&=(l_{2\tau}-l_{1\tau})^2\text{sgn}(k_y)\mc{S}_0^{c,e},\\
\text{region 2:}~~~~~~l_{2\tau}>l_{1\tau}>0,~l_{2y}<0,~~~~~~~~~~\int dp_\tau dq_\tau\left(-\hat{T}_1\mathsf{S}_1^{c,e}+\hat{T}_2\mathsf{S}_2^{c,e}\right)&=l_{1\tau}^2\text{sgn}(k_y)\mc{S}_1^{c,e},\\
\text{region 4:}~~~~~~l_{1\tau}>l_{2\tau}>0,~l_{2y}<0,~~~~~~~~~~\int dp_\tau dq_\tau\left(-\hat{T}_1\mathsf{S}_1^{c,e}+\hat{T}_2\mathsf{S}_2^{c,e}\right)&=l_{2\tau}^2\text{sgn}(k_y)\mc{S}_1^{c,e},\\
\text{region 6:}~~~~~~l_{2\tau}<0,~l_{1y}>l_{2y}>0,~~~~~~~~~~\int dp_\tau dq_\tau\left(-\hat{T}_1\mathsf{S}_1^{c,e}+\hat{T}_2\mathsf{S}_2^{c,e}\right)&=l_{2\tau}^2\text{sgn}(k_y)\mc{S}_0^{c,e}.\\
\end{split}
\eeq

The expression for single log divergence is very involved, but it is managable to calculate the double log divergence as sketched in Section \ref{subsubapp: general strategies}. 
First, we expand the whole expression in powers of $k_y$. 
As in the calculation of the Benz diagram,
we collect terms at the third order in the expansion, i.e. terms proportional to $k_y^2$. 
Note that $k_y$ mainly appears in two different places, in $\mc{S}$ factors as well as in $D(k/2+l_1)D(k/2-l_2)$\footnote{$k_y$ also appears in theta factors, but there it only contributes to a shift of integration regions in the IR that we can ignore for the calculation in the UV.}. 
To simplify expressions further, we are allowed to set $k_y=0$ inside the boson propagators except for the case $c$ and region 2 and 4 of case $e$, where we have to take into account the second term in the expansion of both $\mc{S}^{c,e}_1$ and $D(k/2+l_1)D(k/2-l_2)$ at small $k_y$, i.e. terms proportional to $k_y$ in two terms respectively
\footnote{Expanding $\mc{S}$ factors at small $k_y$, symbolically we have $\mc{S}\approx x_0 + x_1 k_y + x_2 k_y^2$. 
Moreover, $D(k/2+l_1)D(k/2-l_2)\approx d_0 + d_1 k_y + d_2 k_y^2$. We can explicitly check that the integration of $x_0 d_0$, i.e. the constant term, is cancelled among the six cases. The same calculation also shows that the integration of $x_0 d_2$ is cancelled among the six cases. What we now have are $x_2 d_0$ as well as $x_1 d_1$. We then notice that $x_1=0$ expect for $\mc{S}^c_{0,1}$ and $\mc{S}^e_1$.}.

For the sake of presentation, let us first identify the double log divergence coming from the dynamical region \eqref{eq: double log region}\footnote{Here we need to distinguish between integration region 1-6, and dynamical region on the boundary of respective integration region that contributes mostly to the double log divergence.}. We just need to collect terms at the third order of the expansion that are proportional to $1/l_\tau^4$ and contribute to the  IR divergence in the integration of $l_{1\tau},l_{2\tau}$. These terms only come from case $e$, hence we write down the terms proportional to $k_y$ in the series expansion of $\mc{S}^e$,
\beq\label{eq: 3String case e expansion}
\begin{split}
\mc{S}^e_0\Bigg|_{k_y~\text{linear term}} = & \left( \frac{l_{2y}^2 (l_{1y}-l_{2y})^3}{16(l_{1y}l_{2\tau}-l_{1\tau}l_{2y})^4} - \frac{(l_{1y}-l_{2y})(l_{1\tau}-l_{2\tau})^2}{64(l_{1y}l_{2\tau}-l_{1\tau}l_{2y})^4}\right)k_y^2,\\
\mc{S}^e_1 \Bigg|_{k_y~\text{linear term}} =&\left(\frac{l_{2y}^2 l_{1y}^3}{16(l_{1y}l_{2\tau}-l_{1\tau}l_{2y})^4} - \frac{l_{1y}(l_{1\tau}^2+l_{2\tau}^2)+2l_{2y}l_{1\tau}l_{2\tau}}{64(l_{1y}l_{2\tau}-l_{1\tau}l_{2y})^4}\right)k_y^2.\\
\end{split}
\eeq
We see that $k_y$-linear terms of $\mc{S}^e$ are  separated into two terms $\mc{S}^{e, (1)} + \mc{S}^{e, (2)}$. 
The first terms $\mc{S}^{e, (1)}$ of the two expansions are proportional to $1/l_\tau^4$ while the second terms $\mc{S}^{e, (2)}$ are not, and we isolate the first terms,
\beq\label{eq: isolated terms case e 3 String}
\begin{split}
\mc{S}^{e, (1)}_1&= \frac{l_{1y}^3 l_{2y}^2}{16(l_{1y}l_{2\tau}-l_{1\tau}l_{2y})^4}k_y^2,\\
\mc{S}^{e, (1)}_2&=\frac{l_{2y}^3 (l_{2y}^2-3l_{1y}l_{2y}+3l_{1y}^2)}{16(l_{1y}l_{2\tau}-l_{1\tau}l_{2y})^4}k_y^2,\\
\mc{S}^{e, (1)}_0&=\frac{l_{2y}^2 (l_{1y}-l_{2y})^3}{16(l_{1y}l_{2\tau}-l_{1\tau}l_{2y})^4}k_y^2.
\end{split}
\eeq
Just considering these terms coming from case $e$, we see that, to our surprise, Eq. \eqref{eq: 3String Master Equation ce} becomes identical to Eq. \eqref{eq: Benz master Equation} after interchanging $l_{1\tau},l_{1y}$ with $l_{2\tau},l_{2y}$. Namely, after substituting $\mc{S}^e$ in \eqref{eq: 3String Master Equation ce} with $\mc{S}^{e, (1)}$ and interchanging label $l_{1\tau},l_{1y}$ and $l_{2\tau},l_{2y}$, we have
\beq\label{eq: double log master Equation}
\begin{split}
\left(-\hat{T}_1\mathsf{S}_1^{e}+\hat{T}_2\mathsf{S}_2^{e}\right)^{(1)} = & \frac{\sgn(k_y)k_y^2}{16}\left(\hat{T_1}\theta\left(\frac{l_{1\tau}}{l_{1y}}\right) - \hat{T_2}\theta\left(\frac{l_{2\tau}}{l_{2y}}\right) + (\hat{T_2}-\hat{T_1})\theta\left(\frac{l_{1\tau}-l_{2\tau}}{l_{1y}-l_{2y}}\right)\right)\\
&\times\left(-\left(\theta(l_{1y})-\theta(l_{2y})\right)\text{sgn}\left(l_{1y} l_{2\tau}-l_{1\tau} l_{2y}\right)\frac{l_{1y}^2 l_{2y}^3}{(l_{1y}l_{2\tau}-l_{1\tau}l_{2y})^4}+ \left(l_1\rightarrow - l_1,~~~l_2\rightarrow l_2-l_1\right)\right),
\end{split}
\eeq
which is identical to the expression we obtain after substituting \eqref{eq: ky2 term for B2a} into \eqref{eq: Benz master Equation}, except for an extra factor $\frac{\sgn(k_y)}{16}$ which can be accounted for from different factors in front of the whole expression as shown in \eqref{eq: Benz results overview} and \eqref{eq: 3String results overview}. 
Moreover, both contribute to photon self-energy with an extra factor of $4$ as shown in \eqref{eq: Benz case assembling} and \eqref{eq: 3String case assembling}. 
Therefore, we calculate the double log divergence of the 3-String diagram  originating from the same source as the Benz diagram, i.e. coming from the dynamical region Eq. \eqref{eq: double log region},
\beq\label{eq: double log 3String}
\Pi_{\text{3String}}(k)\rightarrow\frac{|k_y|}{e^2}\frac{2\alpha^4}{3\pi^2}\left(\ln\left(\frac{\Lambda_y}{k_y}\right)\right)^2.
\eeq

After subtracting these terms out, now we aim to show that the double log divergences of the rest of the 3-String diagram cancel among the six cases. These double-log divergences that are canceled among themselves come from dynamical regions with large $l_{1\tau}$ or $l_{2\tau}$. 
The UV divergences from these high-energy modes can only renormalize local terms in the action.
As a result, both single-log and double-log, which require non-local counter terms,  are expected to be canceled.
One interesting conclusion we can draw from this surprising result is that the double log divergence actually does not appear at this order in the $\epsilon=0$ version of Ising-nematic transition in a metal \cite{Metlitski2010}, which is very closely related to the nFL theory we discussed here except $\lambda_+=\lambda_-=+1$. Here $\lambda_+=\lambda_-$ because the critical Ising-nematic order parameter, which is the analog of the field $a$ discussed here, couples to a fermion bilinear operator whose symmetry form factor requires $\lambda_+=\lambda_-$.

Let us do the calculation in region 4 of case $e$ as an example. 
To unearth the double-log divergence from the second term in the expansion \eqref{eq: 3String case e expansion}, we do the following change of parameters:
\beq
l_{2y}=-l_{1y}\delta,~~~~~~l_{1\tau} = l_{1y}^2\Delta_1,~~~~~~l_{2\tau} = l_{1y}^2\Delta_2,
\eeq
and integrate $|l_{1y}|$ from $k_y$ to $\Lambda_y$. The logarithmic divergent piece in this region from the second term is then written as follows,
\beq\label{eq: 3Sstring pre log example}
\begin{split}
\left(\Pi^{e}(k)\right)^{(2)}\Bigg|_{\text{region~4}} &\approx \frac{|k_y|}{(2\pi)^6}\int_0^\infty dl_{1y} \int_{-\infty}^0 d l_{2y} \int_0^\infty d l_{1\tau} \int _0^{l_{1\tau}} d l_{2\tau}~l_{2\tau}^2 \mc{S}_1^{e, (2)}D(k/2+l_1)D(k/2-l_2)D(l_2-l_1)\\ &=-\frac{|k_y|}{e^2}\frac{\alpha^4}{16\pi^2}\Bigg(\int\frac{d l_{1y}}{l_{1y}}\int_0^\infty d\delta\int_0^\infty d\Delta_1 \int_0^{\Delta_1} d\Delta_2~\frac{\Delta_2^2(\Delta_1^2+\Delta_2^2+2\delta\Delta_1\Delta_2)}{(\Delta_2 + \delta \Delta_1)^4}\\
&\times\frac{1}{1+\alpha\Delta_1}\frac{\delta}{\delta^2+\alpha\Delta_2}\frac{1+\delta}{(1+\delta)^2+\alpha(\Delta_1-\Delta_2)}\Bigg).
\end{split}
\eeq
The above integration is IR divergent when $\Delta_2$ and $\delta$ go to zero simultanesouly, which can be translated to the dynamical region where
\beq\label{eq: double log region another}
|l_{1y}|\gg |k_y|,~~~|l_{1\tau}|\gg |k_y l_{1y}|,~~~~~~ |l_{2y}| \approx |k_y|,~~~ |l_{2\tau}|\approx |k_y l_{1y}| .
\eeq
To simplify the expression further, we change $\Delta_2$ to $\Delta_2' \delta$, then at small $\delta$ the integrand is approximately
\beq
\frac{\Delta_2'\Delta_1^2}{\alpha(\Delta_1+\Delta_2')^4(1+\alpha\Delta_1)^2}\frac{1}{\delta},
\eeq
which gives the following double log divergence in the bracket
\beq
\frac{1}{6\alpha^2}\int_{k_y}^{\Lambda_y}\frac{dl_{1y}}{l_{1y}}\int_{k_y/l_{1y}}^1\frac{d\delta}{\delta}=\frac{1}{12\alpha^2}\left(\ln\left(\frac{\Lambda_y}{k_y}\right)\right)^2.
\eeq
The calculation in the other integration regions and other cases are as straightforward, and we see that the double-log divergent terms in region 2, 4, 5 and 6 of case $a$ and $b$ are all the same, i.e.
\beq
\Pi^{a,b}(k)\Bigg|_{\text{region~2,4,5,6}}\rightarrow\frac{|k_y|}{e^2}\frac{\alpha^2}{192\pi^2}\left(\ln\left(\frac{\Lambda_y}{k_y}\right)\right)^2,
\eeq
while the double log divergent terms in region 4 and 6 of case $c$ and the second term of $e$ are all the same, i.e.
\beq
\Pi^{c}(k), (\Pi^{e}(k))^{(2)}\Bigg|_{\text{region~4,6}}\rightarrow-\frac{|k_y|}{e^2}\frac{\alpha^2}{192\pi^2}\left(\ln\left(\frac{\Lambda_y}{k_y}\right)\right)^2,
\eeq
and in all the other regions there is no additional double log divergence. Adding them all up we see that the additional double log divergence cancels among the six cases, as promised.

It is worth pointing out that the calculation of the 3-String diagram relies on the ``local divergence conjecture'' we make in Section \ref{subsubapp: general strategies} as well as the correctness of the expansion trick with correct dynamically generated IR cutoff. It will be very interesting if a more rigorous calculation of these coefficients, either analytically or numerically, can be done. We defer this to future work.

\bibliography{supplemental.bib}